\documentclass{aa}

\usepackage{amsmath}
\usepackage{booktabs}
\usepackage{amssymb}
\usepackage[varg]{txfonts}
\usepackage{booktabs}
\usepackage{array}
\usepackage{graphicx}
\usepackage{nicefrac}
\usepackage{dcolumn}
\usepackage{natbib,twoopt}
\usepackage[normalem]{ulem} 
\newcolumntype{L}{D{+}{\,\pm\,}{1,1}}
\usepackage{hyperref, xcolor}
\hypersetup{colorlinks, citecolor=blue}
\bibpunct{(}{)}{;}{a}{}{,} 
\newcommandtwoopt{\citeads}[3][][]{\href{http://adsabs.harvard.edu/abs/#3}%
{\citealp[#1][#2]{#3}}}
\newcommandtwoopt{\citepads}[3][][]{\href{http://adsabs.harvard.edu/abs/#3}%
{\citep[#1][#2]{#3}}}
\newcommandtwoopt{\citetads}[3][][]{\href{http://adsabs.harvard.edu/abs/#3}%
{\citet[#1][#2]{#3}}}
\newcommandtwoopt{\citeyearads}[3][][]%
{\href{http://adsabs.harvard.edu/abs/#3}{\citeyear[#1][#2]{#3}}}

\newcommand{\HI}{\textsc{Hi }}
\newcommand{\HII}{\textsc{Hii }}

\begin{document}

\title{The magnetized disk-halo transition region of M51
\thanks{The reduced FITS images of this paper are available in electronic form at the CDS via anonymous ftp to cdsarc.u-strasbg.fr (130.79.128.5) or via http://cdsweb.u-strasbg.fr/cgi-bin/qcat?J/A+A/}
}
\author{M.~Kierdorf{\inst{\ref{inst1}}
\thanks{\email{kierdorf@mpifr-bonn.mpg.de}}}
\and S.~A.~Mao\inst{\ref{inst1}}
\and R.~Beck\inst{\ref{inst1}}
\and A.~Basu\inst{\ref{inst2}}
\and A.~Fletcher\inst{\ref{inst3}}
\and C.~Horellou\inst{\ref{inst4}}
\and F.~Tabatabaei\inst{\ref{inst5}}
\and J.~Ott\inst{\ref{inst6}}
\and M.~Haverkorn\inst{\ref{inst7}}
}

\institute{Max-Planck-Institut f\"ur Radioastronomie, Auf dem H\"ugel 69, 53121 Bonn, Germany\label{inst1}
\and Fakult\"at f\"ur Physik, Universit\"at Bielefeld, Postfach 100131, 33501 Bielefeld, Germany\label{inst2}
\and School of Mathematics, Statistics and Physics, Herschel Building, Newcastle University, Newcastle-upon-Tyne NE1 7RU, U.K.\label{inst3}
\and Dept. of Space, Earth and Environment, Chalmers University of Technology, Onsala Space Observatory, 439 92 Onsala, Sweden\label{inst4}
\and School of Astronomy, Institute for Research in Fundamental Sciences, P.O. Box 19395-5531, Tehran, Iran\label{inst5}
\and National Radio Astronomy Observatory, 1003 Lopezville Road, Socorro, NM 87801, USA\label{inst6}
\and Department of Astrophysics/IMAPP, Radboud University Nijmegen; P.O. Box 9010, 6500 GL Nijmegen, Netherlands\label{inst7}
}

\date{Received 28 February 2020 / Accepted 24 June 2020}
\titlerunning{The magnetized disk-halo transition Region of M51}
\authorrunning{M. Kierdorf et al.}

\abstract
{
The grand-design face-on spiral galaxy M51 is an excellent laboratory for studying magnetic fields in galaxies. 
Due to wavelength-dependent Faraday depolarization, linearly polarized synchrotron emission at different radio frequencies yields a picture of the galaxy at different depths: observations in the L-band (1\,--\,2\,GHz) probe the halo region, while 
at 4.85\,GHz (C-band) and 8.35\,GHz (X-band),
the linearly polarized emission mostly emerges from
the disk region of M51. We present new observations of M51 using the Karl G. Jansky Very Large Array (VLA) at the intermediate frequency range of the S-band (2\,--\,4\,GHz), where previously no high-resolution broadband polarization observations existed, to shed new light on the transition region between the disk and the halo. 

We present the S-band radio images of the distributions of the total intensity, polarized intensity, degree of polarization, and rotation measure (RM).
The RM distribution in the S-band shows a fluctuating pattern without any apparent large-scale structure.
We discuss a model of the depolarization of synchrotron radiation in a multi-layer magneto-ionic medium and compare the model predictions to the multi-frequency polarization data of M51 between 1\,--\,8\,GHz. The model makes distinct predictions of a two-layer (disk – halo) and three-layer (far-side halo – disk – near-side halo) system. Since the model predictions strongly differ within the wavelength range of the S-band, the new S-band data are essential for distinguishing between the different systems.
A two-layer model of M51 is preferred.

The parameters of the model are adjusted to fit to the data of polarization fractions in a few selected regions.
In three spiral arm regions, the turbulent field  in the disk dominates with strengths between $18\,\mu$G and $24\,\mu$G, while the regular field strengths are $8-16\,\mu$G. In one inter-arm region, the regular field strength of $18\,\mu$G exceeds that of the turbulent field of $11\,\mu$G.
The regular field strengths in the halo are $3-5\,\mu$G.
The observed RMs in the disk-halo transition region are probably dominated by tangled regular fields, as predicted from models of evolving dynamos, and/or
vertical fields, as predicted from numerical simulations of Parker instabilities or galactic winds. Both types of magnetic fields have frequent reversals on scales similar to or larger than the beam size ($\sim550$\,pc) that contribute to an increase of the RM dispersion and to distortions of any large-scale pattern of the regular field.
Our study
devises new ways of analyzing and interpreting broadband multi-frequency polarization data that will be applicable to future data from, for example, the Square Kilometre Array.
}


\maketitle

\section{Introduction}
\label{sec:intro}

Magnetic fields play an important role in the formation and evolution of spiral galaxies (e.g., \citealt{2009ApJ...696...96W,2018MNRAS.473.4077P,2019MNRAS.489.3368B}), but knowledge of their structure, strength, and origin remains limited.
Large-scale patterns of ordered magnetic fields have been observed in multiple nearby spiral galaxies. In face-on galaxies the magnetic field structure shows a spiral pattern, usually following the gaseous spiral arms \citep[e.g.,][]{2016A&ARv..24....7B,2013pss5.book..641B}. In edge-on galaxies, magnetic fields in the disk are observed mostly parallel to the disk plane,
while vertical components are found in the halo
(e.g., \citealt{2012SSRv..166..133H,2015AJ....150...81W,2019Galax...7...54K,2020arXiv200414383K}).

The exchange of material between disk and halo seems to be a crucial process in the evolution of spiral galaxies.
The interaction is believed to be driven by gas flows from so-called galactic fountains (e.g., \citealt{1976ApJ...205..762S,1980ApJ...236..577B}). 
Large-scale halo fields could result from advection of disk fields into the halo, for example, via winds \citep[e.g.,][]{1991A&A...245...79B,1995A&A...297...77E, 2018MNRAS.481.4410P,2020MNRAS.494.4393S}, or from a dynamo operating in the halo \citep{1990Natur.347...51S, 2010A&A...512A..61M,2010A&A...514A..42B}. Due to the lack of simultaneous measurements of both disk and halo field structures in galaxies, the origin of large-scale halo fields and how they are connected to the underlying galactic disk remains poorly understood.


Radio polarization observations are ideal to study the structure of magnetic fields in the interstellar medium (ISM) of spiral galaxies.
Synchrotron emission traces the magnetic field component B$_{\perp}$ in the plane of the sky, perpendicular to the line-of-sight.
To obtain a three-dimensional picture of the magnetic field, 
the effect of Faraday rotation 
can be used to 
infer 
the magnetic field component along the line-of-sight, B$_{\parallel}$. The plane of polarization of an electromagnetic wave is rotated when the wave passes through a magnetized plasma containing thermal electrons: 
\begin{equation}\label{eq:PA}
\psi=\psi_0 + \text{RM}\,\lambda^2 \, ,
\end{equation}
where $\psi$ is the measured polarization angle at wavelength $\lambda$, 
$\psi_0$ is the polarization angle of the emitted electromagnetic wave, and RM is the rotation measure. 
RM is related to the Faraday depth $\Phi$, a physical quantity dependent on the line-of-sight integral of the thermal electron density, $n_\text{e}$, 
times 
the magnetic field component along the line-of-sight, $B_\parallel$:
\begin{equation}\label{eq:RM}
\left(\frac{\Phi}{\text{rad\,m}^{-2}}\right) = 0.81\int\displaylimits^\text{observer}_\text{source}\left(\frac{n_\text{e}}{\text{cm}^{-3}}\right)\,\left(\frac{B_\parallel}{\mu\text{G}}\right)\,\left(\frac{\text{d}L}{\text{pc}}\right) \, ,
\end{equation}
where $\text{d}L$ denotes the infinitesimal path length through the Faraday-rotating medium. By convention, $\Phi$ is positive (negative) for magnetic fields pointing towards (away from) the observer. 
For a simple Faraday-rotating screen located between the synchrotron-emitting source and the observer, RM is equivalent to $\Phi$. 
There are, however, more complicated cases 
(e.g., \citealt{2012MNRAS.421.3300O}; \citealt{2016ApJ...820..144A}; \citealt{2019MNRAS.487.3432M}), 
such as several mixed synchrotron-emitting and Faraday-rotating media located within the volume traced by the telescope beam.

The degree of polarization, $p$, given by the ratio of the polarized intensity to the total intensity of the synchrotron emission, is a measure of
the degree of order of the magnetic field.
The observed degree of polarization can be attenuated by depolarization mechanisms: (1) beam depolarization decreases the polarized signal due to tangled magnetic fields on scales smaller than one resolution element of the observing instrument; (2) in bandwidth depolarization, the plane of polarization is rotated by different angles at different frequencies within the observing frequency band. Averaging over the frequency band entails the reduction of the polarized signal; 
(3) and by wavelength-dependent depolarization due to Faraday rotation intrinsic to the source and/or along the line-of-sight. 
One differentiates between differential Faraday rotation and external and internal Faraday dispersion \citep{Sokoloff98}. 
With broadband polarization data, wavelength-dependent Faraday depolarization can be used as a powerful probe of the 3-D structure of magnetic fields in galaxies (``Faraday tomography'').




The grand-design face-on spiral galaxy M51 provides an excellent laboratory to simultaneously probe its disk and halo\,\footnote{We adopt the notification `halo' for a physical layer between the synchrotron emitting disk and the observer (containing baryonic matter -- not to be confused with a dark matter halo).} fields using wideband polarimetry.
Polarization studies of M51 have shown that different configurations of the large-scale regular magnetic field exist in the disk (probed by observations at 4.85 and 8.35\,GHz, \citealt{Berk97,Fletcher11}) and in the halo (probed by observations at 1\,GHz, e.g., \citealt{1992A&A...265..417H,1996ASPC...97..592N,2015ApJ...800...92M}). 
According to \citet{Fletcher11}, the large-scale regular field in the disk is best described by a superposition of two azimuthal modes (axisymmetric plus quadrisymmetric, $m\,=\,0$ and 2),
whereas the halo field has a strong bisymmetric azimuthal mode ($m\,=\,1$). 
This difference in the magnetic field configurations in the disk and the halo of M51 is still an unresolved mystery. 
 
\citet{Fletcher11} suggested that interactions with M51's companion galaxy NGC\,5195 could be responsible for the configuration in the halo by driving a different mean-field $\alpha$-$\Omega$ dynamo, for example, through tidal forces. Another possibility is that the halo field could have been generated during early evolutionary stages of M51 in the disk and later transported into the halo, while a different dynamo action built the present-day disk field. 
Independent dynamo action in disk and halo of the same galaxy is possible, under the condition that the disk and halo fields are generated by a mean-field $\alpha$-$\Omega$ dynamo that is based on differential rotation and turbulence \citep{1990Natur.347...51S}. However, \citet{2010A&A...512A..61M} argued that in the presence of a galactic wind the halo component of the field may ``enslave'' that of the disk, making different field patterns improbable.


A better understanding of M51's mysterious magnetic field configuration in the disk and halo and the underlying dynamo mechanism(s) will come from observations of the transition region, which are provided by observations at an intermediate frequency range. 
Our new broadband S-band polarization data fill the frequency gap between data obtained by the Karl G. Jansky Very Large Array (VLA) at the L-band (1\,--\,2\,GHz) by \citet{2015ApJ...800...92M} probing the halo, and the C-band (4.85\,GHz) and the X-band (8.35\,GHz) data (VLA + Effelsberg) by \citet{Fletcher11} probing mostly the disk. With the combined high quality and broad frequency coverage dataset we are able to investigate the magneto-ionic properties in different layers of M51. 


One of the main motivations of this work was to compare the observed degree of synchrotron polarization in the S-band to an analytical multi-layer depolarization model
with different magnetic field configurations developed by \citet{Shneider14}. 
This model is more advantageous compared to `classical' depolarization models, which only handle depolarization in a single-layer system (e.g., \citealt{2012MNRAS.421.3300O, 2016ApJ...820..144A}).
\citet{Shneider14} modeled the degree of polarization as a function of wavelength assuming different magnetic field configurations to be present in different layers along the line-of-sight in terms of regular, isotropic turbulent, and anisotropic turbulent fields (for a detailed description of the nomenclature please see \citealt{2019Galax...8....4B}). 
Since the model predictions strongly differ within the wavelength range of the S-band, our new data are crucial to evaluate whether the model predictions are adequate.

We present new Karl G. Jansky Very Large Array (VLA) S-band observations of M51 in total and polarized intensity. In Section\,\ref{sec:obs}, we present details of the observations and data reduction. Section\,\ref{sec:resultsI} gives an overview on the total intensity results. In Section\,\ref{sec:resultsPI}, we summarize the results of polarization analysis (we present maps of polarized intensity, degree of polarization, and rotation measure). In Section\,\ref{sec:modelSection}, we compare the observed degree of polarization across a frequency range of 1\,--\,8\,GHz (a wavelength range of $\lambda$\,3\,--\,30\,cm) to the \citet{Shneider14}  model.
In Section\,\ref{sec:discussion_main}, we discuss our main physical findings, while Section\,\ref{sec:summary} summarizes the paper including an overview on future prospects. 
Most of the material contained in this paper was published in a PhD thesis at the University of Bonn \citep{KierdorfThesis2019}. Preliminary results
were published in \citet{2018arXiv181003638K}.

We adopt a distance to M51 of 7.6\,Mpc \citep{2002ApJ...577...31C}, an inclination of the disk of $l=-20\degr$ ($0\degr$ is face-on),
and a position angle of the disk's major axis of $PA=-10\degr$ \citep{1974ApJS...27..437T}.

\section{Observations and data reduction}
\label{sec:obs}

The observations in the S-band (2\,--\,4\,GHz) were performed using the VLA operated by the National Radio Astronomy Observatory (NRAO) in New Mexico, USA. The observations were done in C- and D-array configuration in November and December 2014, and in October 2015. We observed M51 in full polarization, while the sources J1313+5458 and J1407+2827 (assumed to be unpolarized) were observed for phase and polarization leakage calibration, respectively. The calibrator 3C\,286 was used as the total flux density scale and polarization angle calibrator \citep{2013ApJS..204...19P, 2013ApJS..206...16P} with a polarization angle of $+33\degr$ across the S-band.
The measurement sets contain 16 spectral windows, each with 64  2-MHz channels. M51 was observed two times 90 minutes in C-configuration and 90 minutes in D-configuration to fill the missing spacings in the C-configuration data. The resolution of the concatenated observations is $10''\times7''$ at 3\,GHz (using robust weighting). At the assumed distance of M51, 1$''$ corresponds to a linear scale of about 37\,pc. Therefore, one beam of $10''\times7''$ has a linear size of about 370$\,\times\,260$\,pc.
Observational parameters are summarized in Table\,\ref{tab:obs_summary}.

\begin{table}[t]
\centering
\caption{Radio continuum observational parameters of M51.}
  \begin{tabular}{lccc}
    \toprule\toprule
Frequency (GHz)				& 2\,--\,4 (2.6\,--\,3.6 after flagging)	\\
Bandwidth (MHz)				& 2000 (1000 after flagging)		\\
No. of spectral windows			& 16	(9 after flagging)		\\
Spectral resolution (MHz)			& 2	\\
Central frequency (GHz)			& 3.06	\\
Array configurations				& C; D	\\
Observing time	(minutes)			& 180; 90	\\
Observing dates				& 26 Nov/14 Dec 2014; 	\\
							& 09/10 Oct 2015\\
Total flux density calibrator		& 3C\,286 \\
Polarization angle calibrator		& 3C\,286 \\
\midrule\midrule
Resolution in final maps			&$10''\times7''$ \qquad 15$''$ \\
rms in Stokes $I$ 			&\phantom{0} 30 \qquad \phantom{001}60  \\
($\mu$Jy\,beam$^{-1}$)			&\\
rms in Stokes $Q$ and $U$	& \phantom{00}6   \qquad \phantom{0001}9\\
($\mu$Jy\,beam$^{-1}$)			&\\
    \bottomrule
  \end{tabular}
  \label{tab:obs_summary}
  \end{table}

	\begin{figure*}[htbp]
	\centering
\includegraphics[scale=0.48, trim=1.0cm 7.5cm 4cm 1cm, clip]{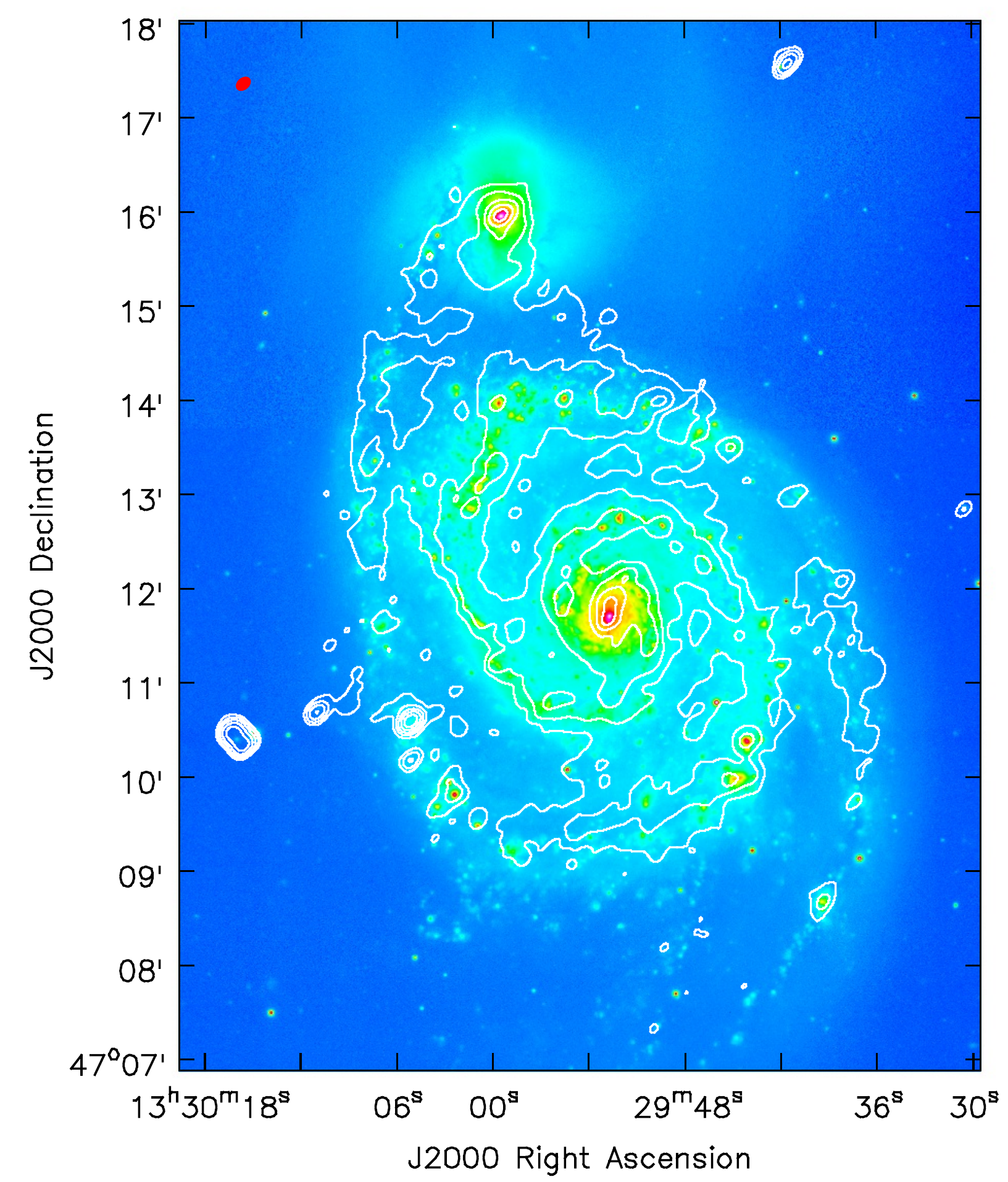}\hspace{5mm}\includegraphics[scale=0.58, trim=1.5cm 10cm 3.8cm 2.cm, clip]{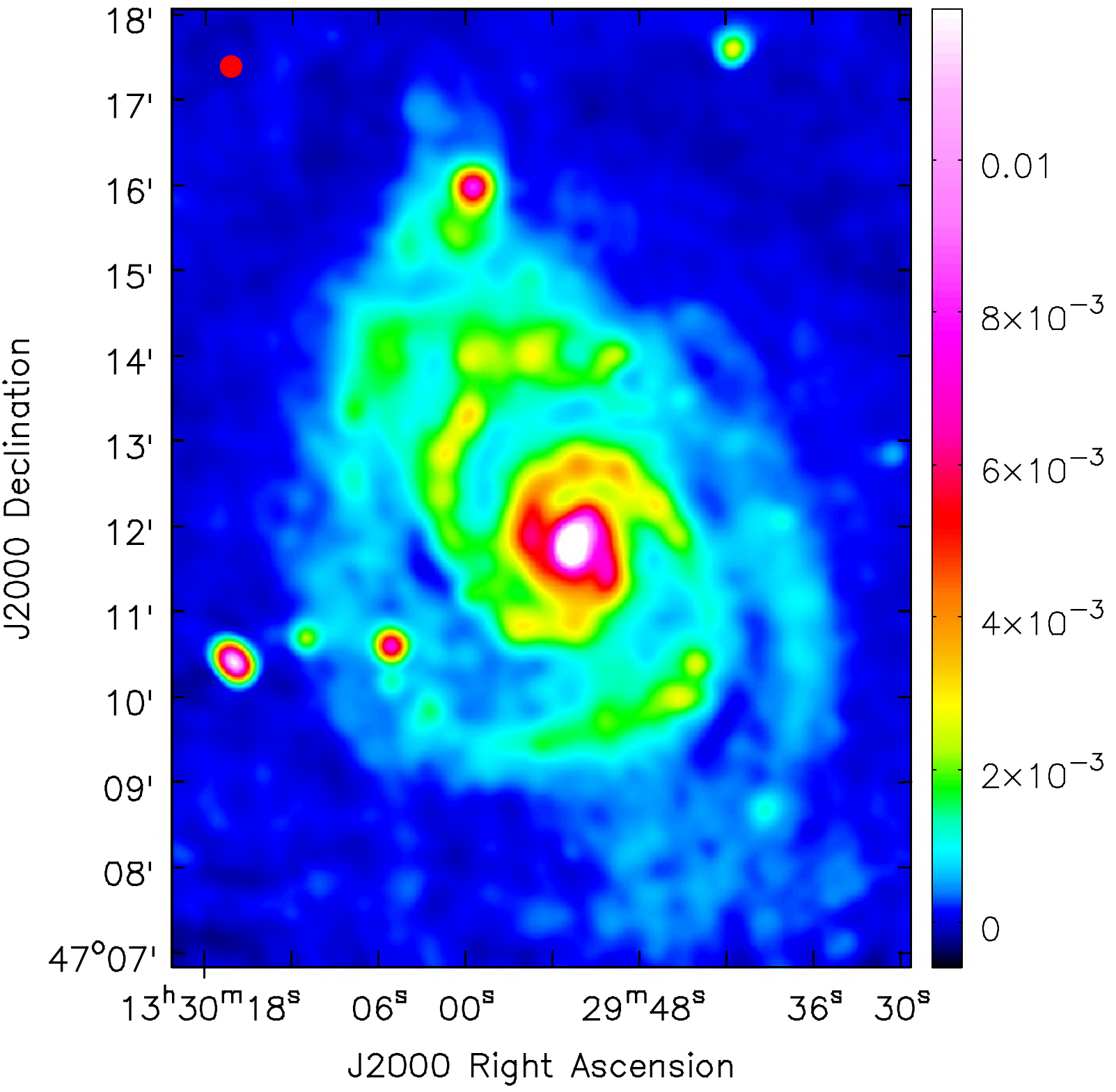}
\caption{Total intensity image of M51 at 3\,GHz with a resolution of $10''\times7''$, shown as contours, overlaid onto a H$\alpha$ image \citep{2003PASP..115..928K} (left) and with a resolution of 15$''$ in color scale in units of Jy\,beam$^{-1}$ (right). The contours are drawn at [8, 16, 32, 64, 128, 256]$\,\times\,30\,\mu$Jy\,beam$^{-1}$ (left). The beam size is shown in the top left corner. We note that this is the Stokes $I$ image from only one spectral window that has a bandwidth of about 100\,MHz (see Section\,\ref{sec:resultsI} for explanation).}
	\label{fig:StokesI}
	\end{figure*}

The calibration and data reduction were done using the NRAO Common Astronomy Software Applications\,\footnote{\url{http://casa.nrao.edu}} (\textsf{CASA}) package \citep{2007ASPC..376..127M}. 
To maximize the effectiveness of automatic flagging, a preliminary bandpass calibration was applied to the data of the flux density calibrator. 
After automatic flagging using \textsf{RFlag} and flagging the beginning and end of each spectral window due to decreasing sensitivity towards  the edges, the visibilities of the calibrators and M51 were carefully inspected for further RFI excisions manually.  After flagging, the usable frequency band was reduced to 1000\,MHz (2.56\,--\,3.56\,GHz) with a central frequency of 3.06\,GHz. 
After an a priori antenna position correction, the right flux density scale of the total intensity calibrator 3C\,286 of 10.9\,Jy at 2.565\,GHz \citep{2013ApJS..204...19P} was set into the model column of the calibrator measurement set using the task \texttt{setjy}. The error of the calibration procedure is 0.2\%. The uncertainty of the flux density of 3C286 of $\pm\,1\%$ \citep{2013ApJS..204...19P} adds to the calibration error.
Self-calibration for phase only was performed to the target visibilities to improve the image quality in terms of less imaging artifacts and lower rms noise (by more than a factor of 10). 
Images in Stokes $I$, $Q$, and $U$ were created using the \texttt{clean} algorithm in \textsf{CASA} \citep{1974A&AS...15..417H} with a cell size of 1$''$. We used Briggs weighting with a robust parameter of 0.0 \citep{1995AAS...18711202B}. 
Multi-scale cleaning was applied to decompose the emission in the field of view into scales with different angular sizes \citep{2008ISTSP...2..793C}. We used scales ranging from 0$''$ (which corresponds to point sources) over 6$''$ (which corresponds to the size of about one beam) to 200$''$, which is about half of the size of the galaxy at 3\,GHz\,\footnote{We used scales of 0, 6, 9, 18, 30, 45, 60, 100, and 200 arcsec.}. 
The multi-term multi-frequency synthesis algorithm by \citet{2011A&A...532A..71R} was attempted using two Taylor terms (using nterms\,$=\,2$ in \texttt{clean}). However, this method degraded the total integrated flux density presumably due to too steep in-band spectral indices computed by the algorithm. 
This was also found in other observational studies using VLA S-band data  \citep[e.g.,][]{2015arXiv150205616C,2017MNRAS.464.1003B,2017A&A...600A...6M}. 
Due to the attenuated total flux density in the total band image, we used the Stokes $I$ map of the central spectral window at 3.05\,GHz instead of the multi-frequency synthesized Stokes $I$ image for further analysis (for example, for computing the map of the degree of polarization). 

\section{Total intensity analysis}

\subsection{Total intensity and in-band spectral index}
\label{sec:resultsI}

\begin{table}[t]
  \centering
\caption{Integrated total radio continuum flux densities of M51.}
  \begin{tabular}{lccc}
    \toprule\toprule
Frequency		&Flux density		& Reference\\
GHz     			&Jy				& \\ 
  \midrule
0.151	&  8.1\,$\pm$\,0.6		&  \citet{2014AA...568A..74M}\\
0.408	&  3.5\,$\pm$\,0.1		&  \citet{1980AAS...41..329G}\\
0.610	&  2.63\,$\pm$\,0.06		&  \citet{1977AA....54..703S}\\
1.4		&  \phantom{00}1.4\,$\pm$\,0.1\phantom{00}		&  \citet{2011AJ....141...41D}\\
2.6		&  0.771\,$\pm$\,0.05\phantom{0}	&  \citet{1984AA...135..213K}\\
2.56 		& 0.822\,$\pm$\,0.002 	&$I^{\text{spw}}$ (this work)\\
2.69 		& 0.779\,$\pm$\,0.002 	&$I^{\text{spw}}$ (this work)\\
2.82 		& 0.759\,$\pm$\,0.002 	&$I^{\text{spw}}$ (this work)\\
2.95 		& 0.731\,$\pm$\,0.002 	&$I^{\text{spw}}$ (this work)\\
3.05 		& 0.703\,$\pm$\,0.001 	&$I^{\text{spw}}$ (this work)\\
3.18 		& 0.688\,$\pm$\,0.001 	&$I^{\text{spw}}$ (this work)\\
3.31 		& 0.661\,$\pm$\,0.002 	&$I^{\text{spw}}$ (this work)\\
3.43 		& 0.644\,$\pm$\,0.002 	&$I^{\text{spw}}$ (this work)\\
3.56 		& 0.628\,$\pm$\,0.002 	&$I^{\text{spw}}$ (this work)\\
4.85		&  0.420\,$\pm$\,0.080	&  \citet{2009ApJ...693.1392S}\\
8.35		&  0.308\,$\pm$\,0.103	&  \citet{2011AJ....141...41D}\\
10.7		&  0.241\,$\pm$\,0.014	&  \citet{1981AA....94...29K}\\
14.7		&  0.190\,$\pm$\,0.020	&  \citet{1984AA...135..213K}\\
22.8		&  0.142\,$\pm$\,0.015	&  \citet{1984AA...135..213K}\\
    \bottomrule
  \end{tabular}
   \tablefoot{The listed flux densities are plotted in Figure\,\ref{fig:spectrum}. The errors reported for our integrated flux densities in the S-band are given by the noise contribution in the individual images.
   The calibration error of about 1\% is the same for all spectral windows and is not taken into account here.
   $I^{\text{spw}}$ stands for the total intensity obtained from the spectral window (spw) images. 
   }
   \label{tab:spectrum}
 \end{table}

The total intensity images of the central spectral window of the S-band at 3.05\,GHz (with a bandwidth of about 100\,MHz) with $10''\times7''$ and 15$''$ resolution (which corresponds to a physical scale of about 370$\,\times\,260$\,pc and 550\,pc, respectively) are shown in Figure\,\ref{fig:StokesI}. The left panel shows the total intensity as 
contours, overlaid onto the 
H$\alpha$ image of \citet{2003PASP..115..928K}. The right panel shows the total intensity at 15$''$ in rainbow color scale. The 15$''$ image is used for the scientific analysis in this paper.  
The two spiral arms and the irregular dwarf companion galaxy NGC\,5195 at the northern end of M51 are well discernible in both images. The high resolution Stokes $I$ emission shows a close correspondence with the optical spiral arms and central region of M51 as already discussed in detail in \cite{Fletcher11}. Also detailed structures of the gas, for example compact \HII regions, are visible. The lower resolution image shows emission at slightly larger radii, due to better signal-to-noise ratio.

\begin{figure}[t]
\centering
\includegraphics[scale=0.49, trim=0.5cm 0cm 0cm 0cm, clip]{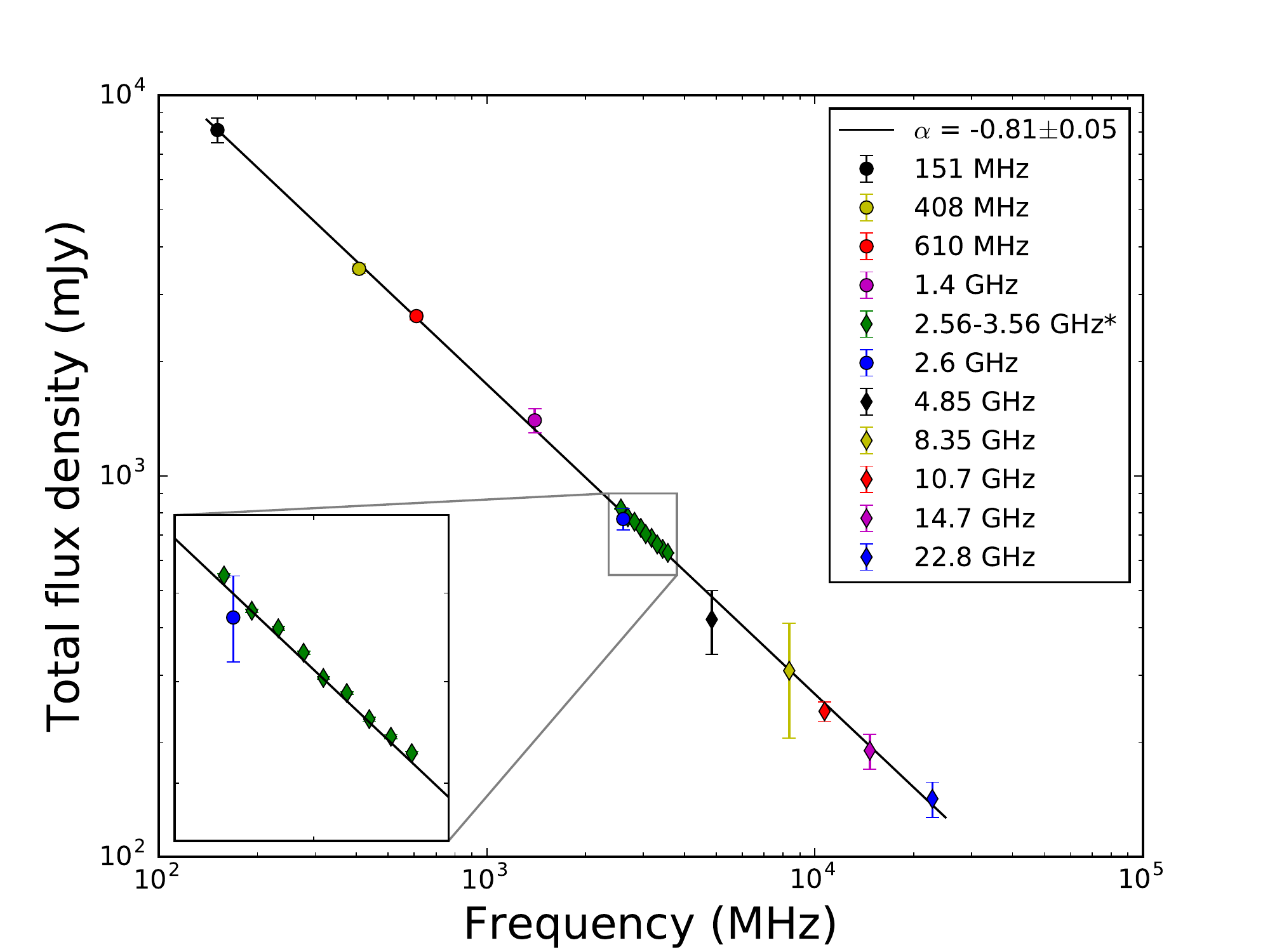}
  \caption{Total integrated radio continuum spectrum of M51 with a fitted power law, giving a total spectral index $\alpha_{\text{tot}}=-0.81 \pm 0.05$. The flux densities and references are listed in Table\,\ref{tab:spectrum}. In the bottom left corner, a zoom in to the S-band frequency range with the integrated flux densities of the spectral window images is shown. The data points marked with * in the legend are from this work.
  } 
\label{fig:spectrum}
  \end{figure}

The integrated total flux density of M51 amounts to $703 \pm 1$\,mJy at 3.05\,GHz.
Table\,\ref{tab:spectrum} lists flux densities of M51 observed between 151\,MHz and 22.8\,GHz and the corresponding references. Figure\,\ref{fig:spectrum} shows the total integrated radio continuum spectrum of M51 between 151\,MHz and 22.8\,GHz. 
The green diamonds in Figure\,\ref{fig:spectrum} show the integrated total intensity from the nine spectral window images across the S-band. 
The flux densities of the spectral window images are in excellent agreement with the power-law fit performed using the archival Stokes $I$ data at multiple frequencies, where the spectral index $\alpha$ is defined as $I_\nu\,\propto\,\nu^\alpha$. This shows that our observations recover the right flux density level, and thus the data do not suffer from missing large-scale flux densities due to lack of short antenna spacings. 
The rms noise level in the total intensity spectral window image at 3.05\,GHz amounts to about 30\,$\mu$Jy\,beam$^{-1}$ at $10''\times7''$ resolution and 60\,$\mu$Jy\,beam$^{-1}$ at 15$''$ resolution.

\subsection{Separation of thermal and non-thermal emission}
\label{sec:thermal}

In star-forming galaxies, such as M51, the radio
continuum emission originates from a mix of synchrotron (non-thermal) and thermal free-free emission. The relative
contribution of the free-free emission increases towards higher
frequencies making it important to subtract its 
contribution from the total intensity at frequencies of our
observations for the determination of the degree of
polarization of synchrotron emission. To estimate the free-free emission in spatially
resolved M51, we use the mid-infrared emission
as its tracer.
Following \citet{murph08}, the free-free flux density
($S_{\text{th},\nu}$) at a frequency $\nu$ is related to the
flux density at $24\,\mu$m as
\begin{equation}
	\left(\frac{S_{\text{th},\nu}}{\text{Jy\,beam}^{-1}}\right) = 7.93 \times 10^{-3}\,
	\left(\frac{T_{\text{e}}}{10^4\,\text{K}} \right)^{0.45}\,\left(\frac{\nu}{\text{GHz}}\right)^{-0.1}\,
	\left(\frac{S_{24\,\mu\text{m}}}{\text{Jy\,beam}^{-1}}\right).
	\label{eq:thermal}
\end{equation}
Here, $T_{\text{e}}$ is the electron temperature, assumed to be $10^4$\,K.


We used the {\it Spitzer}-MIPS $24\,\mu$m map of M51, observed as a part
of the SINGS \citep{kenni03}. This map is available at $6''$
angular resolution in units of MJy\,sr$^{-1}$. We first converted the
$24\,\mu$m emission to Jy\,beam$^{-1}$ units and then applied
Equation\,\eqref{eq:thermal} to obtain the map of free-free emission at the desired
frequency, in our case, at 3.05\,GHz. The free-free emission map was then subtracted from the total intensity map
(Figure\,\ref{fig:StokesI}, right-hand panel) by aligning it
to the same coordinate system and convolving it to $15''$.

Using this method,
the galaxy-integrated thermal fraction\,\footnote{The thermal fraction at frequency $\nu$ is defined as $f_{\rm th, \nu} = S_{\rm th, \nu}/S_{\rm tot, \nu}$, where $S_{\rm tot, \nu}$ is the total radio continuum flux density.} at 3\,GHz ($f_{\rm th,3\,GHz}$) of M51 is found to be $0.14 \pm 0.01$, which
corresponds to $f_{\rm th,1\,GHz} = 0.11 \pm 0.07$. This is consistent with
\citet{klein18} and locally agrees with the $f_{\rm th, 1\,GHz}$ map of
\citet{heese14} within about 10\% in bright star-forming regions
and $\lesssim5\%$ in the low surface brightness inter-arms. With a relative error\,\footnote{The relative
error of $f_{\rm th}$ is defined as $a = \,\pm\,\text{d}f_{\rm th}\,/f_{\rm
th}$, where $\text{d}f_{\rm th}$ is the absolute error of $f_{\rm th}$.} of
$\pm\,a$ in the estimated $f_{\rm th}$, the relative error in the synchrotron
emission fraction ($\Delta\,f_{\rm nth}$, where $f_{\rm nth} = 1 - f_{\rm th}$)
is given by $\Delta f_{\rm nth} = [1 - (1\,\pm\,a)\,f_{\rm th}]\,/\,[1 - f_{\rm th}]$
\citep{basu17}. This means that an error of up to 20\% ($a = 0.2$) in a region with
$f_{\rm th} = 0.4\,(0.2)$ yields a relative error in the estimated synchrotron emission of about 15\,(5)\%. This will not significantly affect 
the 
results presented in the rest of this paper.

The $24\,\mu$m emission traces mostly the ionized gas in star-forming regions. The estimate of the thermal radio emission based on the $24\,\mu$m emission can be uncertain
by a factor as large as two on kpc scales \citep{2012AJ....144....3L}, especially in the inter-arm regions and in the outer disk. However, even such a large uncertainty would modify the degrees of non-thermal polarization in the S-band only within the typical measurement errors.

\section{Polarization analysis}
\label{sec:resultsPI}
\subsection{RM-Synthesis application}
\label{sec:Bpara-synth}

To obtain the maps in polarized intensity ($PI$), polarization angle ($\psi$), and RM,
we applied RM-Synthesis to the polarization Stokes $Q$ and $U$ data \citep{2005A&A...441.1217B},
using the python-based code developed by Michael Bell\,\footnote{\url{http://www.github.com/mrbell/pyrmsynth}}.
With the usable frequency band (after flagging) we reach a resolution in Faraday depth of 522\,rad\,m$^{-2}$, a maximum detectable scale of 229\,rad\,m$^{-2}$, and a maximum detectable Faraday depth of 1357\,rad\,m$^{-2}$.
RM-Clean \citep{2009A&A...503..409H}, a technique to deconvolve the complex polarization from
the RM Spread Function, similar to the clean algorithm used in interferometric imaging, was included in the python package and applied to the data as well.  
We used a step size of 2\,rad\,m$^{-2}$ in a range from $-2000$ to $+2000$\,rad\,m$^{-2}$ and cleaned down to a cut-off of $6\,\sigma_{QU}$, where $\sigma_{QU}$ is the average rms noise in Stokes $Q$ and $U$ across the band (given in Table\,\ref{tab:obs_summary}). The specifications used for RM-Synthesis and the resulting parameters are summarized in Table\,\ref{tab:RMsynth_param}. 
We extracted the peak polarized intensity and the corresponding Faraday depth from the Faraday spectrum at each pixel across the galaxy to obtain the maps of polarized intensity and Faraday depth. 
Given the poor resolution in Faraday depth space, the Faraday spectra from different lines-of-sight show no complex nature, just a single peak, not broader than the resolution $\delta\Phi$.
In this case, we may assume RM$\,\equiv\Phi$.
Therefore, we adopt the notation ``peak RM'' (or just ``RM'') throughout this paper. However, we note that due to the poor resolution in Faraday depth, we cannot rule out that there are several components in the Faraday spectrum hidden in the broad peak, resulting from different components along the line-of-sight contributing to the observed RM.

\begin{table}[t]
  \centering
\caption{RM-Synthesis parameters and specifications in the S-band.}
  \begin{tabular}{lcc}
    \toprule\toprule
Parameter							& Value				& Explanation 				\\
  \midrule
$\Phi_{\text{min}}$					&  $-2000$\,rad\,m$^{-2}$		&  Minimum Faraday depth	\\
$N_{\Phi}$							&  2000						&  Number of steps			\\
$d\Phi$							&  2\,rad\,m$^{-2}$			&  Step size				\\
cutoff							&  6$\,\sigma_{\text{QU}}$	&  RM-Clean cutoff			\\
$\lambda^2_{\text{min}}$ 				&  0.0137\,m$^2$				&  Minimum wavelength		\\
$\delta\lambda^2$ 					&  0.0013\,m$^2$				&  Channel width			\\
$\Delta\lambda^2$ 					&  0.0066\,m$^2$				&  Wavelength-coverage		\\
$\delta\Phi$ 						&  522\,rad\,m$^{-2}$		&  Resolution in $\Phi$-space	\\
$\vert\vert{\Phi_{\text{max}}}\vert\vert$	&  1357\,rad\,m$^{-2}$	&  Maximum detectable $\Phi$	\\
max-scale							&  229\,rad\,m$^{-2}$		&  Maximum detectable scale	\\
    \bottomrule
  \end{tabular}
   \tablefoot{$\sigma_{\text{QU}}$ is the average rms noise in the Stokes $Q$ and $U$ images, given in Table\,\ref{tab:obs_summary}.}
   \label{tab:RMsynth_param}
 \end{table}

To ensure that the polarization analysis is not affected by bandwidth depolarization, we examined the amount of depolarization within the observational bandwidth.
The reduction of the degree of polarization by bandwidth depolarization
depends on the amplitude of the observed RM, the observational frequency, and the bandwidth. The effect is strongest at low frequencies. 
For the S-band (with subbands of 128\,MHz width in the spectral window images), a |RM| of 500\,rad\,m$^{-2}$ would reduce the degree of polarization systematically by about 5\,\%, but the maximum observed |RM| in the S-band is a factor of about two smaller (see Section\,\ref{sec:Bpara}). Also at higher frequencies (X- and C-bands), the amplitude in |RM| does not exceed the limit that reduces the degree of polarization by more than 5\,\%.  
Therefore, bandwidth depolarization does not significantly affect our 
analysis.

\subsection{M51's magnetic field in the plane of the sky revealed by the S-band data}

In this section, the magnetic field component perpendicular to the line-of-sight across M51 is discussed. Information on this component is given by the spatial distribution of the polarized intensity $PI$ and the intrinsic polarization angle $\psi_0$\,\footnote{assuming a single Faraday depth component along the line-of-sight} 
that, rotated by 90 degrees, shows the magnetic field orientation in the plane of the sky. 


	\begin{figure*}[htbp]
	\centering
\includegraphics[scale=0.57, trim=3cm 10cm 5cm 2cm, clip]{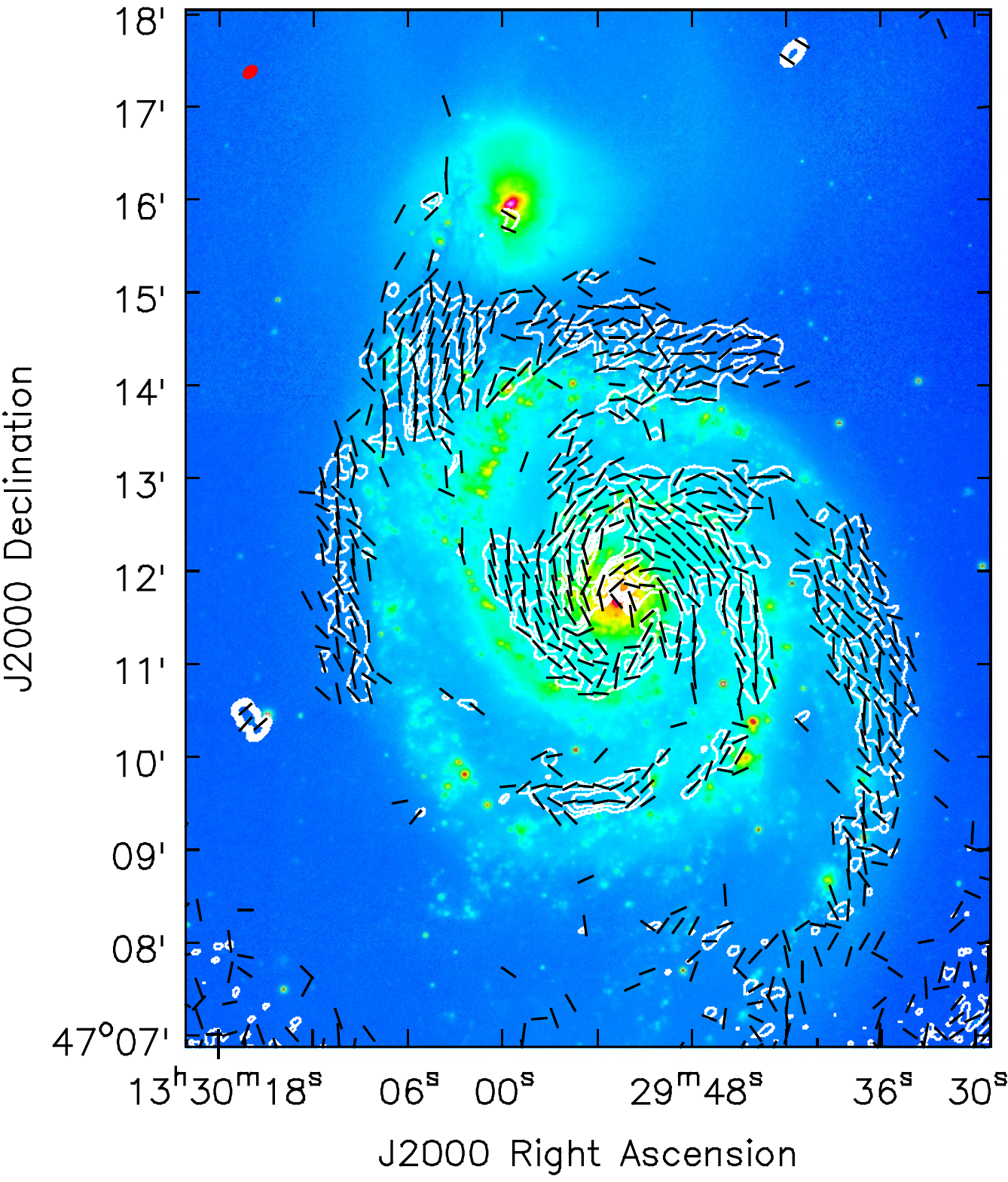}\hspace{10mm}\includegraphics[scale=0.57, trim=1.5cm 10cm 3.8cm 2cm, clip]{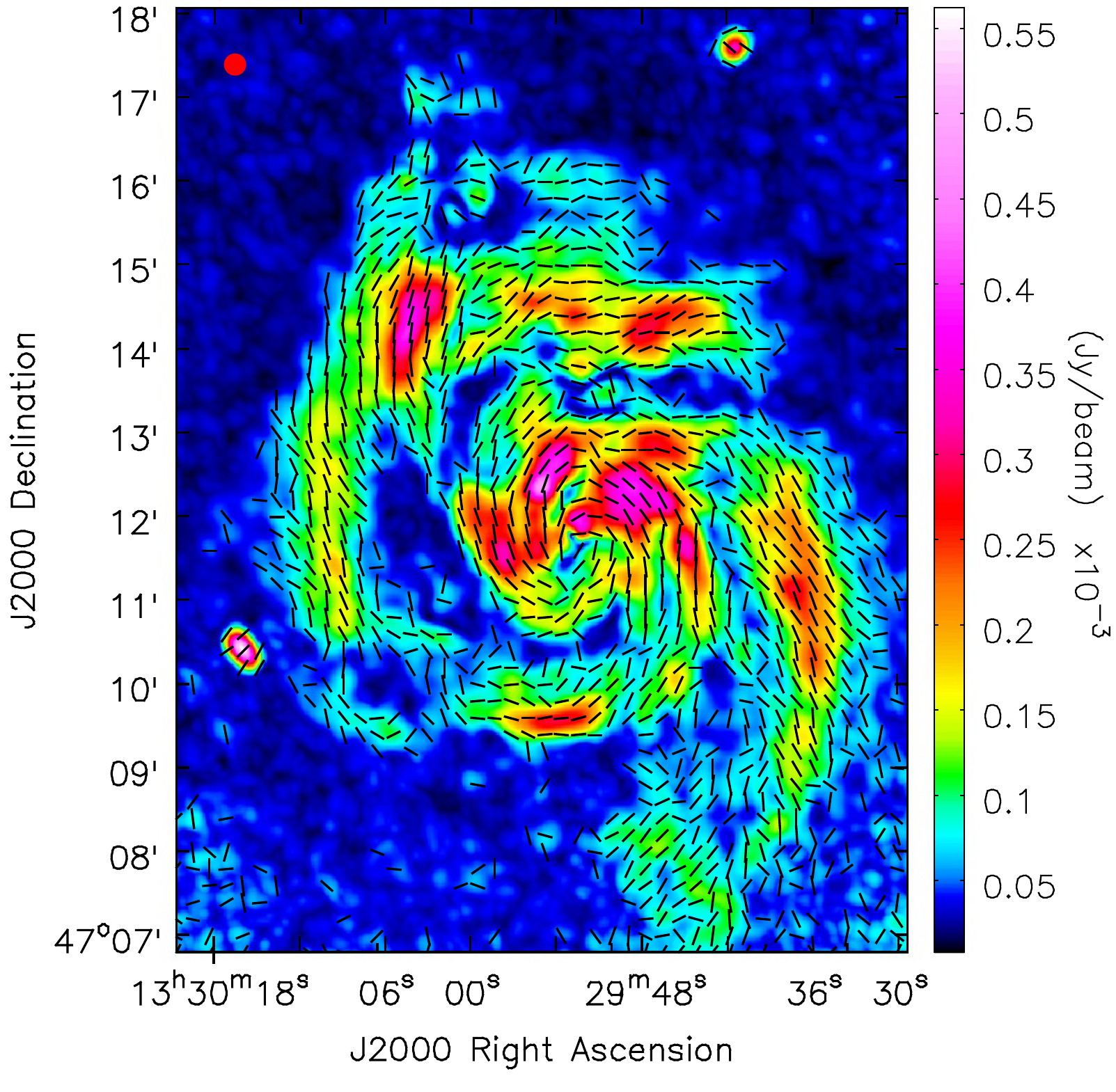}
\caption{Linearly polarized intensity image of M51 at 3\,GHz with a resolution of $10''\times7''$ overlaid onto a H$\alpha$ image \citep{2003PASP..115..928K} (left) and with a resolution of 15$''$ in color scale in units of Jy\,beam$^{-1}$ (right). The contours in the left panel are drawn at [8, 12, 16, 24, 32]$\,\times\,6\,\mu$Jy\,beam$^{-1}$. The polarized intensity maps are overlaid with the polarization $E+90\degr$-orientations,  corrected for Faraday rotation to show the magnetic field structure. Data of the polarization orientations are only shown where the signal-to-noise ratio in polarized intensity exceeds five. The beam size is shown in the top left corner.}
	     \label{fig:PI+PA}
	\end{figure*}

\subsubsection{Polarized intensity and magnetic field structure}
\label{sec:PI}

The polarized intensity in the S-band overlaid with the polarization angles rotated by 90 degrees indicating the magnetic field orientation is shown in Figure\,\ref{fig:PI+PA}. The left panel shows the polarized intensity as contours overlaid on the H$\alpha$ image from \citet{2003PASP..115..928K} at $10''\times7''$ resolution. The right panel shows the polarized intensity at $15''$ resolution as rainbow color scale. The polarized intensity at $15''$ resolution is used in this paper for analysis. The polarization angles are corrected for Faraday rotation via $\psi_0\,=\,\psi - \text{RM}\,\lambda_\text{c}^2$, where $\psi_0$ is the intrinsic polarization angle, $\lambda_\text{c}^2=0.0097$\,m$^2$ is the weighted average of the observed range in $\lambda^2$, and $\psi$ is the observed polarization angle at the average wavelength of the S-band.

The magnetic field structure shows a spiral pattern. Compared to the total intensity, which shows a clear correspondence with the optical spiral arms, the spatial distribution of the polarized intensity across the galaxy is more complicated. 
Some parts of polarized emission coincide well with the optical spiral arms seen in H$\alpha$, but at some locations the peak of polarized emission is located in the inter-arm regions, as discussed in studies at other frequencies
(e.g., \citealt{Fletcher11}). 

\subsubsection{Field ordering}
\label{sec:p}

\begin{figure*}[htbp]
	\centering
\hspace*{-7mm}\includegraphics[scale=0.55, trim=0.5cm 10cm 4cm 2cm, clip]{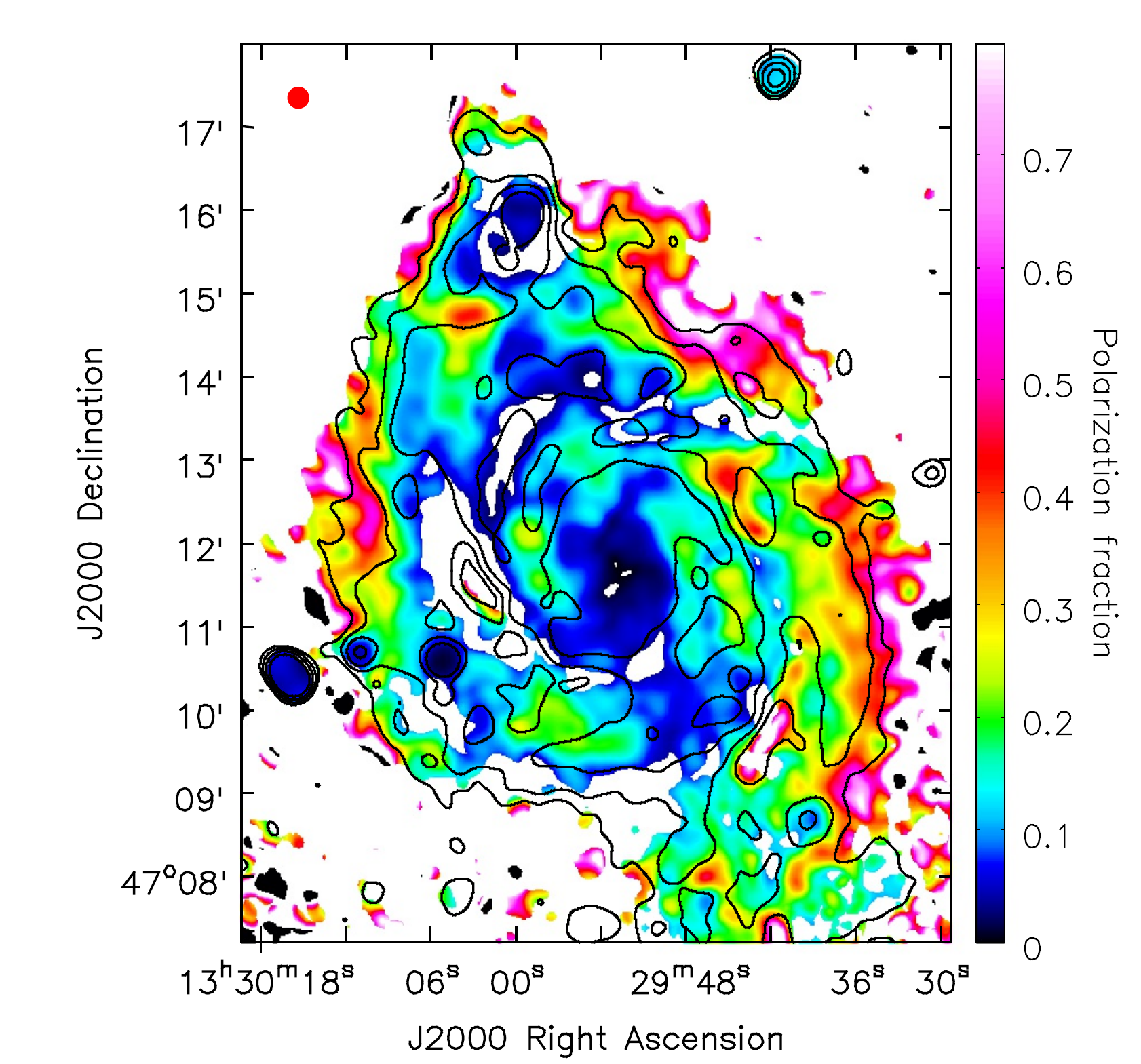}\includegraphics[scale=0.55, trim=0.5cm 10cm 4cm 2cm, clip]{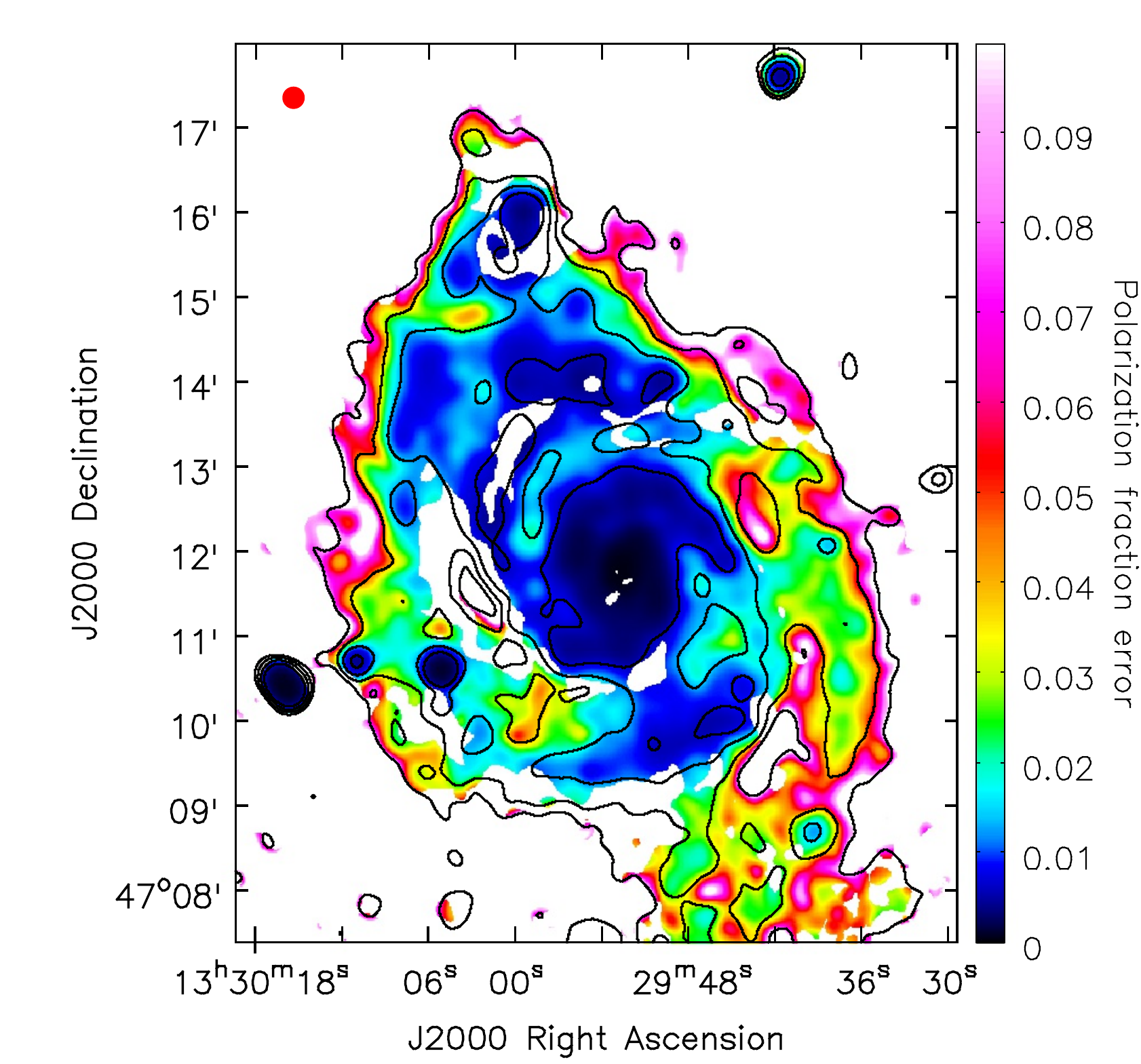}
	     \caption{Map of the observed degree of polarization of the non-thermal emission at 3\,GHz at $15''$ resolution (left) and the corresponding error map (right), overlaid with the total intensity contours at [4, 8, 16, 32]$\,\times\,60\,\mu$Jy\,beam$^{-1}$. Data are only shown where the signal-to-noise ratio in polarized intensity exceeds five. The synthesized beam is shown in the top left corner. 
	     }
	     \label{fig:pmap}
	\end{figure*}

To visualize the degree of order of the magnetic field in M51, we compute a map of the degree of polarization by dividing the polarized intensity by the non-thermal intensity map at 3.05\,GHz. The fractional polarization map is shown in the left panel of Figure\,\ref{fig:pmap}, while the error map is shown in the right panel. 
The observed degree of polarization varies from a few percent in the inner spiral arms up to about 40\,--\,50\,\% in the inter-arm regions. The low values ($< 10$\,\%) in the central region and at the locations of the gas spiral arms probably originate
from star-forming activity, generating small-scale fields on the turbulence scale (50\,--\,100\,pc) and/or fields tangled on larger scales, but smaller than the size of the telescope beam.
The extremely high values of up to 100\,\% in the outskirts of M51 arise from a low a signal-to-noise ratio in the non-thermal intensity map, have large errors (50\,--\,100\,\%), and are not physical. 

\begin{figure}[htbp]
	\centering
\includegraphics[scale=0.43]{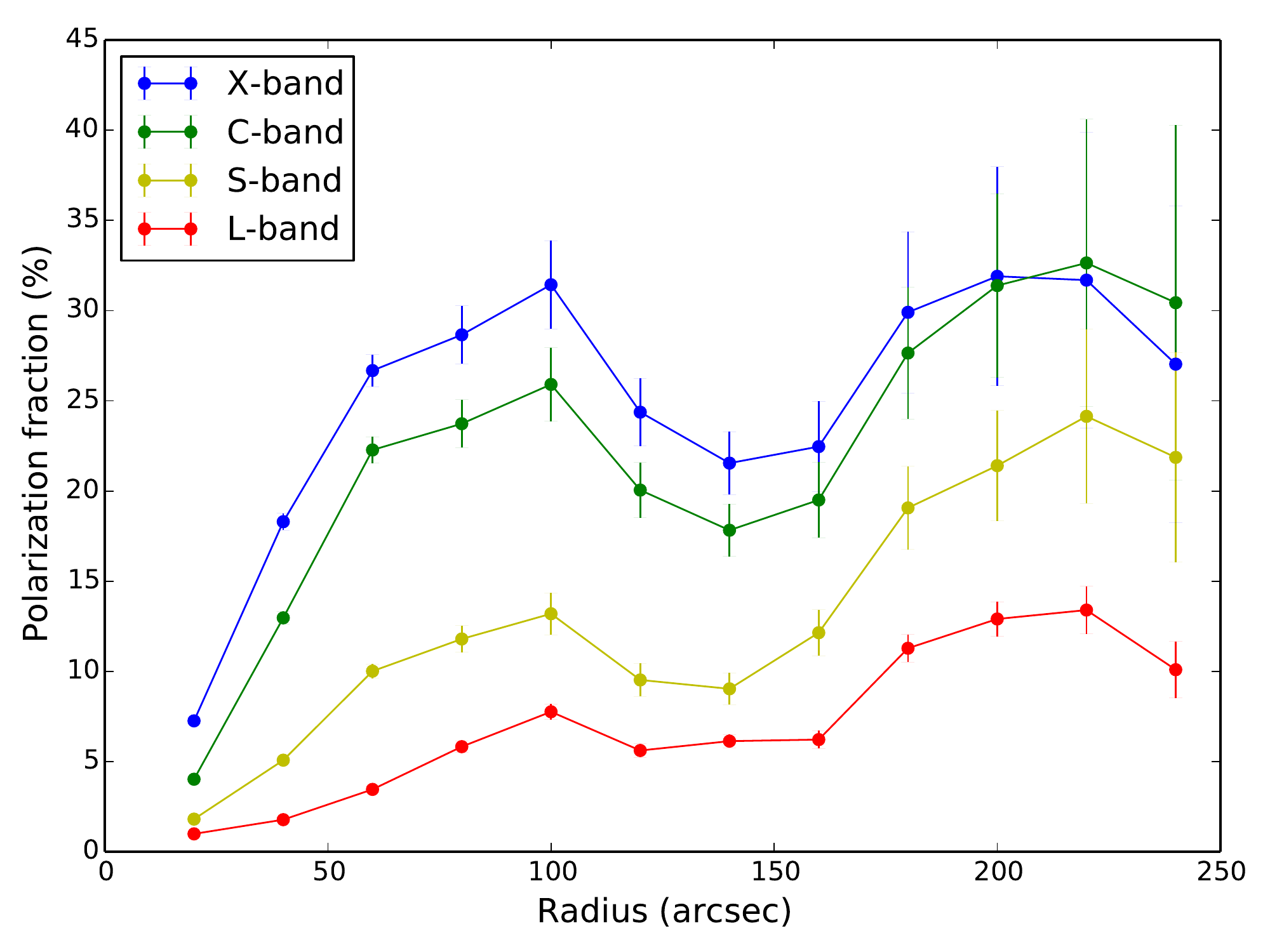}
	     \caption{Azimuthally averaged degree of polarization of the non-thermal emission as a function of radius in M51,
	     computed from our radio maps at 15$''$ resolution in rings of 20$''$ radial width in the plane of the galaxy (with inclination $l\,=\,-20\degr$ and position angle PA$\,=\,-10\degr$) out to a maximum radius of 240$''$ (this refers to the middle of the ring). The error bars are calculated from the rms noise in the maps from which the degrees of polarization are derived. 
	     }
	     \label{fig:radius}
	\end{figure}

A closer look at the observed degree of polarization in Figure\,\ref{fig:pmap} reveals a radial increase from about 2\,\% in the center up to about 40\,\% at the outer spiral arms in the S-band. 
We computed the degree of polarization as a function of the radius determined from the average non-thermal and polarized intensities at 15$''$ resolution in radial rings (Figure\,\ref{fig:radius}). The error bars in Figure\,\ref{fig:radius} are derived from the rms noise in the maps in the non-thermal intensity, Stokes $Q$, and $U$ maps. 

The degree of polarization increases from a few percent at small radii up to about 20\,\% at larger radii. Additionally, we show the degree of polarization as a function of radius at higher (the X- and C-bands, from \citealt{Fletcher11}) and lower (the L-band, from \citealt{2015ApJ...800...92M}) frequencies. 
The trend of an increasing degree of polarization towards larger radii is similar at all frequencies, whereas the amplitudes are significantly different (about a factor of three larger at higher frequencies compared to the S-band and the L-band).
The lower amplitudes in the S-band and the L-band arise from the fact that we see less deeply into the disk than at higher frequencies. The S-band polarization data probe the halo and a layer above the disk where the disk emission is partly depolarized. The L-band polarization data probe only the front part of the halo, while the disk emission is completely depolarized.

There are two bumps in the degree of polarization as a function of radius plot at 100$''$ ($\sim\,$3.7\,kpc) and about 200$''$ ($\sim\,$7.4\,kpc). The rings at these radii well coincide with the radius of the inter-arm regions between the two well-pronounced gas spiral arms in M51. 
The inter-arm regions are believed to host well-ordered magnetic fields, which results in a high degree of polarization.  If this is the case, we expect the minima to appear at the position of the gas spiral arms where the
turbulent field is stronger and hence the degree of polarization is lower. Indeed, the minima in Figure\,\ref{fig:radius} occur at the galaxy center and at a radius of about $140''$ ($\sim\,$5.2\,kpc), which is
approximately
the radius at which both spiral arms are located. The degree of polarization at all frequencies changes by
a factor of $\sim\,$1.4 between the arm and inter-arm regions. We note that the rings do not
coincide with the spiral arms due the pitch angles of the spiral arms\,\footnote{The pitch angle is defined as the angle between the tangent to the spiral arm and the tangent to a circle in the galaxy plane, measured at the point where the arm and the circle intersect \citep[e.g.,][]{1996ima..book.....C}.}.
Still, because the strongest emission from both prominent spiral arms and the weakest emission from inter-arm regions, appear within the same rings, this approximation is sufficient for the purpose of this study.

\subsection{M51's magnetic field along the line-of-sight revealed by the S-band data}\label{sec:Bpara}

\begin{figure*}[htbp]
	\centering
\hspace*{-5mm}\includegraphics[scale=0.55, trim=0.5cm 10.5cm 3.5cm 1cm, clip]{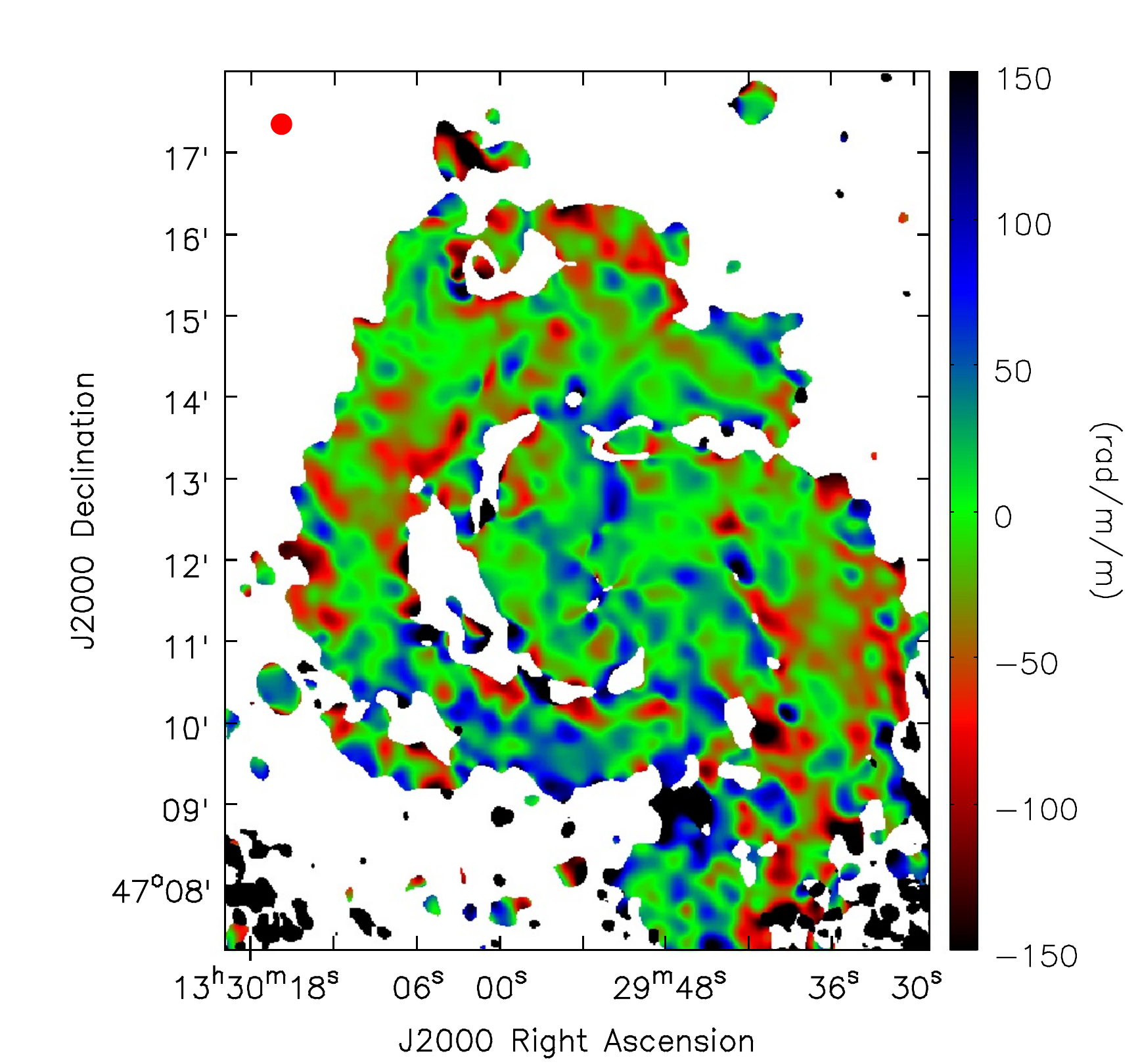}\includegraphics[scale=0.55, trim=0.5cm 10.5cm 3.5cm 1cm, clip]{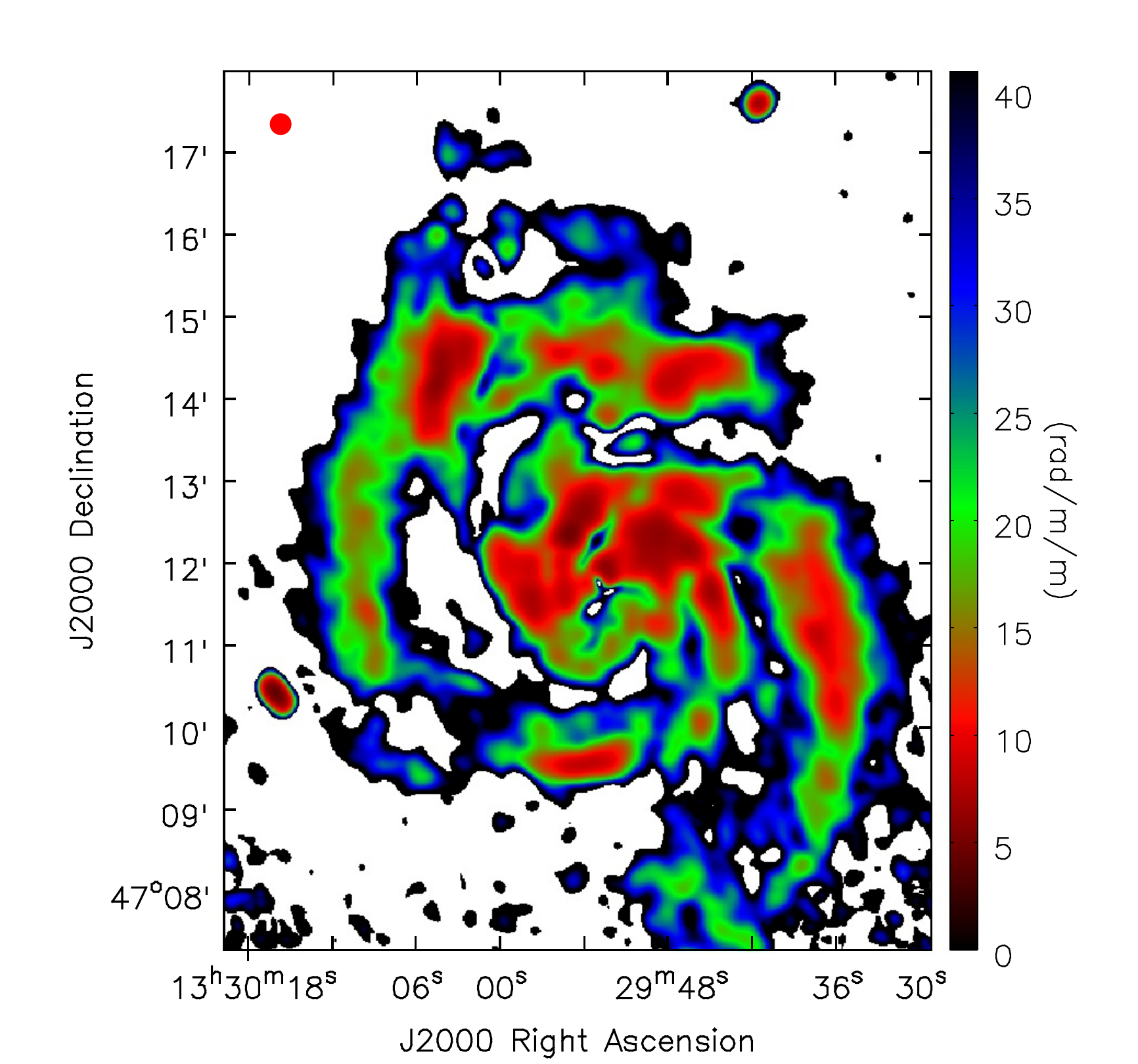}
	     \caption{RM map of M51 at $15''$ resolution (left) and the corresponding error map (right), calculated using Equation\,\eqref{eq:RMerror}. Data are only shown where the signal-to-noise ratio in polarized intensity exceeds five. The beam size is shown in the top left corner.} 
\label{fig:RMmap}
	\end{figure*}

The RM map of M51 in the S-band is shown in Figure\,\ref{fig:RMmap}. It is clipped by the polarized intensity map using five times the average rms noise in Stokes $Q$ and $U$, $\sigma_{QU}$. The error map is shown as well. The error in RM is calculated 
as 
\begin{equation}\label{eq:RMerror}
\Delta_\text{RM}=\frac{0.5\,\delta\Phi}{ \text{S/N}_{PI}}\,,
\end{equation}
where S/N$_{PI}$ is the signal-to-noise ratio in polarized intensity and $\delta\Phi$ is the resolution in Faraday depth \citep[e.g.,][]{2013A&A...549A..56I}. The RM map is corrected for the Milky Way foreground via RM$\,=\,\text{RM}_{\text{obs}}-\text{RM}_{\text{MW}}$ assuming a contribution of RM$_{\text{MW}}=+13 \pm 1\,$rad\,m$^{-2}$ \citep{2015ApJ...800...92M}. 

\begin{figure}[t]
\centering
\includegraphics[scale=0.4, trim=0.5cm 0cm 0cm 0cm, clip]{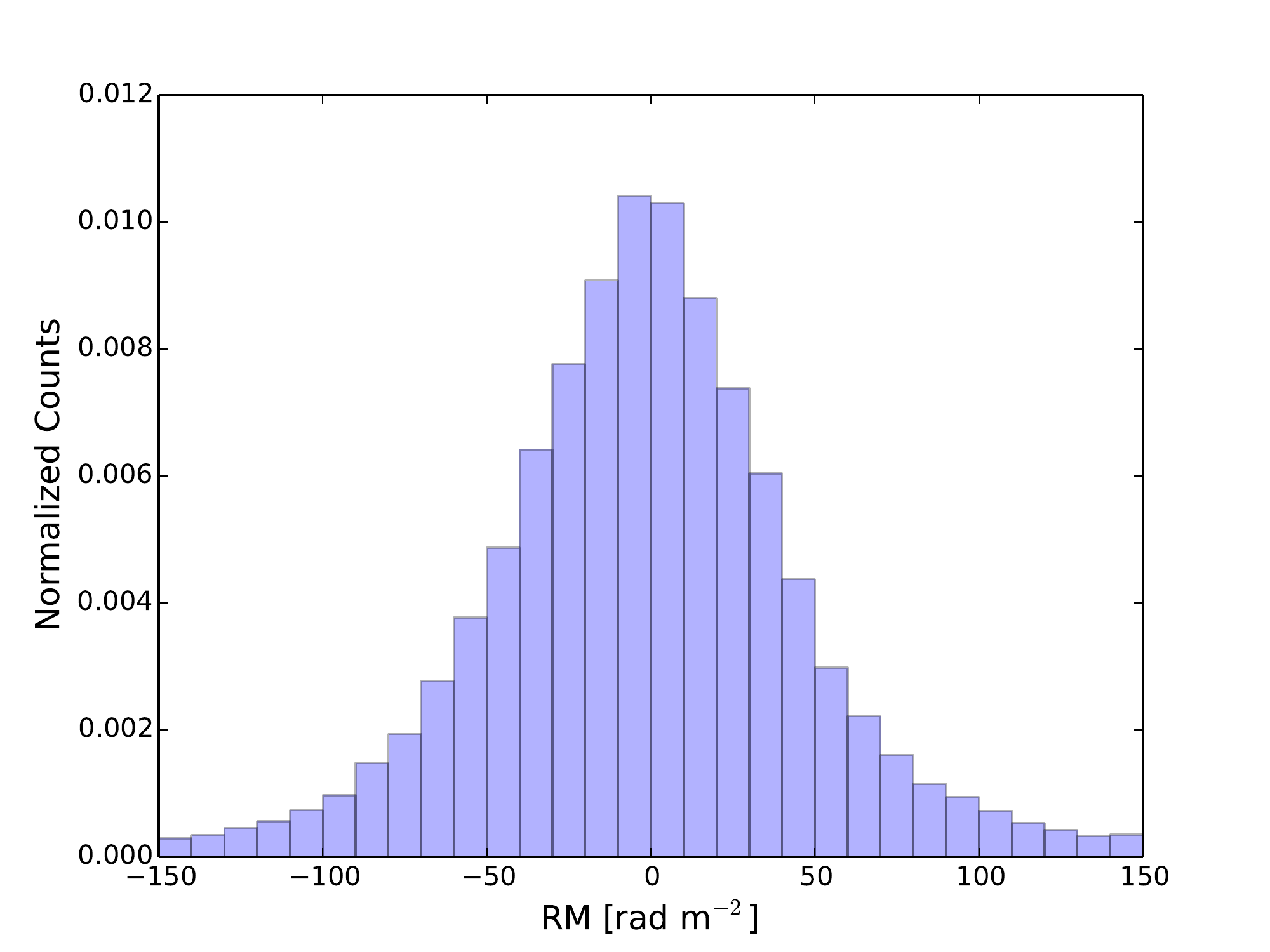}
\caption{Histogram of the RM map at 15$''$, shown in Figure\,\ref{fig:RMmap}. Data were only used where the signal-to-noise ratio in polarized intensity exceeds five.
The distribution  
has a mean of  $-$3\,rad\,m$^{-2}$ and a standard deviation of 46\,rad\,m$^{-2}$.} 
\label{fig:RMhist}
  \end{figure}
  
The RM distribution shows a mean of about $-$3\,rad\,m$^{-2}$ with a standard deviation of 46\,rad\,m$^{-2}$ (Figure\,\ref{fig:RMhist}, see also Table\,\ref{tab:RMvalues}).
The RM values range from about $-150$\,rad\,m$^{-2}$ to $+150$\,rad\,m$^{-2}$. This is comparable to the range of RM found between the C- and X-bands (4.85\,GHz and 8.35\,GHz) by \cite{Fletcher11}, but a factor of five larger than the range found in the L-band (1\,--\,2\,GHz) by \cite{2015ApJ...800...92M} of about $\pm\,30$\,rad\,m$^{-2}$ (compare Table\,\ref{tab:RMvalues}). The varying ranges in RM found at different frequencies result from polarized emission probing different physical depths: the polarized signal at short wavelengths originates from the disk and experiences strong Faraday rotation on the way through the galaxy along the line-of-sight when passing through the turbulent mid-plane where large fluctuations in electron densities and magnetic fields produce a much larger RM. 
At longer wavelengths (around 20 cm) the signal from the disk is almost completely depolarized by wavelength-dependent Faraday depolarization effects (including differential Faraday rotation and internal and external Faraday dispersion) and the remaining signal originates in the halo, experiencing only little Faraday rotation, which is reflected in the small amplitude of RM detected in the L-band. 

Also the RM dispersions (given by the standard deviation of the RM distribution in Figure\,\ref{fig:RMhist}) have different values at different frequencies: the RM dispersion between 4.85\,GHz and 8.35\,GHz, hence in the disk, have the highest value ($\sigma_{\text{RM, X/C-band}}=50\,$rad\,m$^{-2}$)\,\footnote{We note that this new revised value of $\sigma_\text{RM}$ should be used instead of the incorrect value in the \cite{Fletcher11} paper.}, whereas the standard deviation in the L-band, hence in the halo, is significantly smaller ($\sigma_{\text{RM, L-band}}=14\,$rad\,m$^{-2}$).

\begin{table}[t]
  \centering
\caption{Mean RM and RM dispersion values from different frequency bands.}
  \begin{tabular}{lccc}
    \toprule\toprule
Quantity (in rad\,m$^{-2}$)					& X/C-band	& S-band & L-band 				\\
  \midrule
Maximum |RM|					&  $\sim$200	&  $\sim$150 	& $\sim$30	\\
Mean RM from distribution			& 6 			&  $-3$ 		& 11	\\
$\sigma_\text{RM}$ from distribution 	&  50			&  46	  		& 14	\\
    \bottomrule
  \end{tabular}
   \tablefoot{The values of the distribution are computed from RM maps, which are not corrected for the Milky Way foreground.}
   \label{tab:RMvalues}
 \end{table}

Due to the mild inclination of M51 of $-20\degr$ and the large-scale spiral structure of the magnetic field indicated by the polarization angles seen in Figure\,\ref{fig:PI+PA}, one would expect to see a large-scale signature in the RM map if the large-scale magnetic field is regular/coherent.
However, the spatial RM distribution observed in the S-band in Figure\,\ref{fig:RMmap} has a fluctuating nature and no obvious large-scale pattern can be recognized.
\cite{Fletcher11} found a weak large-scale RM pattern only after spatial filtering.
When averaging our new S-band data in sectors of several rings, the RMs are compatible with the model
by \citet{Fletcher11} (see Appendix\,\ref{sec:mode_analysis}).

The apparent paradox between polarization angles and RMs was already discussed in \cite{Fletcher11} who proposed that 
the large-scale spiral pattern seen in the polarization angles is dominated by the components of anisotropic turbulent fields on the plane of the sky, while
the line-of-sight components cancel and hence do not contribute to RM.

The model of the large-scale regular field in the disk by \cite{Fletcher11} for the radial range of 2.4\,--\,4.8\,kpc predicts RM amplitudes of $\pm\,18$\,rad\,m$^{-2}$ and $\pm\,10$\,rad\,m$^{-2}$ for the $m=0$ and $m=2$ modes in the disk and the inclination of $-20\degr$.
The much larger RM values of $\pm\,150$\,rad\,m$^{-2}$ seen in Figure\,\ref{fig:RMmap} indicate that strong regular magnetic fields occur on scales of a few times the beam size of $\sim 550$\,pc.
This fluctuating nature of the RM could be a signature of
vertical fields emerging from the disk, dominating the signal in Faraday rotation.
Another explanation could be tangled regular fields
that contribute to RM and to polarized emission
(see Sections\,\ref{sec:Bparadiscussion} and \ref{sec:complex} for discussions).



\section{Application of an analytical depolarization model}
\label{sec:modelSection}


In this section,
we explain the general features of the analytical multi-layer depolarization model developed by \cite{Shneider14} and compare a representative sample of model configurations to the observations between 1
and 8\,GHz (between $\lambda$\,30
and $\lambda$\,3.5\,cm). 
\cite{Shneider14} compared the model predictions with
three data sets at 8.35\,GHz, 4.85\,GHz, and 1.4\,GHz ($\lambda\lambda\lambda$ 3.5\,cm, 6.2\,cm, and 20\,cm) from \cite{Fletcher11, 1992A&A...265..417H, 1996ASPC...97..592N}.
The new S-band data at 2\,--\,4\,GHz ($\lambda\lambda$ 7\,--\,15\,cm) lie in a critical wavelength range between the previous data sets where the model predictions strongly differ and thus will help to better constrain the depolarization model.


\subsection{A multi-layer depolarization model}
\label{sec:shneider}

\citet{Shneider14} developed a model of the depolarization of synchrotron radiation in a multi-layer magneto-ionic medium, applied specifically to M51. 
They developed model predictions for the degree of polarization as a function of wavelength and distinguished between a two-layer system with a disk and a near-side halo, and a three-layer system with a far-side halo, a disk and a near-side halo. 
Details are given in \citet{Shneider14} and \cite{KierdorfThesis2019}.

The model distinguishes between scenarios of different magnetic field set-ups (in terms of regular magnetic fields, isotropic turbulent magnetic fields, and anisotropic turbulent magnetic fields), where different model configurations include either one of those or a mixture of the different magnetic field set-ups,
which are considered to be present in different layers (disk and/or halo). 

We model the degree of polarization in normalized form $\left(p/p_0\right)$\,\footnote{We note that there is a typo in Equation\,(25) of \cite{Shneider14}: The term $\cos(D)$ in the third row needs to be replaced by $\cos(C_\text{h})$ and in the fourth row the term $\cos(C)$ needs to be corrected to $\cos(C_\text{d})$.} as a function of wavelength under the following assumptions:


\begin{itemize}


\item The thermal electron density is assumed to be spatially constant in each layer, but with different values in the disk and in the halo. The magnetic field strengths are also assumed to be spatially constant in each layer, with different values for the regular and turbulent fields (isotropic and/or anisotropic).



\item The model is based on the two low (large-scale) azimuthal modes of the regular field found in M51 by \citet{Fletcher11} (see Appendix\,\ref{sec:mode_analysis}) and hence it neglects higher modes (azimuthal variations on smaller scales) and vertical components of the regular field in the disk and the halo.


\item The intrinsic degree of polarization at $\lambda\,=\,0$ is assumed to be $p_0=70$\,\% everywhere in the galaxy. This corresponds to the theoretical injection spectrum for electrons accelerated in supernova remnants with $\alpha_{\text{syn}}=-0.5$
and was observed by \citet{Fletcher11} in the spiral arms.
In the inter-arm regions,
the average $\alpha_{\text{syn}}$ of $-1.1$
gives an intrinsic degree of polarization of $p_0=76$\,\%. Assuming $p_0=70$\,\% instead would give an overestimation of $\left(p/p_0\right)$ by about 8\,\%. 

\item The anisotropy of turbulent magnetic fields is considered as the result of amplification of the components of the turbulent field, caused by compression in spiral arms and by shear from differential rotation, whereas the vertical field component remains unchanged. 
Anisotropic turbulent fields on scales smaller than the telescope beam contribute to polarized intensity and to depolarization (via Faraday dispersion), but not to RM.

\item Several wavelength-dependent depolarization mechanisms were considered:
(1) differential Faraday rotation caused by regular magnetic fields, (2) internal Faraday dispersion caused by turbulent magnetic fields (isotropic or anisotropic) within the emitting region, and (3) external Faraday dispersion caused by turbulent magnetic fields (isotropic or anisotropic) in front of the emitting region.

\item In the case where turbulent magnetic fields are ``switched on'', wavelength-independent depolarization (beam depolarization) is considered as well.


\item The case of Faraday depolarization caused by gradients of RM across the telescope beam \citep{Sokoloff98} is not taken into account. This could underestimate the amount of depolarization.

\item
Depolarization effects from the Galactic foreground are assumed to be negligible.
\end{itemize}

The model parameters and their values used by \citet{Shneider14} are given in Table\,\ref{tab:parameters_start}.
Most parameters were fixed, while for the strengths of the regular and turbulent fields (constant along azimuthal angle in one radial ring) physically reasonable values were chosen to make the model consistent with the observations available at that time in one example region located in that ring. Surprisingly, the strengths of the regular field $B$ in Table\,\ref{tab:parameters_start} are larger than those estimated by \cite{Fletcher11} from the model of the large-scale field for the same radial ring, which are $1.4\pm0.1\,\mu$G and $1.3\pm0.3\,\mu$G for the disk and halo, respectively. Similarly large regular field strengths in disk and halo were found by fitting
these two parameters to the data in 18 azimuthal sectors per radial ring for four different rings \citep{Shneider14II}. Potential reasons of this discrepancy will be discussed in Section\,\ref{sec:complex}.

\begin{table}[t]
\centering
\caption{Model parameter values used by \citet{Shneider14}.}	
\begin{tabular}{l c c}
\toprule
\toprule
 Parameter (unit)					&Disk	& Halo  	\\ 
\midrule
$n_{\textrm{e}}$ (cm$^{-3}$)		&0.11	&	0.01	\\
$n_{\textrm{CRE}}$ (arbitrary)		&0.1	&	0.1 \\
$L$ (pc)						&800		&	5000 \\
$d$ (pc)						&55		&	370 \\ 
$\alpha$ (anisotropy)			&2.0		&	1.5 \\
$B_{\textrm{}}$  ($\mu$G)		&5		&	5	\\ 
$b_{\textrm{}}$  ($\mu$G)		&14		&	4	\\
\bottomrule
\end{tabular} 
\tablefoot{Parameters in the disk and halo used by \citet{Shneider14} to model the degree of polarization as a function of wavelength in the example region ``A'' in M51, located in the radial ring $2.4 - 3.6$\,kpc and centered at the azimuthal angle of $100\degr$. Fixed parameters were: the electron densities $n_{\textrm{e}}$ and path lengths $L$ (adopted from \citealt{Berk97}), the turbulence cell size $d$ (computed using Equation\,\eqref{eq:d} with the given parameter values), the number density of CREs (in arbitrary units; its value is not relevant since it cancels out when calculating the depolarization $\left(p/p_0\right)$), and the degree of anisotropy $\alpha$ of the turbulent field  ($\alpha=1$ meaning a purely isotropic field). The values of the regular and turbulent magnetic field strengths $B_{\textrm{}}$ and $b_{\textrm{}}$ (assumed to be constant along azimuthal angle in the radial ring) were chosen in \citet{Shneider14} to be consistent with the observed degrees of polarization $p$ available at that time.
}
\label{tab:parameters_start}
\end{table}


\begin{table}[t]
\centering
\caption{Model configurations for a two-layer system.}	
\label{tab:checkmarks}
\begin{tabular}{l c c c c c}
\toprule
\toprule
 & \multicolumn{3}{ c }{Disk} &  \multicolumn{2}{ c }{Halo} \\ 
\midrule 
 		& Reg.		& Iso. 		& Aniso. 		& Reg. 		& Iso. 	\\ 
\midrule 
DH		&$\checkmark$&			&			&$\checkmark$&  						\\ 
DAH		&$\checkmark$&			&$\checkmark$&$\checkmark$&			 			\\	
DIH		&$\checkmark$&		$\checkmark$	&&$\checkmark$&			 			\\	
DIHI		&$\checkmark$&$\checkmark$&			&$\checkmark$&$\checkmark$				\\
DAIHI	&$\checkmark$&$\checkmark$&$\checkmark$	&$\checkmark$&$\checkmark$			\\
\bottomrule
\end{tabular} 
\tablefoot{Different model configurations discussed in this paper. The nomenclature is as follows: Capital letters ``D'' and ``H'' stand for regular fields in the disk and the halo, respectively.
Capital letters ``I'' and ``A'' denote isotropic and anisotropic turbulent fields. For example, ``DAIHI’' stands for a configuration with regular fields together with isotropic and anisotropic turbulent fields in the disk (DAI), and regular and only isotropic turbulent fields in the halo (HI).
In case of a three-layer system, for example, ``HDH'' means ``regular (far side) halo field + regular disk field + regular (near-side) halo field'' (see bottom panel of Figure\,\ref{fig:shneider_plots}). The checkmarks show which fields are switched ``on'' and ``off''.
Anisotropy of the halo field is neglected in our study to make the models simpler.}
\end{table}

\begin{figure}[t]
\centering
\includegraphics[scale=0.42]{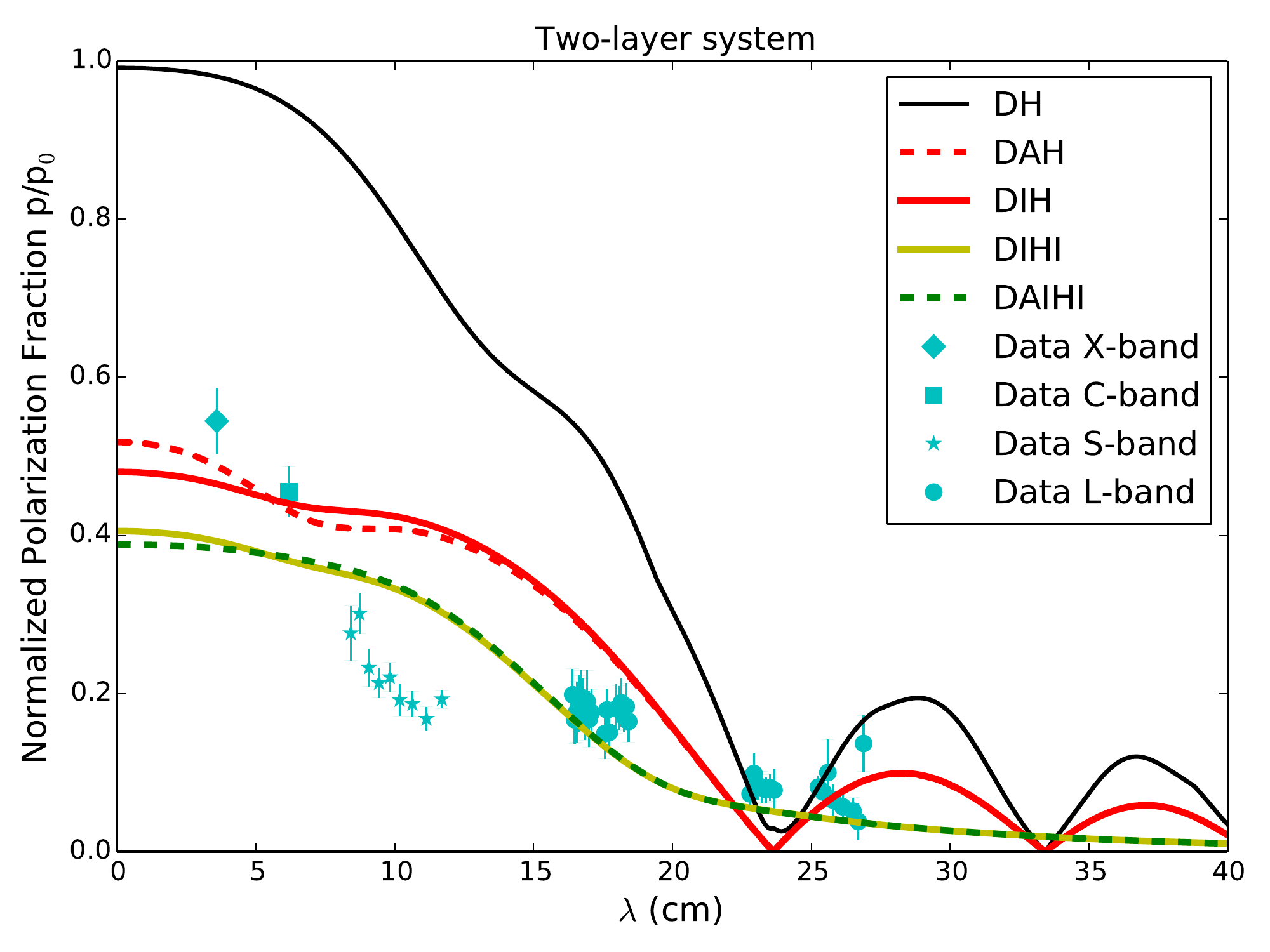}
\includegraphics[scale=0.42]{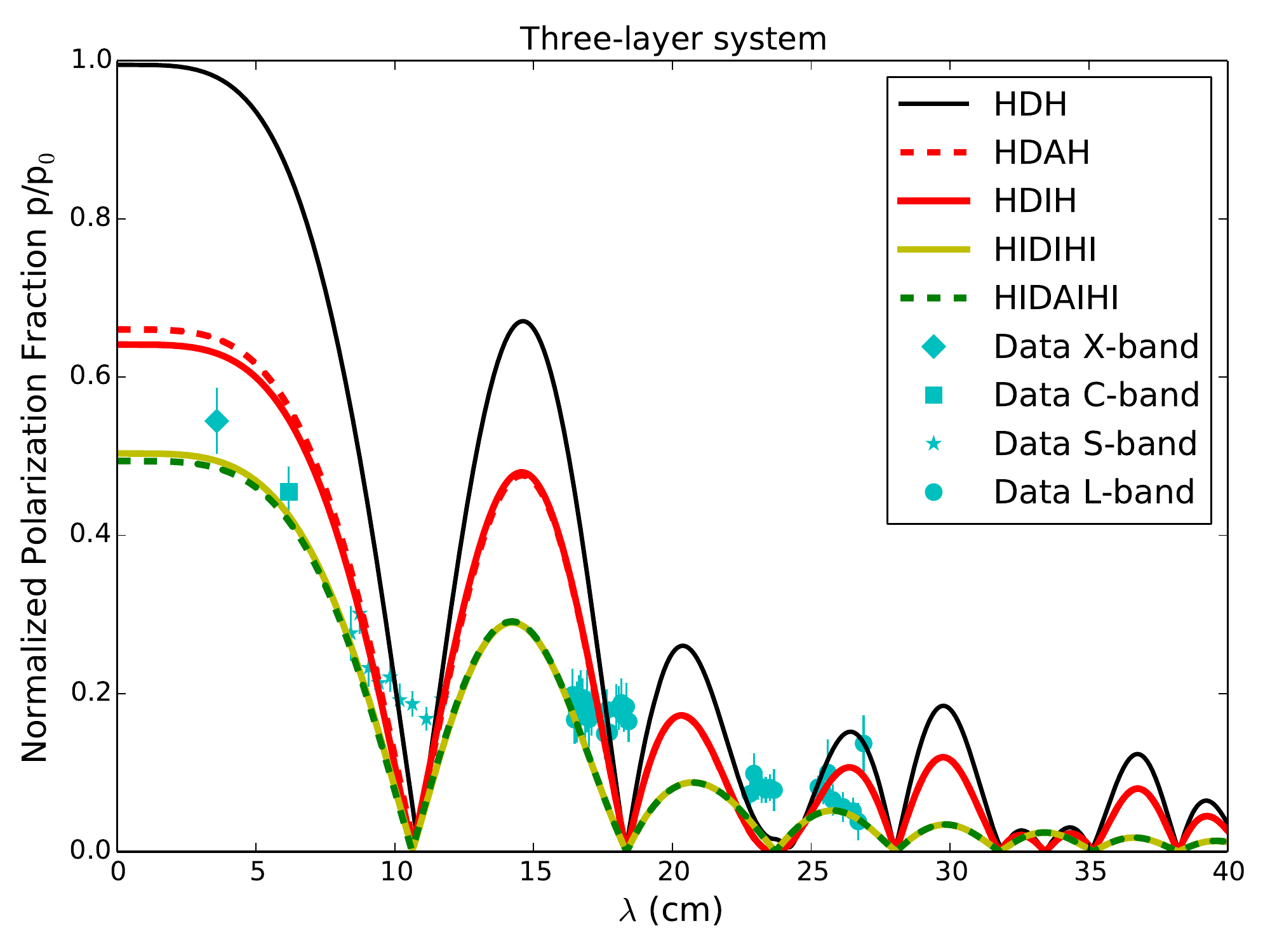}
\caption{Normalized degree of polarization as a function of wavelength for a two-layer system (top) and a three-layer system (bottom) in M51. The plots show some representative model configurations from \citet{Shneider14}.
The observed degrees of polarization in region ``A'' are displayed with error bars. 
None of the model configurations can fit the data in the S-band (around $\sim10$\,cm). 
All model profiles featured have been constructed from the set of parameters given in Table\,\ref{tab:parameters_start} (but using $\alpha=1$ for the halo): a regular field strength of 5\,$\mu$G in the disk and in the halo, a disk turbulent field of 14\,$\mu$G, and a halo turbulent field of 4\,$\mu$G (see Table\,\ref{tab:checkmarks} for nomenclature and description of the model types listed in the legend).}

\label{fig:shneider_plots}
  \end{figure}

\begin{figure*}[t]
\centering
\includegraphics[scale=0.4, trim=4cm 0.3cm 3cm 0.2cm, clip]{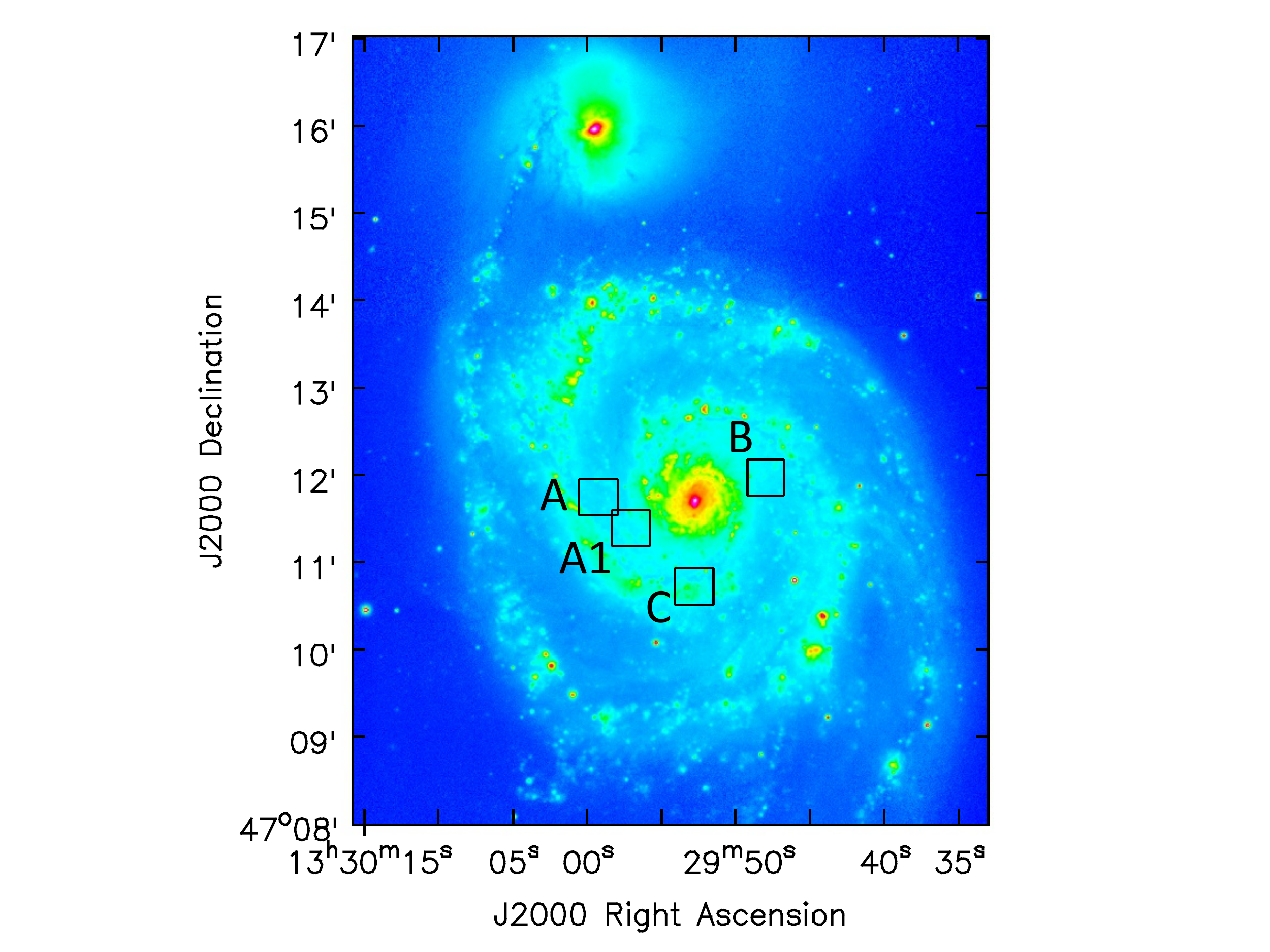}\hspace*{-0.8cm}\includegraphics[scale=0.4, trim=5.2cm 0.3cm 2cm 0.2cm, clip]{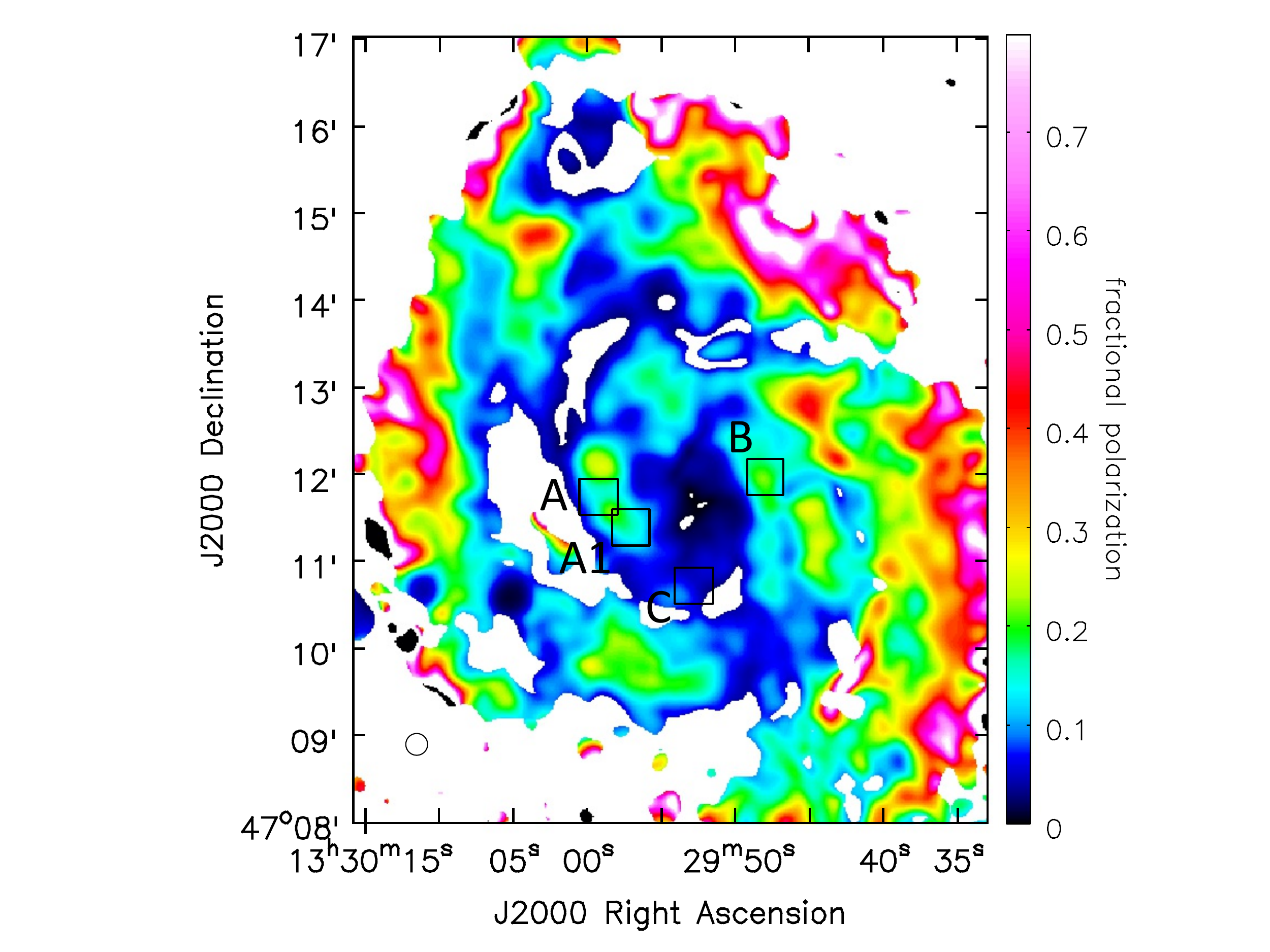}
\caption{Optical image (left) and degree of polarization at 3\,GHz (right) of M51 with the regions (marked with black boxes) in which we compared the observed degree of polarization as a function of wavelength to the depolarization model of \cite{Shneider14}.}
\label{fig:M51_sectors}
\end{figure*}

In this paper, we discuss some representative depolarization model configurations from \cite{Shneider14}, as summarized in Table\,\ref{tab:checkmarks}.
In Figure\,\ref{fig:shneider_plots}, those model predictions for a two-layer (top panel) and three-layer (bottom panel) system are shown for region ``A'' (marked in Figure\,\ref{fig:M51_sectors}), the same that was analyzed by \cite{Shneider14}. We discuss model configurations with purely regular fields in disk and halo and configurations that contain regular plus turbulent magnetic fields. We note that for the three-layer model the near-side and far-side halo are assumed to have identical properties, that is, the halo is symmetric with respect to the disk.  

In the case of purely regular magnetic fields (model DH\,--\,black solid line), the intrinsic degree of polarization (at $\lambda=0)$ starts at its theoretical maximum (chosen to be 70\,\%). The overall trend is a $sinc$-like function, as expected for a uniform slab (a uniform slab is a volume containing uniformly distributed thermal electrons and regular magnetic field lines, causing differential Faraday rotation).
All other model configurations contain turbulent magnetic fields (DAH, DIH, DIHI, and DAIHI; see Table\,\ref{tab:checkmarks}). 
Compared to scenarios with purely regular magnetic fields, turbulent fields significantly decrease the intrinsic degree of polarization at $\lambda=0$ due to wavelength-independent depolarization by turbulent fields (beam depolarization). 
Comparing models DH and DAH, we find that the nulls appear at the same wavelengths, namely at $\sim$\,$\lambda\lambda$\,23 and 33\,cm (in case of the two-layer system). The nulls appear at wavelengths depending on the RM of the layer. 
RM is however dependent on the regular magnetic field strength, which is assumed to be equal in the disk and halo. The turbulent field in model DAH
attenuates the amplitude of the degree of polarization. 
Since the regular magnetic field strength is equal in the disk and the halo and therefore the RM is the same for models DH and DAH, the nulls appear at the same wavelength.

Replacing the anisotropic turbulent fields in the disk by isotropic turbulent fields (compare red solid and red dashed line in Figure\,\ref{fig:shneider_plots}) only decreases the intrinsic degree of polarization at short wavelengths by a few \%. The reason is that, in addition to polarized emission from regular fields, anisotropic turbulent fields contributes
to the polarized signal whereas for purely isotropic turbulent fields the polarized signal vanishes.

\subsection{Adapting the model to the observations between 1\,--\,8\,GHz}
\label{sec:shneider_M51_application}

We compared the observed degree of polarization as a function of wavelength with the model predictions in several representative regions in the galaxy, marked in Figure\,\ref{fig:M51_sectors}. 
Before we discuss the results in all selected regions (Section\,\ref{sec:sectors}), we explain the procedure on the basis of the original region for which \cite{Shneider14} compared the model with the data (region marked with ``A'' in Figure\,\ref{fig:M51_sectors}). This region has 
an azimuthal angle centered at $\phi=\,$100$\degr$ 
and radial boundaries of 2.4\,--\,3.6\,kpc.
This region was chosen because it has a high signal-to-noise ratio in polarized and non-thermal intensity. 
The values of the degree of polarization in all available frequency bands are given in Table\,\ref{tab:datapoints}. 
We use the mean values of Stokes $I$ and  $PI$ in the region to calculate the 
fractional polarization. The errors are calculated from the mean error of Stokes $Q$ and $U$ in the region. 
To get the non-thermal polarization fraction, we subtracted the thermal emission (see Section\,\ref{sec:thermal}) in Stokes $I$ from the total intensity by using the mean value of the thermal fraction in each considered region. Combining all data, we end up with 45 data points across 1\,--\,8\,GHz ($\lambda\lambda$ 3\,--\,27\,cm). 

The size of the region is chosen such that there are
enough independent turbulent cells
within the considered region
to have deterministic expressions for the observed degree of polarization. 
The turbulence cell size $d$ (in pc) is given as \citep{Fletcher11}
\begin{align}\label{eq:d}
 d \simeq \left[\frac{D\,\,\sigma_{\text{RM,D}}}{0.81 \,n_{\text{e}}\, b_{\parallel}\, L^{\nicefrac{1}{2}}}\right]^{\nicefrac{2}{3}}\,,
 \end{align}
where $\sigma_{\text{RM,D}}$ is the RM dispersion in the disk (assumed to be 15\,rad\,m$^{-2}$ in the model), $D$ is the beam size (in pc), $n_{\text{e}}$ is the thermal electron density (in cm$^{-3}$), $b_{\parallel}$ the turbulent field strength along the line-of-sight (in $\mu$G), and $L$ the path length through the medium (in pc).
For a turbulence cell size of 55\,pc, our beam of 15$''$ ($\sim550$\,pc) contains about 100 turbulent cells. The considered regions contain about 5 beams, hence about 500 turbulent cells. This is sufficient to be deterministic \citep{Sokoloff98}. 

By comparing the observed degrees of polarization over a larger number of frequencies to the various model predictions in Figure\,\ref{fig:shneider_plots}, we can rule out model configurations with only regular magnetic fields in the disk and halo (DH) since the observed data deviate the most (a factor of four lower in the S-band). This is in agreement with observations of turbulent magnetic fields in the ISM of spiral galaxies \citep[e.g.,][]{2016A&ARv..24....7B}. Especially at short wavelengths model DH has a degree of polarization close to the intrinsic value without wavelength-independent depolarization (at $\lambda=0$). If turbulent fields are present, beam depolarization is always present because the beam of our observation does not resolve the scale of turbulence, so that the intrinsic degree of polarization is about half of the theoretical maximum.

Surprisingly, none of the model predictions with the parameter values given in Table\,\ref{tab:parameters_start} are in agreement with the observed data in the S-band. For the two-layer system, the data points deviate by a factor of up to about two from the model predictions, whereas for the three-layer case, some data points are well reproduced by the model predictions, but the lines drop to zero at $\lambda\approx11$\,cm ($\approx\,$2.7\,GHz) and at $\lambda\approx17$\,cm ($\approx\,$1.8\,GHz), which is clearly ruled out by the observed data. In any case, our new S-band data
are crucial to evaluate whether the model predictions fit the observations.

The \cite{Shneider14} model contains many
parameters. Some of them, specifically the thermal electron densities, the path lengths $L$, and the turbulence cell sizes in disk and in halo (Table\,\ref{tab:parameters_start}), as well as the fitted parameters of the different magnetic Fourier modes in disk and halo (pitch angle, azimuth, and amplitudes of the Fourier modes), were constrained using prior studies \citep{Berk97,Fletcher11}. 
The most uncertain parameters
are the regular and turbulent magnetic field strengths and the thermal electron densities in the disk and in the halo,
for which the local values could be significantly different from the global estimates.

To understand the model and the influence of the parameters,
we developed an interactive tool in \textsf{Python} where some selected model parameter values (such as the regular and turbulent magnetic field strengths and also the thermal electron densities in the disk and halo) can be varied to visually inspect how well the model matches the data with physically reasonable parameter values (see Figure\,\ref{fig:shneider_interactive}).
We vary these parameters over the whole range of physically reasonable values until we find the optimum combination.
To judge the quality of this ``eye-ball fit'', we calculate a reduced $\chi^2$ value
(with 45 data points and five free parameters). 
We do not perform an automated least-square fit because the convergence of the fit highly depends on the initial parameters. 
Furthermore, the interactive tool makes it easy to visualize 
the changes in the degree of polarization as a function of wavelength when various parameters are modified. 

\begin{figure}[t]
\centering
\includegraphics[scale=0.42]{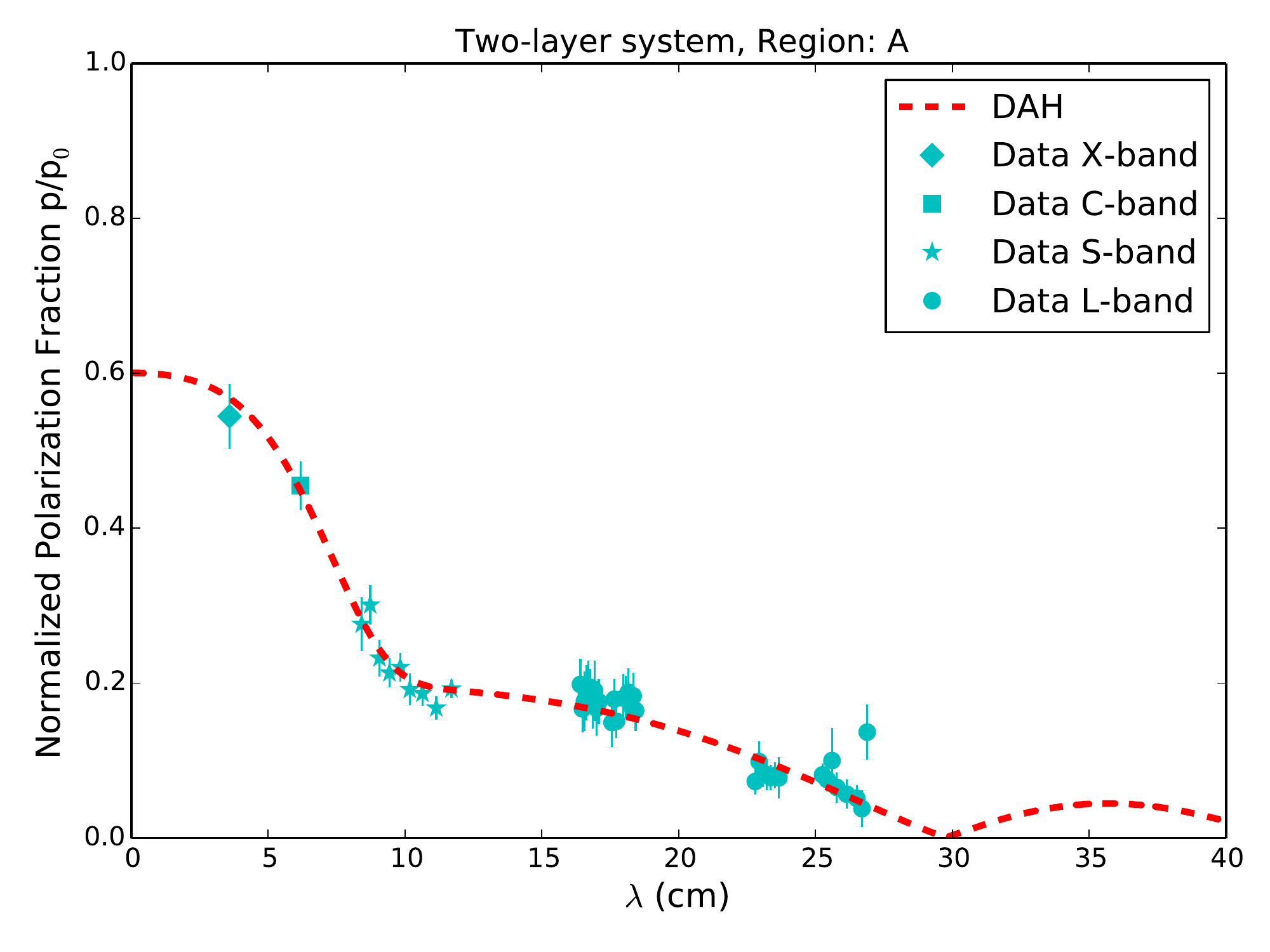}
\includegraphics[scale=0.42]{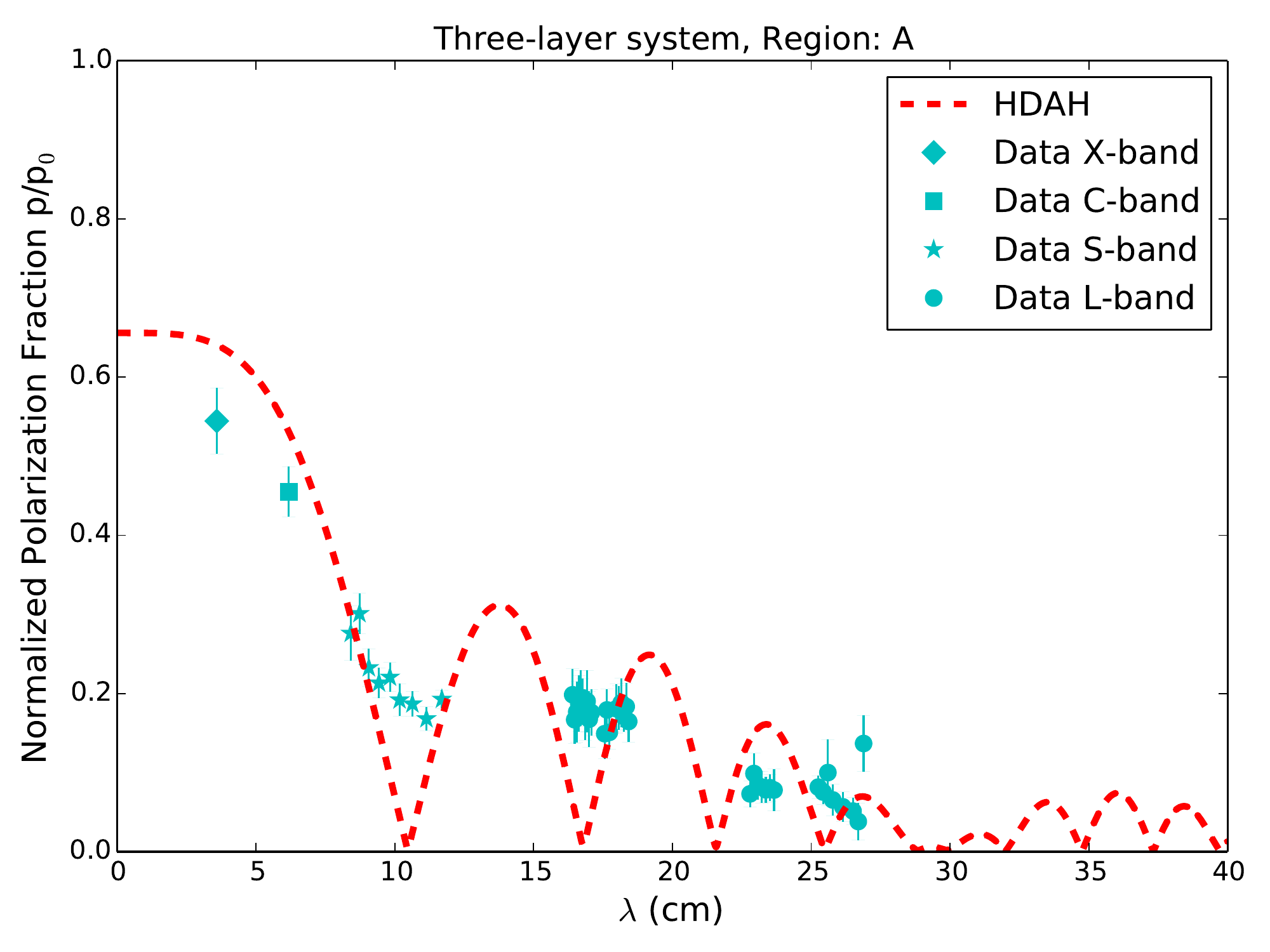}
\caption{Model DAH (red dashed line) for a two-layer system (top panel) and a three-layer system (bottom panel) in M51 with the optimum set of parameters to represent the observed degree of polarization at multiple wavelengths in the region marked ``A'' in Figure\,\ref{fig:M51_sectors}. The parameter values are given in Table\,\ref{tab:parameters}.
} 
\label{fig:shneider_fit}
  \end{figure}
 
Figure\,\ref{fig:shneider_fit} shows
the DAH model configuration with the optimum set of parameters for region ``A'' (red dashed line). 
The
magnetic field strengths and electron densities are listed in Table\,\ref{tab:parameters} (columns 3 and 4) and are all physically plausible values. The uncertainties are obtained by varying the optimum value of each parameter found for the minimum reduced $\chi^2_\text{min}$ (keeping the others fixed) towards smaller and larger values until the reduced $\chi^2$ becomes larger by 50\% compared to $\chi^2_\text{min}$. This 
criterion is the same as that used by \citet{Shneider14II}.
Although our procedure cannot reveal the detailed shape of the five-dimensional surface in parameter space where $\chi^2$ increases by 50\%, it measures the diameters along the five parameter axes.

Our values differ significantly from those used by \cite{Shneider14} (Table\,\ref{tab:parameters}, column 2): the field strengths in the disk are larger, while the electron densities are smaller. Compared to \cite{Shneider14}, the product $B \cdot n_{\textrm{e}}$ for the two-layer model is larger for the disk and smaller for the halo, hence Faraday depolarization by regular disk fields is more important in our study. On the other hand, Faraday depolarization by turbulent disk fields (depending on $b_{\textrm{d}} \cdot n_{\textrm{e,d}}$) is smaller than in  \cite{Shneider14}.

\begin{table*}[ht]
\centering
\caption{Parameter values adopted by \cite{Shneider14} (column 2) and optimum sets of parameters
from our study for the DAH model configuration (columns 3\,--\,7).}	
\begin{tabular}{l |c|c c |c c c c}
\toprule
\toprule
 Parameter 			& Region A 		&  Region A	&  Region A	& Region A1 & Region B & Region C      \\ 
 					& (spiral arm)	& two-layer   	&  three-layer & (spiral arm) & (inter-arm) & (spiral arm)\\ 
\midrule
$B_{\textrm{d}}$ ($\mu$G)&5.0		  &$15.5^{+1.9}_{-2.5}$		&$9.5\pm0.7$		 &$13.2^{+2.0}_{-2.3}$		&$18.3^{+1.7}_{-1.4}$    &$8.2^{+5.7}_{-8.2}$ \\[7pt]
$b_{\textrm{d}}$ ($\mu$G)	&14.0	&$17.7\pm1.7$		&$18.7\pm2.8$		 &$19.9\pm1.7$		&$10.6^{+2.0}_{-2.3}$    &$24.1^{+3.4}_{-2.7}$ \\[7pt]
$B_{\textrm{h}}$ ($\mu$G)&5.0		  &$4.6\pm0.4$		&$4.1\pm0.5$		  &$3.8\pm0.6$		&$3.6\pm0.3$     &$2.9\pm0.4$ \\ [7pt]
$b_{\textrm{h}}$ ($\mu$G)	&4.0		   &-			&-			   &-		&-&- \\ [7pt]
$n_{\textrm{e,d}}$ (cm$^{-3}$)	&0.110	 &$0.057^{+0.02}_{-0.01}$		& $0.038\pm0.003$		 &$0.026\pm0.005$	&$0.12\pm0.03$    &$0.07^{+0.08}_{-0.02}$ \\ [7pt]
$n_{\textrm{e,h}}$ (cm$^{-3}$)	&0.010	 &$0.0068\pm0.0007$	 &$0.0053\pm0.0010$		 &$0.0068\pm0.0009$	&$0.0085\pm0.0013$    &$0.011^{+0.005}_{-0.008}$ \\[7pt] 
\midrule
Reduced $\chi^2_\text{min}$				&	-- 	&0.81		& 2.4		 &0.98	& 0.82 & 1.5 \\
\midrule
$B_{\textrm{tot,d}}$ ($\mu$G) &14.9  &$23.5\pm2.0$		&$21.0\pm2.5$		 &$23.9\pm1.8$		&$21.1\pm1.7$    &$25.5\pm3.6$ \\ 
\bottomrule
\end{tabular} 
\tablefoot{Parameters used to model the degree of polarization as a function of wavelength, derived with our interactive tool (see Figure\,\ref{fig:shneider_interactive}). The third and fourth columns give
the optimum parameter sets of the DAH model for a two-layer and three-layer system for the degree of polarization observed in region ``A''. The last three columns give the
optimum parameter sets of the DAH model (two-layer system) for the degree of polarization observed in regions ``A1'', ``B'', and ``C''. $B_{\textrm{tot,d}}$ is the total field strength in the disk computed from these values ($B_{\textrm{tot,d}}^2 = B_{\textrm{d}}^2 + b_{\textrm{d}}^2)$.
}
\label{tab:parameters}
\end{table*}

The nulls in the three-layer system (bottom panels of Figs.~\ref{fig:shneider_plots} and \ref{fig:shneider_fit}) appear at wavelengths where polarized emission from the far-side halo and the near-side halo are canceled by differential Faraday rotation. This is possible because the physical properties of the far-side and near-side halos are assumed to be identical and the polarized emission from the halo strongly dominates over that of the disk.

For the three-layer system it is not possible to 
remove the nulls in the model by changing any
of the free parameters
within its physically viable range. 
The same holds for the other three-layer model configurations tested in our study.
This suggests that we do not detect polarized emission in the S-band and the L-band from the far-side halo because 
the signal emitted by the far-side halo gets almost completely depolarized by the disk, or that the dominance of the halo emission is not valid (see Section\,\ref{sec:Shneider_discussion} for a discussion).


In their second paper, \citet{Shneider14II} fitted two parameters, the average field strengths of the regular field in the disk and halo, to the observed degrees of polarization at $\lambda\lambda\lambda$ 3, 6, and 20\,cm, averaged in 18 azimuthal sectors in each of four rings across the galaxy M51. For each ring they assumed fixed values of the thermal electron density in disk and halo and a set of nine fixed values of the turbulent field strength.
They used the magnetic field model for the four rings from their first paper \citep{Shneider14} and performed a statistical comparison via $\chi^2$ analysis of predicted to observed polarization data to get the best-fit regular magnetic field strengths in the disk and halo at each ring. 
They found that a two-layer system provides better fits compared to the three-layer model, although the best-fit magnetic field strengths for a three-layer system are comparable.

Our new S-band data
confirm
that the two-layer model is preferred over the three-layer model by directly comparing the model predictions with observations by visual inspections, without performing the fitting procedure as it was done by \cite{Shneider14II}, but also allowing the electron density and turbulent field strength to vary. 
Including the new S-band data, our analysis gives stronger evidence for this statement because in this wavelength range the model predictions differ most. 


We have also tested two-layer model configurations that include turbulent fields in the halo (compare Table\,\ref{tab:checkmarks}).
Basically for all considered model configurations
optimum parameter sets can be found with similarly good reduced $\chi^2$ values. It is not surprising that
the optimum parameter sets with an additional parameter, which means adding another degree of freedom,
represent the data well since we can already find
good representations
for a simpler case. Hence, we exclude model configurations with turbulent fields in the halo.
Although we do see signatures of random magnetic fields in the halo from the L-band data \citep{2015ApJ...800...92M}, their effect on the
modeling procedure is likely weak.

The model configurations DAH and DIH can be
represented with almost equal field strengths and electron densities. Therefore, we cannot distinguish whether the turbulent field in the disk is isotropic or anisotropic 
in region ``A''.

\subsection{Results from different regions}
\label{sec:sectors}

\begin{figure}[t]
\centering
\includegraphics[scale=0.42]{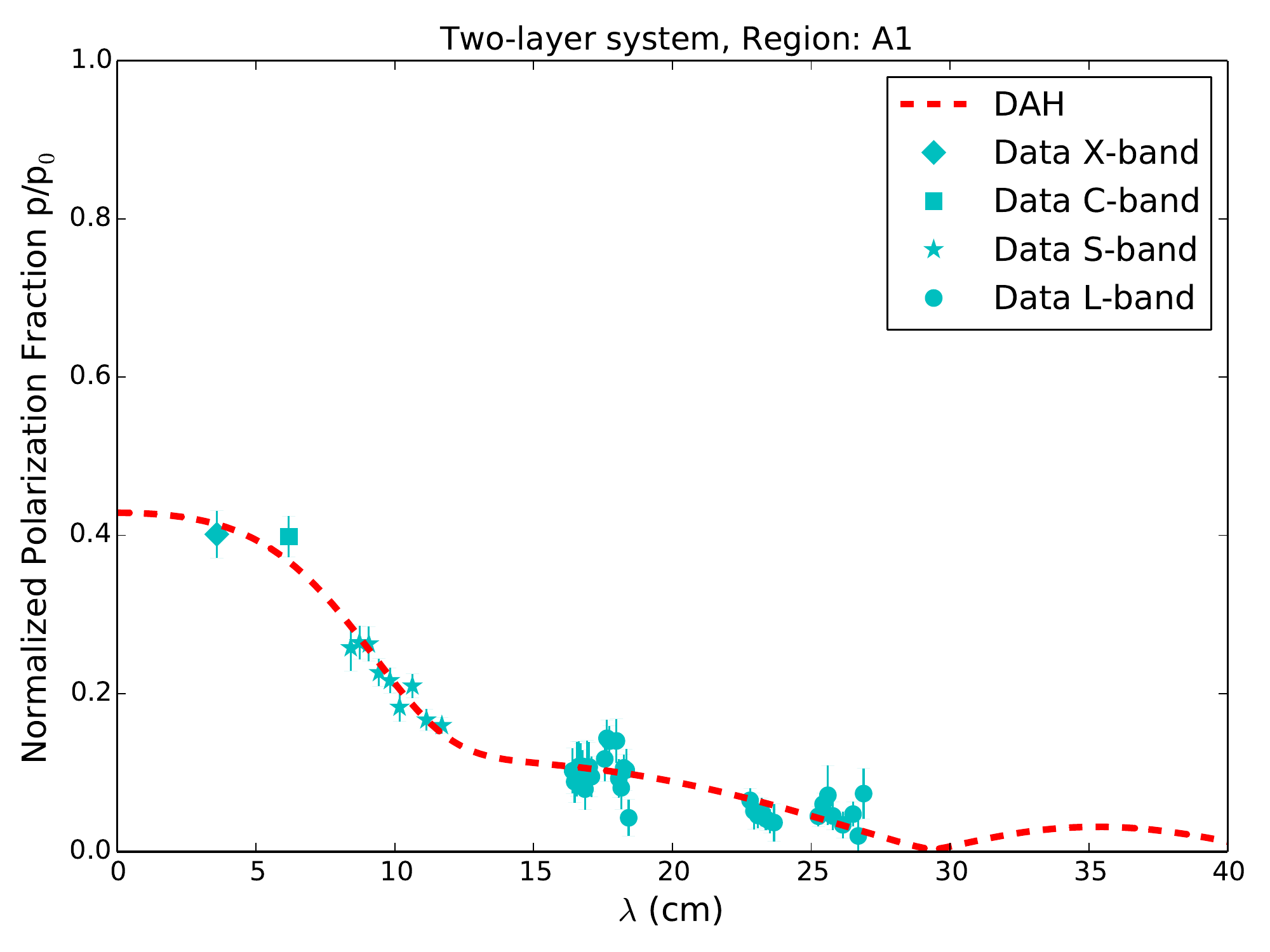}
\includegraphics[scale=0.42]{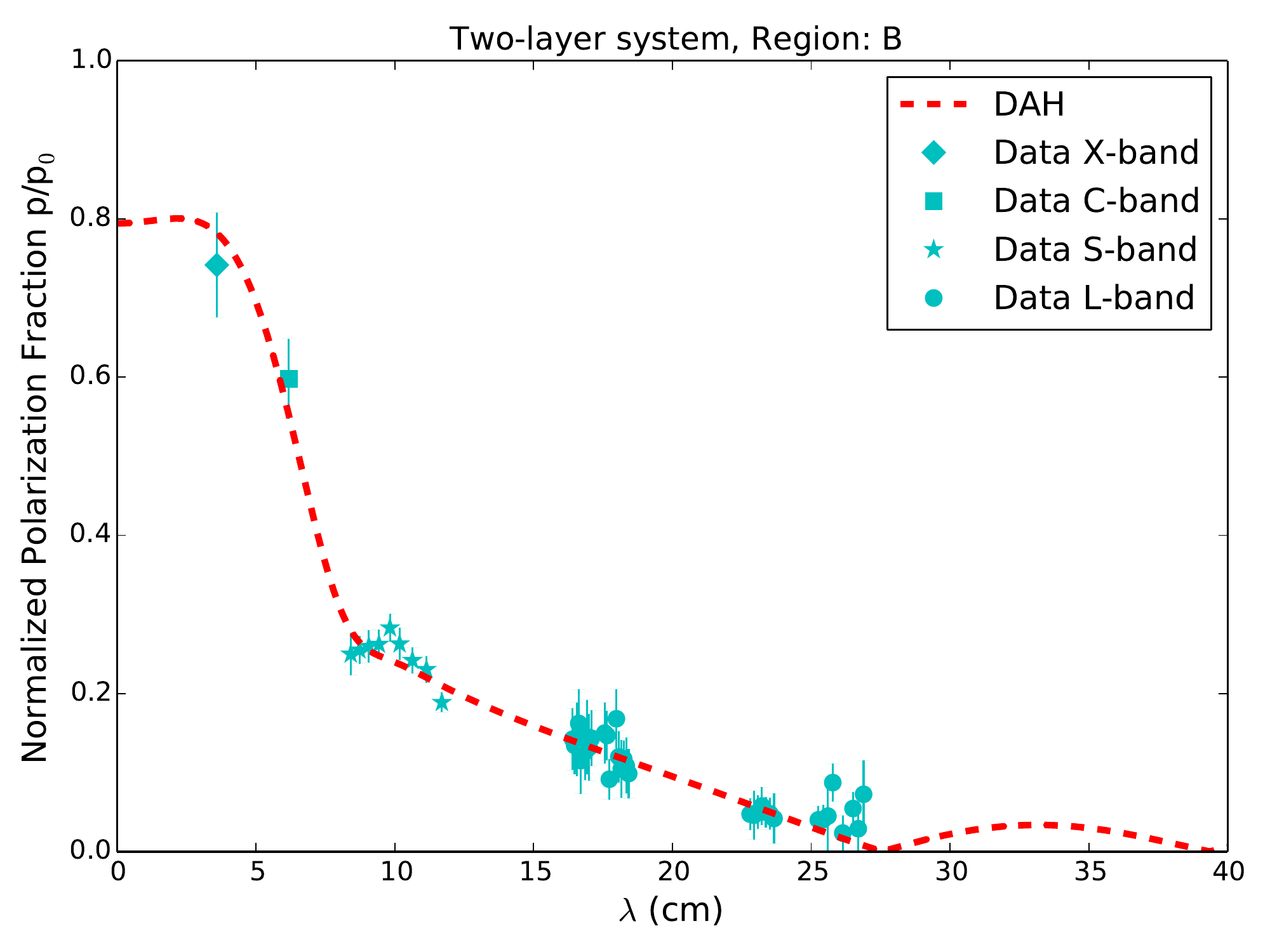}
\includegraphics[scale=0.42]{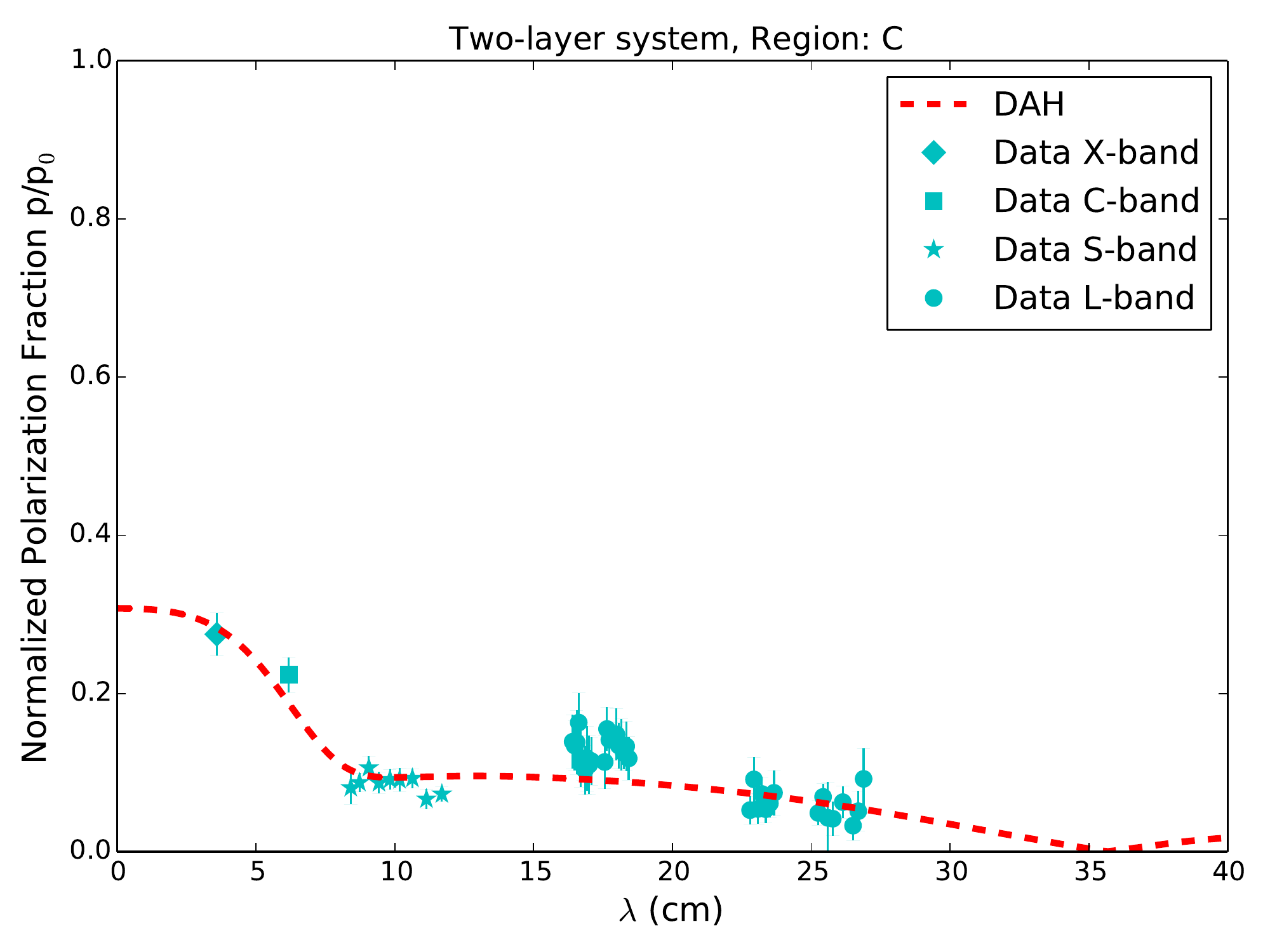}
\caption{Model DAH (red dashed line) for a two-layer system in M51 with the optimum set of parameters to represent the observed degree of polarization at multiple wavelengths in region ``A1'' (top), region ``B'' (middle), and region ``C'' (bottom).
} 
\label{fig:shneider_fit_otherRegions}
  \end{figure}

Using the interactive tool, we found optimum sets of parameters of the DAH model also for the observed polarization fractions
in three regions located in the same radial ring of 2.4\,--\,3.6\,kpc: A neighboring region with respect to the original region (marked with ``A1'' in Figure\,\ref{fig:M51_sectors}, at azimuthal angle $140\degr$), in a region located 
in the inter-arm 
on the western side of the galaxy (region ``B'', at azimuthal angle $285\degr$), and in a region located at a spiral arm with low signal-to-noise in polarized intensity (region ``C'', at azimuthal angle $190\degr$).

All discussed regions have low RM values in the X-, C-, to L-band (up to maximum of 25\,rad\,m$^{-2}$), hence, the contribution from a possible vertical magnetic field component is negligible.




The results
are listed in Table\,\ref{tab:parameters}, while Figure\,\ref{fig:shneider_fit_otherRegions} shows the DAH model configurations with the optimum parameter sets for the degrees of polarization observed in regions ``A1'', ``B'', and ``C''.
The relative uncertainties in the regular field strengths in disk and halo, $B_{\textrm{d}}$ and $B_{\textrm{h}}$, in Table\,\ref{tab:parameters} are between 8\% and 16\% (except for $B_{\textrm{d}}$ in region ``C'' that cannot be constrained well).
With the availability of the new S-band data, the model parameters are constrained better, with relatively lower uncertainty, as compared to \citet{Shneider14II}. 
The relative uncertainties in the other parameters in Table\,\ref{tab:parameters} are of the same order (except for the thermal densities in region ``C'').

The field strengths in region ``A1'' are very similar to those found in region ``A''.
In the inter-arm region ``B'' the degree of polarization is very large at high frequencies (up to 50\,\% at 8.5\,GHz or 3.5\,cm), close to the theoretical maximum.
The highly regular magnetic field is
even a factor of about two stronger than the turbulent magnetic field.
This confirms our current understanding of strong regular fields to be present in the inter-arm regions.
The low observed RM of $-6$\,rad\,m$^{-2}$ suggests that the regular field component along the line-of-sight, which means in the vertical direction, is small in thís region.
Our results for the spiral arm region ``C'' gives the smallest regular and the highest turbulent field strength, as expected for a region located in the spiral arm that contains only weak regular fields but a strong turbulent component. As for region ``A'', the model configurations DAH and DIH can be
represented with almost equal parameter values in all other regions.

For the ring of 4\,--\,3.6\,kpc, \cite{Shneider14II} found (also for a two-layer model) regular field strengths of $B_{\textrm{d}}\approx9\,\mu$G and
$B_{\textrm{h}}\approx4\,\mu$G and turbulent field strengths of
$b_{\textrm{d}}\approx11\,\mu$G and  $b_{\textrm{h}}\approx5\,\mu$G.
Our analysis shows that the field strengths vary significantly from one to another region (Table\,\ref{tab:parameters}).
As all four regions are located in the same radial ring, the model assumption of constant values within one radial ring is obviously not valid.
We find regular field strengths of $B_{\textrm{d}} = 8-18\,\mu$G in the disk and $B_{\textrm{h}} = 3\,-\,5\,\mu$G in the halo, while the turbulent field $b_{\textrm{d}}$ in the disk varies between 11\,$\mu$G and 24\,$\mu$G.

Our values yield a range of total field strengths in the disk of $B_{\textrm{tot,d}} = 21-26\,\mu$G in the four regions (Table\,\ref{tab:parameters}), which agrees well with the equipartition values in this radial range derived from the total synchrotron emission (see Figure\,8 of \citealt{Fletcher11}). This indicates that most of the total synchrotron emission emerges from the disk, while the halo hardly contributes.

The thermal electron densities also vary considerably between the four regions of our analysis. The average value in the disk of $n_{\textrm{e,d}}\sim 0.07$\,cm$^{-3}$ is significantly smaller than the constant value assumed by \cite{Shneider14} and \cite{Shneider14II} for this ring. This supports our approach to include the thermal electron densities
as free parameters and allow them to vary within a ring.


\section{Discussion}
\label{sec:discussion_main}

\subsection{Decreasing turbulent field strength in M51's outskirts}

\begin{figure}[t]
\centering
\includegraphics[scale=0.4]{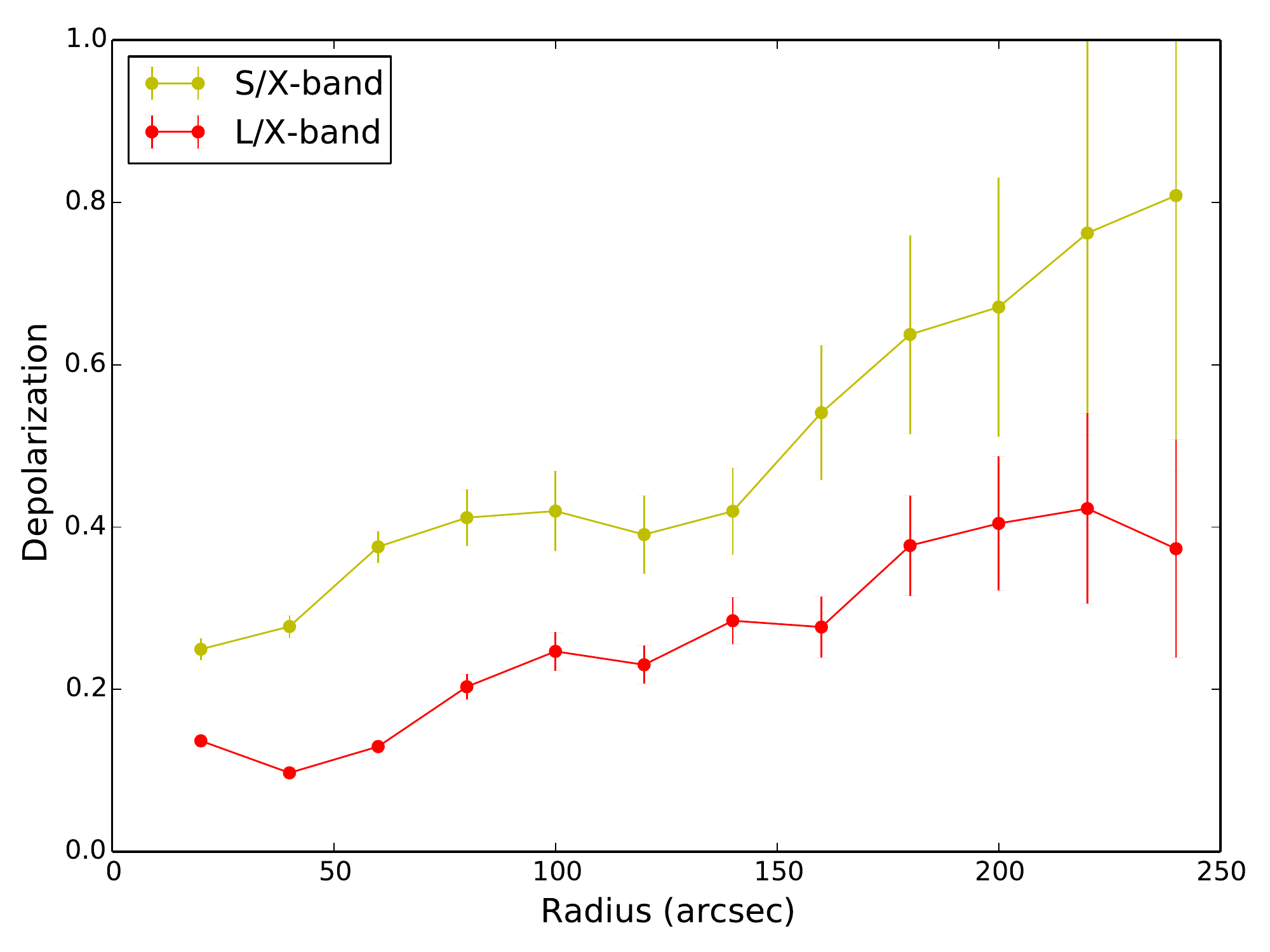}
\caption[Faraday depolarization as a function of radius in M51]{Faraday depolarization between the S-band (2\,--\,4\,GHz) and the X-band (8.35\,GHz), and between the L-band (1\,--\,2\,GHz) and the X-band in M51 as a function of radius in rings of 20$''$ radial width. The depolarization was calculated by the ratio of the degrees of polarization at the corresponding frequencies, where DP = 1 means no depolarization and DP = 0 means total depolarization.
} 
\label{fig:ring_depol}
  \end{figure}

The degree of polarization in M51 
increases towards larger radii (Section\,\ref{sec:p}),
which could be caused by a decrease of Faraday depolarization as a function of radius.  
To verify this,
we calculate the depolarization between different bands. Figure\,\ref{fig:ring_depol} shows the depolarization (DP\,=$p_{\nu_1}/p_{\nu_2}$) between the S-band (3.05\,GHz)
and the X-band (8.35\,GHz) and between the L-band (1.5\,GHz) and the X-band.
Indeed, DP increases towards larger radii (from 0.2\,--\,0.8 between the X-band and the S-band and from 0.1\,--\,0.4 between the X-band and the L-band), which shows that the depolarization effect becomes weaker at larger radii.

Decreasing Faraday depolarization (due to decrease in Faraday dispersion) towards larger radii can be attributed to a decreasing (isotropic) turbulent magnetic field strength $b$, probably also accompanied by a decreasing thermal electron density (compare Equation\,\eqref{eq:sigmaRM}), both of which are a consequence of decreasing star-formation rate (SFR) towards larger radii. $b$ and SFR are closely related \citep[e.g.,][]{2014AJ....147..103H}. A decreasing turbulent field strength is evident from the total synchrotron emission that decreases towards larger radii (Figure\,\ref{fig:StokesI}). 
The regular magnetic field can also contribute to Faraday depolarization (differential Faraday rotation, see Section\,\ref{sec:shneider}) and, because its strength is expected to decrease towards larger radii, this depolarization effect is also expected to decrease.

The model fits performed by \cite{Shneider14II} (see end of Section\,\ref{sec:shneider_M51_application}) in four rings over the radial range 2.4\,--\,7.2\,kpc (65\,--\,195\arcsec) did not indicate significant radial decreases of the turbulent and regular field strengths and hence cannot explain the results shown in Figure\,\ref{fig:ring_depol}.
The relative roles of the two above Faraday depolarization mechanisms should be investigated by an improved depolarization model.

At large radii, the observed degree of polarization is about 40\% in the X-band (left-hand panel of Figure\,\ref{fig:pmap}),
which is below the theoretical maximum degree of polarization of 76\%\,\footnote{With a typical synchrotron spectral index of $\alpha_\text{syn}=-1.1$ at the inter-arm regions \citep{Fletcher11}.},
suggesting that another
depolarization mechanism is 
at work.
The same turbulent fields responsible for depolarization by Faraday dispersion at smaller radii can also cause considerable wavelength-independent depolarization.
Isotropic turbulent fields have scales smaller than the scale probed by the telescope beam and hence cannot be resolved, so that
their synchrotron emission is unpolarized. If Faraday depolarization is negligible (for example, at the high frequency of the X-band), the observed degree of polarization $p$ depends on $B_{\textrm{ord},\perp}$, the strength of the ordered field in the sky plane, and $b$, the isotropic turbulent field (\citealt{1966MNRAS.133...67B}, corrected by \citealt{1996ASPC...97..457H}):
\begin{equation}
\frac{p}{p_0}=\frac{B_{\textrm{ord},\perp}^2}{B_{\textrm{ord},\perp}^2+\frac{2}{3}\,b^2}\, ,
\label{eq:B_order}
\end{equation}
with $p_0\,=\,0.76$ being the maximum degree of polarization. This allows us to determine the ratio of the isotropic turbulent field strength compared to the strength of the ordered field (regular plus anisotropic turbulent):
\begin{equation}
\frac{b}{B_{\textrm{ord},\perp}}=\sqrt{\frac{3}{2}\,\left(\frac{p_0}{p}-1\right)\,}\,\,.
\label{eq:B_ratio}
\end{equation}


Using the $p$ values observed in the X-band (Figure\,\ref{fig:radius}), the ratio $b/B_{\textrm{ord},\perp}$ decreases drastically from the central region towards larger radii. 
(Noteworthy, in the inner parts of M51 the X-band emission comes from disk, mostly from spiral arms. In the outer parts (at radii of about $>200\arcsec$) spiral arms are faint and the disk emission is weak, so that disk and halo contribute similarly.) We get $b/B_{\textrm{ord},\perp}\sim 3.6$ at a radius of 20$\arcsec$. At larger radii ($>\,50\arcsec$), we get an average ratio between $\sim 1.4$ in inter-arm regions (at 100$\arcsec\sim $3.7\,kpc) and in the outer disk (at 200$\arcsec\sim 7.4$\,kpc), whereas $\sim 1.8$ in regions of spiral arms at $140\arcsec$ ($\sim 5.2$\,kpc, compare Figure\,\ref{fig:radius}).

The above field ratios cannot be directly compared with those from our two-layer depolarization model (Table\,\ref{tab:parameters}) because Equation\,\eqref{eq:B_ratio} gives a weighted average between disk and halo. As our depolarization model did not deliver values for the turbulent field $b_\textrm{h}$ in the halo, we can compare $b/B_{\text{ord},\perp}$ only with the disk values $b_\textrm{d}/B_\textrm{d}$\,\footnote{Our depolarization model assumes that the regular field $B$ is oriented parallel to the plane, so that we can assume $B_\perp \approx B$ for the following estimate.}, which is valid for radii of about $<200\arcsec$ where the disk emission dominates. Furthermore,
because $B_{\textrm{ord},\perp}^2=B^2+b_\textrm{aniso}^2$, the ratio $b/B_{\textrm{ord},\perp}$ includes anisotropic fields and hence is a lower limit for the ratio $b/B$ that was derived from the depolarization model\,\footnote{We note that we considered the model setting DAH, which includes anisotropic fields. However, for the case with isotropic fields (DIH) the parameter values are almost equal in all four regions.}. 

In spite of the uncertainties, good agreement of the two ratios is found for the inter-arm regions ``A'' and ``A1'' ($b_\textrm{d}/B_\textrm{d} \sim 1.3$), indicating that the contribution of anisotropic turbulent fields to $B_{\textrm{ord},\perp}$ is small, while for the spiral arm region ``C'' we found $b_\textrm{d}/B_\textrm{d}\sim 2.9$ that is larger than $b/B_{\textrm{ord},\perp} \sim 1.8$ for spiral arm regions, as expected for significant anisotropic turbulent fields in such regions. In the inter-arm region ``B'',
hosting a particularly strong regular field,
$b_\textrm{d}/B_\textrm{d} \sim 0.6$ is
significantly smaller compared to the average of $\ge1.4$ for inter-arm regions 
as estimated  above. We conclude that the results applying Equation\,\eqref{eq:B_ratio} are consistent with those obtained in Section\,\ref{sec:sectors}.

\subsection{Traces of vertical fields in the disk-halo transition region of M51}
\label{sec:Bparadiscussion}

The regular magnetic field in the disk-halo transition region of M51
is apparently dominated by fluctuations (Section\,\ref{sec:Bpara}).
Table\,\ref{tab:RMvalues} gives the mean RM and RM dispersion found at the different frequency bands.

The RM dispersion expected from (isotropic) turbulent fields with turbulence size $d$ (``cells'') can be expressed as \citep[e.g.,][]{2016A&ARv..24....7B}:
\begin{align}
\label{eq:sigmaRM}
\begin{split}
\sigma_\text{RM} &= 0.81\,n_\text{e}\,b_\parallel \, d \sqrt{N_\parallel}\,\, / \sqrt{N_\perp} \\
&= 0.81\,n_\text{e}\,b_\parallel \sqrt{L\,d\,} \,\, (d/D)\, ,
\end{split}
\end{align}
which is a different version of Equation\,\eqref{eq:d}. 
$\sigma_\text{RM}$ increases with the square root of the number of cells $N_\parallel$ along each line-of-sight because a large number of cells allows that several of them have similar field directions, so that their RMs can accumulate.
On the other hand, a large number of cells $N_\perp$ across the beam of size $D$ averages out the RM values from individual lines-of-sight and reduces $\sigma_\text{RM}$.
Inserting physically reasonable values of $n_{\text{e}}$, $b_{\parallel}$, $L$, and $d$ into Equation\,\eqref{eq:sigmaRM}, we can calculate the dispersion of the RM distribution for the entire line-of-sight through M51.
We use the results from the \cite{Shneider14} two-layer model, that is, the mean values of the turbulent magnetic field strength of about 18\,$\mu$G (divided by $\sqrt{3}$ to get the parallel component) and thermal electron densities of about 0.07\,cm$^{-3}$ in the four discussed regions, shown in Table\,\ref{tab:parameters}. We use $L=800$\,pc and $d=55$\,pc from Table\,\ref{tab:parameters_start}.
With these values, we get a RM dispersion of only 12\,rad\,m$^{-2}$ in the disk. However, in Table\,\ref{tab:RMvalues}, 
$\sigma_{\rm RM}$ is observed to be about 50\,rad\,m$^{-2}$ in the X/C-bands and in the S-band, corresponding to the disk and disk-halo transition regions, respectively.
If the dispersion is caused by purely isotropic turbulent fields, we would expect to get the same $\sigma_\text{RM}$ from Equation\,\eqref{eq:sigmaRM} as those reported in Table\,\ref{tab:RMvalues}.  


To obtain a $\sigma_\text{RM}$ from Equation\,\eqref{eq:sigmaRM} consistent with observations (Table\,\ref{tab:RMvalues}), we would need a turbulence cell size of about 150\,pc in Equation\,\eqref{eq:sigmaRM}, assuming all other values to be correct and the dispersion caused only by isotropic turbulent fields. From observations we know that the turbulence cell size can vary, especially from the disk to the halo, but a cell size larger by a factor of three than commonly referred to the disk of M51 \citep{2013ApJ...766...49H} is hard to explain. Furthermore, in the Milky Way, different studies consistently give sizes of $d$ between 10\,pc and 100\,pc (e.g., \citealt{1989ApJ...343..760R, 1993MNRAS.262..953O, 2008ApJ...680..362H}), so that turbulent fields can hardly explain the large RM dispersion.

We note that a part of the dispersion of 50\,rad\,m$^{-2}$ in the observed RM distribution
could arise due to the errors
from measurement noise. As can be seen in Figure\,\ref{fig:RMmap} (right-hand panel), RM errors in the majority of the pixels are $\lesssim 15$\,rad\,m$^{-2}$. 
We therefore believe that the discrepancy between the
expected and observed $\sigma_\text{RM}$ is real, but perhaps by only slightly less than a factor of three.

Potential sources of field disturbances are star formation-driven gas flows forming holes in neutral hydrogen (\textsc{Hi}) with a typical size of about 1\,kpc and transporting the local regular magnetic field from the disk into higher layers, causing that RMs
change from one to the other edge of the hole.
\citet{2012ApJ...754L..35H} found a co-location of a sinusoidal variation in RM and a hole observed in neutral hydrogen (\textsc{Hi}) close to a spiral arm of the face-on nearby spiral galaxy NGC\,6946.
Also, \citet{2017A&A...600A...6M} found one RM variation coinciding with a \HI hole in the face-on spiral galaxy NGC\,628 in the S-band.
On the other hand, by comparing the position of \HI hole detections in M51 \citep{2011AJ....141...23B} with our RM map at 15$''$ in the S-band visually, no obvious RM variation coinciding with a \HI hole was 
found.
A reason for the non-detection of RM variations coinciding with \HI holes in M51 could be 
the weak large-scale regular magnetic field in the disk-halo transition region and that the field is dominated by fluctuations and reversals. For such fields we do not expect to see a systematic variation in RM even if a magnetic field loop is formed and coincides with the \HI hole because randomly occurring field reversals along the loop cancel out each other. Hence, a RM variation across a \HI hole may only be observable in the presence of a strong large-scale regular (coherent) magnetic field component.  

Promising reasons for the broadening of the observed RM distribution could be:
(1)
\textbf{Tangled regular fields} have a correlation length similar to or larger than the beam size (see Section\,\ref{sec:complex} for a further discussion).
(2)
\textbf{Vertical fields} present in the disk and halo increase $B_{\parallel}$ and RM.



Vertical fields can be generated by
galactic winds, fountains, supernova remnants, or 
Parker instabilities \citep{1966ApJ...145..811P}.
\citet{2016ApJ...816....2R} performed numerical MHD simulations of the Parker instability in a domain of size 6\,kpc\,$\times$\,12\,kpc\,$\times\,$3.5\,kpc and computed RM maps for face-on view (their Figure\,12). RM were found to reverse on scales of 1\,--\,2\,kpc.
Global numerical MHD simulations of galaxies by \citet{2018MNRAS.481.4410P} revealed reversing signs of the wind-driven vertical field components also on scales of 1\,--\,2\,kpc (see their Figure\,8).
The RM maps of two synthetic galaxies were found to be similar to the RM map presented by \citet{Fletcher11}.

Observations show the presence of vertical magnetic fields in most edge-on spiral galaxies \citep{2015AJ....150...81W,2020arXiv200414383K}.
Variations in field strength and direction and/or in thermal electron density lead to RM dispersion.
For example, the vertical filaments in NGC\,4631 change field directions on a scale of several kpc \citep{2019A&A...632A..11M}.
Reversals on smaller scales may exist, but cannot be detected in RM maps of edge-on galaxies with the resolution of present-day observations.
We conclude that a system of vertical fields with reversing directions could give rise to the large RM dispersion observed in M51.

\subsection{The complex nature of the magnetic fields in M51}
\label{sec:complex}

\textbf{Inconsistency of RMs from regular fields:}
In Section\,\ref{sec:sectors}, we report
the optimum values of regular magnetic field strengths and thermal electron densities in the radial ring 2.4\,--\,3.6\,kpc of M51 derived from the observed degrees of polarization in different regions in this ring (Table\,\ref{tab:parameters}).
Field strengths and electron densities are found not to be constant along azimuthal angle in the ring, as assumed by the model, but show considerable variations. This is strengthened by comparing  
the RM values
derived from the
field strengths and electron densities in Table\,\ref{tab:parameters}
with the observed values of RM in Figure\,\ref{fig:RMmap}.
Using the regular field strength $B$ (in $\mu$G), the electron density $n_\text{e}$ (in cm$^{-3}$) from Table\,\ref{tab:parameters}, the pathlength $L$ (in pc) from Table\,\ref{tab:parameters_start}, and the azimuthal angle $\phi$ of the region, we can estimate the expected RM
from Equations\,\eqref{eq:RM_model} and \eqref{eq:B_model}\,\footnote{We also use the revised version of Equations\,(6.1.4) in \citet{KierdorfThesis2019}, given in the Appendix (Equation\,(\ref{eq:thesis})), to relate $B_\text{d}$ to the mode amplitudes $B_0$ and $B_2$ in the disk and $B_\text{h}$ to the mode amplitudes $B_{\text{h}0}$ and $B_{\text{h}1}$ in the halo.}.
For region ``A'' we expect RM$_\text{d}\approx+140$\,rad\,m$^{-2}$ in the disk and RM$_\text{h}\approx+40$\,rad\,m$^{-2}$ in the halo. For region ``B'', we expect RM$_\text{d}\approx-430$\,rad\,m$^{-2}$ in the disk and RM$_\text{h}\approx+10$\,rad\,m$^{-2}$ in the halo.
In the S-band we trace part of the disk (according to Fig.\,\ref{fig:ring_depol} about 40\% in the considered ring) and the halo, so that a total RM of $\,\approx+100$\,rad\,m$^{-2}$ and $\,\approx-160$\,rad\,m$^{-2}$ should be observed in regions ``A'' and ``B'', respectively. However, we measure only RM$\,=+3\pm20$\,rad\,m$^{-2}$ and RM$\,\sim-6$\,rad\,m$^{-2}$ in these regions
(Figure\,\ref{fig:RMmap}).
For the halo, we measure RM$\,=-3\pm10$\,rad\,m$^{-2}$ in the L-band \citep{2015ApJ...800...92M} in region ``A'', also much lower than expected. (The mean RM in the L-band in region ``B'' is not possible to determine because of low signal-to-noise.)
Hence, the strengths of the regular field derived from the depolarization model are inconsistent with the observed RMs. This indicates a complex three-dimensional structure of the regular fields, as elaborated in the following.


\textbf{Tangled regular fields:} The time scale for a fully developed large-scale regular field generated by the mean-field ($\alpha-\Omega$) dynamo in a spiral galaxy is almost 10\,Gyr \citep{2009A&A...494...21A}.
Interactions with M51's companion galaxy may distort this process and prolong the build-up time, so that the large-scale regular field may not yet have reached its saturation level.
As a result, the evolving regular field could still be ``spotty'' and tangled, as indicated from numerical simulations \citep{2009ApJ...706L.155H,2011AN....332..524A,2012A&A...537A..68M}. In terms of dynamo theory, field amplification occurs simultaneously on all scales, from the energy injection scale of turbulence to the size of the galaxy, with the smallest scales being amplified fastest \citep[e.g.,][]{2012SSRv..169..123B}. The mean-field dynamo is always accompanied by field amplification on smaller scales, which causes tangling \citep[see, e.g., Figure\,3 in][]{2013MNRAS.430L..40G}.

The \citet{Shneider14} model considers the observed degree of polarization as a function of wavelength.  
Faraday depolarization by a regular field (differential Faraday rotation, DFR)
does not depend on the sign of the regular field. In the case of tangled regular fields that change sign along the line-of-sight, DFR in such a reversing field is the same as in the case without reversals, whereas the observed RM is smaller in the first case.
Such reversals may occur on scales similar to or larger than the scale traced by the telescope beam ($\sim 550$\,pc), much larger than the scale of the anisotropic turbulent field ($\sim 55$\,pc).
Tangled regular fields may solve the discrepancy between the values of RM obtained from multi-layer depolarization models and the observed RMs.
Tangled regular fields also contribute to polarized intensity and may reduce the quest for small-scale anisotropic turbulent fields.


Tangled regular fields also contribute to RM dispersion. We propose to express the RM dispersion expected from tangled plane-parallel regular fields of strength $B_\text{tang}$ (similar to Equation\,\eqref{eq:sigmaRM}) as:
\begin{align}
\label{eq:sigmaRM2}
\sigma_\text{RM} = 0.81\,n_\text{e}\,B_\text{tang} \sqrt{L\,a\,} \,\, (a/D)\, ,
\end{align}
where $a$ is the scale of field reversals (correlation scale).\,\footnote{The factor $(a/D)$ applies if the beam size $D$ is larger than $a$.}
Dynamo theory predicts that regular fields with a range of scales $a$ will be present in a galaxy \citep{2012SSRv..169..123B,2013MNRAS.430L..40G}. Observing with higher resolution (hence, with a smaller beam size $D$) should yield larger values of $\sigma_\text{RM}$, to be tested with future observations.


\subsection{Limitations of the \cite{Shneider14} model}
\label{sec:Shneider_discussion}

In Section\,\ref{sec:modelSection}, we discussed the analytical multi-layer depolarization model developed by \cite{Shneider14}, applied to the galaxy M51. 
So far, this model is the only one that distinguishes between isotropic and anisotropic turbulent fields and includes multiple layers along the line-of-sight.
The \citet{Shneider14} model has certain limitations.\\

We propose the refinement of the following
assumptions:

\begin{itemize}
 
\item For the CRE density, \citet{Shneider14} assumed the same value in the disk and halo. However, this is inconsistent with the exponential scale heights of the synchrotron emission in edge-on galaxies \citep{2018MNRAS.476..158H}. A typical exponential scale height of $h_\text{syn} \sim 1.5$\,kpc gives a CRE scale height of $h_\text{CR}\,=\,h_\text{syn}\cdot\,(3+\alpha_\text{syn})\,/2\,\sim 3$\, kpc (assuming energy equipartition between CRs and magnetic fields), so that the CRE density should decrease from the disk to the halo
by a factor of $\sim\,1.4$ at a height of 1\,kpc above the disk plane and up to a factor of $\sim\,5$ at a height of 5\,kpc. 
A different CRE density changes the synchrotron intensity. If the values of the CRE density in the disk and halo are different, they do not cancel out when calculating $\left(p/p_0\right)$.

\item For the three-layer model it is not possible to ``lift up'' the zero drops of the degree of polarization as a function of wavelength by changing any of the free parameters. Even when including turbulent fields in the disk, the Sinc-function trend of the degree of polarization as a function of wavelength remains.
The reason is that in the model the polarized emission from the halo always dominates the trend of the degree of polarization, because the field strength and electron density are assumed to be constant in the halo and the path length through the halo is much larger than that through the disk. These assumptions are hardly realistic and should be modified. Furthermore, model configurations including turbulent fields in the halo should be included in future analysis especially of three-layer systems, in order to suppress the Sinc-function trend caused by depolarization from the regular field.

\item The depolarization model assumes the same synchrotron spectral index in disk and halo, while different values should be used according to observations in edge-on galaxies, as the result of energy losses of CREs
\citep[e.g.,][]{2019A&A...632A..12S}.
This is especially important at high frequencies ($\gtrsim 5$\,GHz), because the polarized
emission from the relatively steeper spectrum halo could be sub-dominant compared to emission
from the flatter spectrum disk.
Therefore, the intrinsic degree of polarization (at $\lambda=0$) should also be different in disk and halo.

\item For calculating the turbulence cell size in M51 via Equation\,\eqref{eq:d}, \citet{Shneider14} used the same $\sigma_{\text{RM,D}}=15\,$rad\,m$^{-2}$ \citep{Fletcher11} in the disk and halo,
which is however incompatible with observations (Table\,\ref{tab:RMvalues}):
In the L-band (tracing only the polarized emission from the halo) we found $\sigma_{\text{RM,D}}\approx14\,$rad\,m$^{-2}$, while at higher frequencies
$\sigma_{\text{RM,D}}\approx50\,$rad\,m$^{-2}$. 
This yields a turbulence cell size of about 90\,pc in the disk
and about 250\,pc in the halo.

\item Assuming a constant size of turbulence cells within the disk (55\,pc) and within the halo (370\,pc) could be too simplistic.
The turbulence cell in spiral arms could be smaller compared to the size of turbulence at inter-arm locations (due to star-forming processes, which mainly takes place in the dense spiral arms, driving the turbulence). For example, for the Milky Way, \citet{2008ApJ...680..362H} found a turbulence cell size of less than 10\,pc in the gaseous spiral arms, while in the inter-arm locations the turbulence cell size amounts to about 100\,pc.


\end{itemize}


The following fundamental improvements are proposed:

\begin{itemize}

\item \citet{Shneider14} only considered the degree of polarization as a function of wavelength. As discussed in Section\,\ref{sec:complex}, the model parameters should be checked for consistency with the observed RM values, allowing for variations of the magnetic field strengths and electron densities in disk and halo along each ring.

\item Another future step would be
to extend the model to \textrm{fitting} the observed Stokes $Q(\lambda)$ and $U(\lambda)$ values. This means considering not only the amplitude of the polarized signal but also the phase between Stokes $Q$ and $U$
and hence the polarization angles.


\item The \citet{Shneider14} model neglects vertical components (with respect to the disk plane) of the regular magnetic field.
From polarization observations of edge-on galaxies \citep[e.g.,][]{2019A&A...632A..11M} we know that vertical field components exist in spiral galaxies. Some face-on galaxies show clear evidence of vertical regular fields in their RM maps, too. For example, \citet{2015ApJ...800...92M} found in their L-band data signature of an overall vertical magnetic field component in the halo of M51 that produces a RM of about -9\,rad\,m$^{-2}$. From dynamo theory we know that
magnetic fields in spiral galaxies has non-vanishing vertical components. For a quadrupolar magnetic field configuration, as it was considered in the discussed depolarization model, vertical field components are required to fulfill the divergence-free condition. 
In a quadrupolar magnetic field of M51, the negative direction of the vertical field component (away from us) must be accompanied by an outward-directed radial field component in the disk, which is indeed observed (see Figure\,14 in \citealt{Fletcher11}). Therefore,
vertical field components need to be implemented in the depolarization model.

\item We can test
if there is a signature of a quadrupole or a dipole halo field in the equations by
flipping the sign of the vertical component. This would provide an important step towards understanding the symmetry properties of different magnetic field configurations.

\end{itemize}

Still, the \cite{Shneider14} model already gives a good approximation for a multi-layer magneto-ionic medium and has strong advantages compared to classical depolarization models: it contains many more -- galaxy specific -- details and is able to decompose different layers along the line-of-sight, which is especially advantageous in case of the complicated magnetic field configuration in M51 (having different configurations in the disk and halo).

The optimum parameter values from our interactive tool described in Section\,\ref{sec:shneider_M51_application} can be used as initial conditions to 
perform an automized least-square fit or, even better, a MCMC simulation to probe the full posterior distribution of the parameter values of the depolarization model
and to determine the values of the parameters accurately (if converging).

\section{Summary and outlook}
\label{sec:summary}

In this paper, we present a radio observational study of the magnetic field properties of the nearby grand-design spiral galaxy M51. 
The observations were performed using the 
VLA providing high spatial resolution and a good image quality at a wide frequency coverage in the range 2\,--\,4\,GHz (S-band) thanks to the broadband high-sensitivity receivers. 
Broadband polarization data allow us to probe the frequency-dependent character of the polarized emission and thus to study depolarization mechanisms caused by different underlying magnetic field configurations.

Studying M51 in the S-band traces a so-far unknown polarized layer that is proposed to probe the transition region between the disk and halo. The goal was to make a major step towards understanding how large-scale regular magnetic fields are generated in the halo of spiral galaxies and how they are connected to the disk field. 
Here, we highlight some of the major findings.


We found an increasing degree of polarization in the S-band as a function of radius (Figure\,\ref{fig:radius}).
The Faraday depolarization calculated between different frequency bands (Figure\,\ref{fig:ring_depol}) implies a decreasing turbulent magnetic field strength and/or a decreasing thermal electron density towards larger radii in M51. 

The observed Faraday rotation measures (RMs) in the disk-halo transition region do not reveal an obvious large-scale pattern (Figure\,\ref{fig:RMmap}), but instead show a large dispersion (Figure\,\ref{fig:RMhist}). 
These results indicate that
the observed RMs in the disk-halo transition region are dominated by \textbf{tangled regular fields} and/or
\textbf{vertical fields} (with respect to the galaxy plane), distorting any signature of a large-scale pattern of the regular field and increasing the RM dispersion, 
as discussed in Sections\,\ref{sec:Bparadiscussion} and \ref{sec:complex}.


To get information on the field strength and structure in different layers,
the new S-band polarization data were combined with VLA + Effelsberg observations in the C- and X-bands at 4.85\,GHz and 8.35\,GHz and with the broadband L-band (1\,--\,2\,GHz) VLA data. 
In Section\,\ref{sec:modelSection}, we compared the observed degrees of polarization as a function of wavelength to a depolarization model from \cite{Shneider14}.
The new S-band data are critical to distinguish between a two-layer (disk\,--\,halo) and a three-layer (halo\,--\,disk\,--\,halo) system.
A two-layer model of M51 is preferred.

In general, we found that the model configurations need to contain regular fields in the disk and the halo, as well as turbulent fields
in the disk.
Anisotropic turbulent fields give a small contribution to the polarized signal and increases the intrinsic degree of polarization at short wavelengths by a few per cent compared to purely isotropic turbulent fields.
The model configurations DAH and DIH (that is, regular fields in the disk and halo and anisotropic or isotropic turbulent fields in the disk) can be
represented equally well
with almost the same field strengths and electron densities. Therefore, we cannot distinguish whether the turbulent field in the disk is isotropic or anisotropic.

Our 
study provides an estimate of the regular and turbulent magnetic field strengths in nearby galaxies,
independent of the widely used assumption of equipartition between the energies of magnetic field and of cosmic rays. 
In the three spiral arm regions we investigated, the turbulent field in the disk dominates with strengths between $18\,\mu$G and $24\,\mu$G, while the regular field strengths are between $8\,\mu$G and $16\,\mu$G. In an inter-arm region, the regular field strength of  $18\,\mu$G exceeds that of the turbulent field of $11\,\mu$G (Table\,\ref{tab:parameters}). The strengths of the regular fields in the halo are $3-4\,\mu$G in all four regions. (The strengths of the turbulent fields in the halo could not be determined.) The relative uncertainties of field strengths are mostly between 8\% and 20\%. The total field strengths in the disk are consistent with the equipartition estimates by \cite{Fletcher11}. The thermal electron density also varies considerably between the four regions of our analysis.

We found a striking discrepancy between the RMs expected from the regular field strengths from the depolarization model and the observed RMs. This indicates frequent field reversals of the regular field with correlation lengths larger than the scale traced by the telescope beam ($\sim550$\,pc), another hint to tangled regular fields (Section\,\ref{sec:complex}), possibly accompanied by vertical filaments of regular fields driven by outflows (Section\,\ref{sec:Bparadiscussion}). Tangled regular fields are predicted by models of evolving large-scale dynamo fields, and vertical fields are predicted by numerical simulations of Parker instabilities or galactic winds.

The depolarization model should be refined and extended (Section\,\ref{sec:Shneider_discussion}).
Its systematical application to the entire galaxy,
and allowing the magnetic field and electron density to vary spatially, would
deliver maps of these quantities over the galaxy.
Vertical fields should be included in the model because these are required to exist from dynamo theory.
Significant vertical field components would leap out as
outliers of the values of one or more of the
parameters. 
This would provide knowledge on the
field properties in the disk and halo of M51
and other nearby spiral galaxies with almost face-on orientation.
We also need an improved dynamo model for M51 that includes tidal forces and a significant halo component \citep[see Chapter~8.2 in][]{2019Galax...8....4B}.


In future observations, higher angular resolution, along with higher signal-to-noise ratio in polarized intensity to reduce the error in RM, would help to investigate the detailed spatial distribution of
regular fields and possible relations to dynamic phenomena in the ISM of the disk (for example, \HII regions, Parker loops, supernova remnants, etc).

We showed that with broadband polarization data, depolarization mechanisms can be used as a powerful
tool to probe the 3-D structure of magnetic fields in galaxies. 
Future capabilities provided by the new Square Kilometre Array (SKA), with dramatically improved
frequency coverage and excellent sensitivity for regions with weak radio surface brightness, will
facilitate new studies of magnetic fields in the ISM of nearby galaxies. For this forthcoming radio astronomy era, our paper shows the path towards analyzing and interpreting broadband polarization data. 



\begin{acknowledgements}
We thank Dr. Olaf Wucknitz for careful reading of the manuscript and the anonymous referee for valuable suggestions for improvements. We thank Drs. Carl Shneider and Anvar Shukurov for helpful discussions. A.B. acknowledges financial support by the German Federal Ministry of Education and Research (BMBF) under grant 05A17PB1 (Verbundprojekt D-MeerKAT). -- The VLA is operated by the National Radio Astronomy Observatory (NRAO). The NRAO is a facility of the National Science Foundation operated under cooperative agreement by Associated Universities, Inc. This work is based (in part) on observations made with the Spitzer Space Telescope, which was operated by the Jet Propulsion Laboratory, California Institute of Technology under a contract with NASA.
\end{acknowledgements}

\bibliography{references}{}
\bibliographystyle{aa}

\clearpage
\appendix 

\section{Consistency of the new S-band data with the model of the large-scale magnetic field}
\label{sec:mode_analysis}

\begin{figure*}[ht!]
\centering
\includegraphics[scale=0.42]{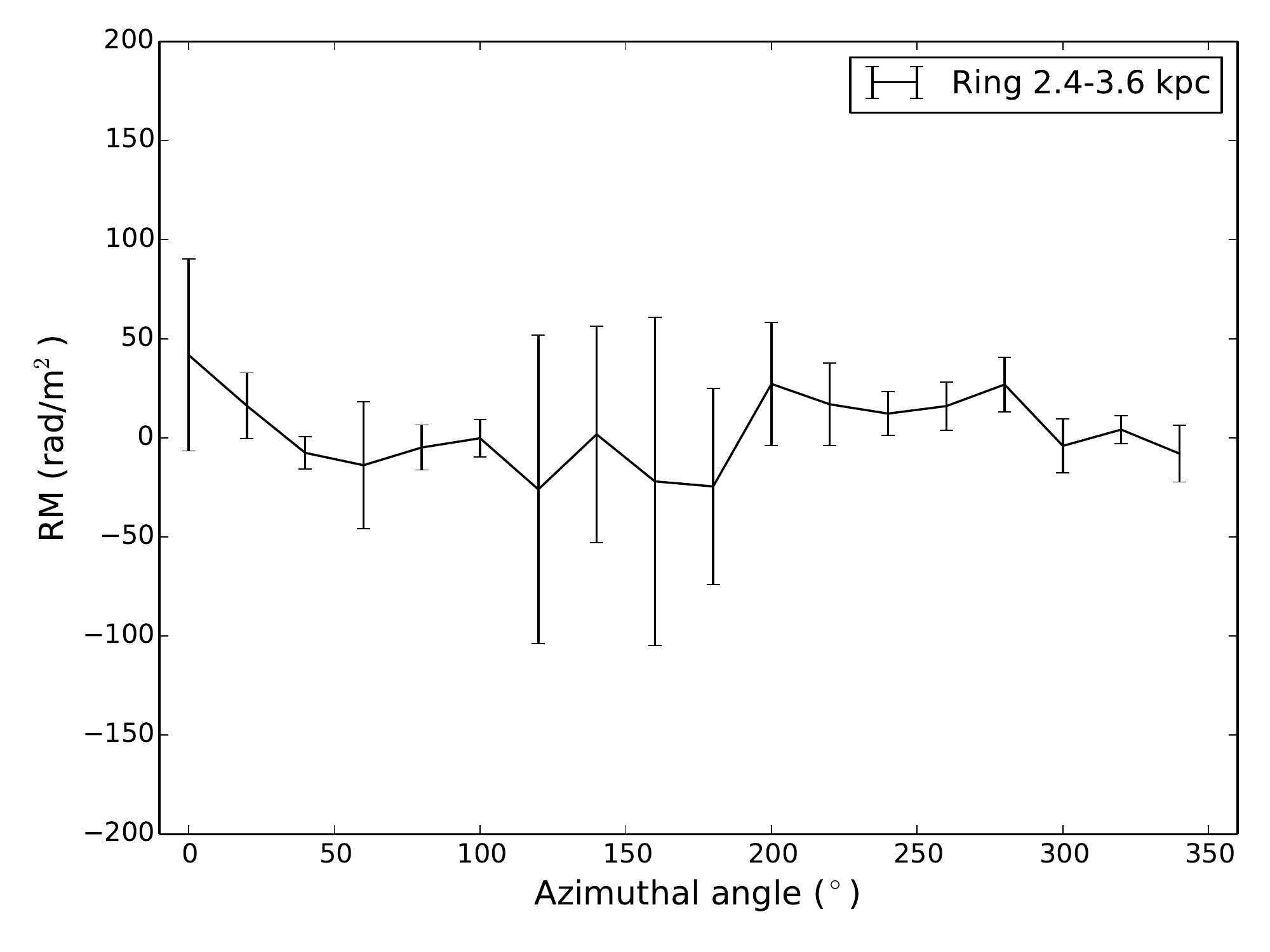}\includegraphics[scale=0.42]{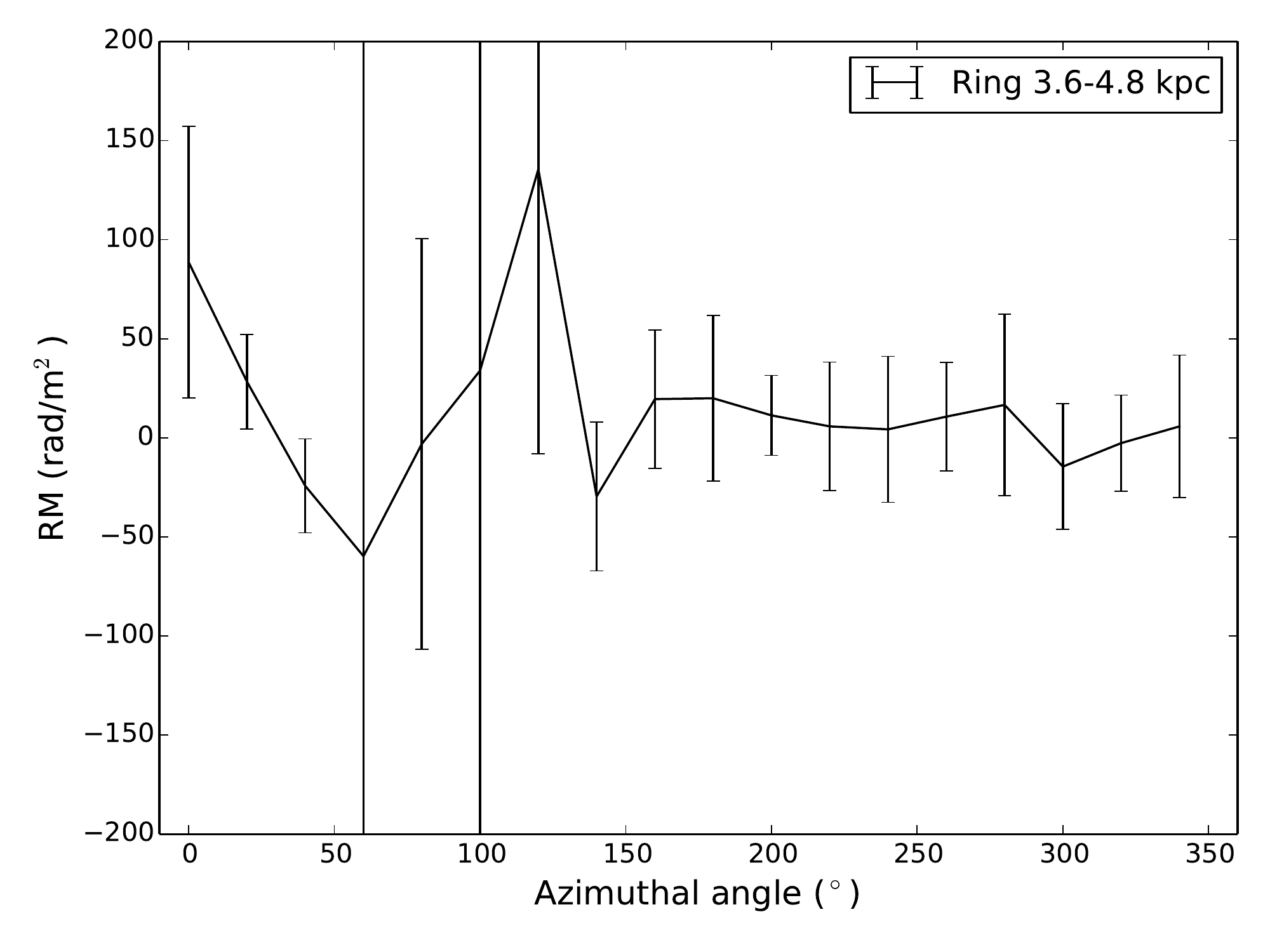}
\includegraphics[scale=0.42]{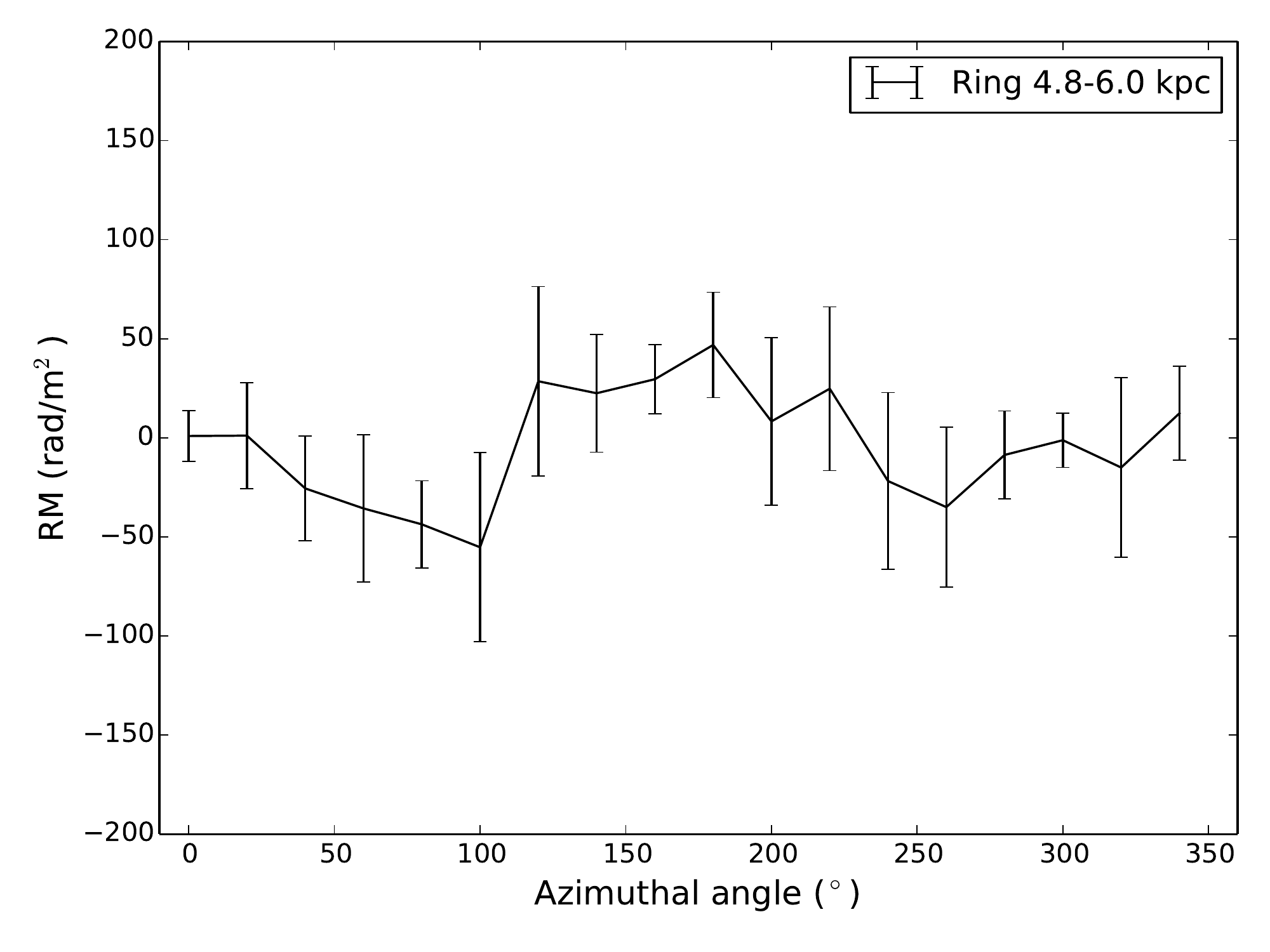}\includegraphics[scale=0.42]{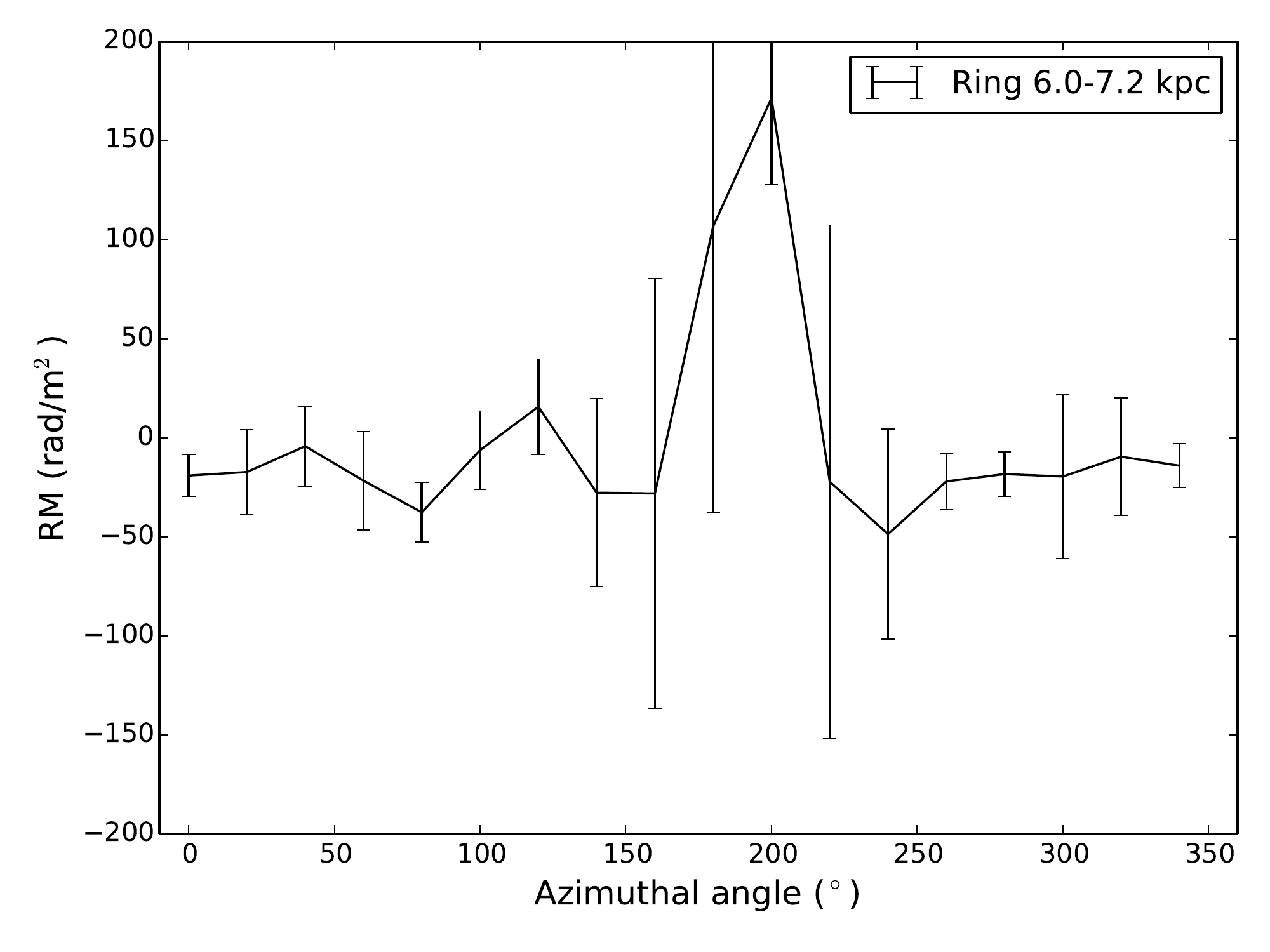}
\caption{Rotation measure against azimuthal angle in the galaxy plane for four rings with 1.2\,kpc width. The data points show the RM averaged in sectors with an opening angle of $20\degr$ with the standard deviation in each sector as the error bar.
} 
\label{fig:mode_fits}
  \end{figure*}

To check whether we probe the transition region in the S-band, we tested if the new S-band data are compatible with the model of the large-scale regular magnetic field by \citet{Fletcher11}
who found azimuthal modes $m=0+2$ in the disk, $m=0+1$ in the halo in the innermost ring, and only a $m=1$ mode in the halo in all other rings. 
To describe the disk and halo in the model, the Faraday rotation from M51 is split into two components:
\begin{align}
\text{RM}(\phi) &= \text{RM}_{\text{fg}} + \xi_\text{D}\, \text{RM}(\phi)_\text{D} + \xi_\text{H}\, \text{RM}(\phi)_\text{H}\, ,
\end{align}
where $\phi$ is the azimuthal angle
in degrees, $\text{RM}_{\text{fg}}$ is the foreground rotation measure from the Milky Way, and $\xi_\text{D}$ and $\xi_\text{H}$ are parameters that allow us to model what fraction of the disk and halo are visible in polarized emission at a given wavelength.
A $\xi$-parameter equal to 0 in the disk means that the disk is invisible in polarized emission, while the value of 1 means that we can observe the whole layer in polarization.
\citet{Fletcher11} used $\xi_\text{D}=0$ and $\xi_\text{H}=1$ in the L-band and $\xi_\text{D}=1$ and $\xi_\text{H}=1$ in the C- and X-bands. 

The RMs in the disk and halo are proportional to the field components along the line-of-sight:
\begin{align}\label{eq:RM_model}
\begin{split}
\text{RM}_\text{D} &\propto -\left[B_r\sin(\phi) + B_\phi\cos(\phi)\right]\sin(l) \\
\text{RM}_\text{H} &\propto -\left[B_{\text{h}r}\sin(\phi) + B_{\text{h}\phi}\cos(\phi)\right]\sin(l)~, 
\end{split}
\end{align}
where
\begin{align}\label{eq:B_model}
\begin{split}
B_r &= B_0\sin(p_0) + B_2\sin(p_2)\cos(2\phi - \beta_2) \\
B_\phi &=  B_0\cos(p_0) + B_2\cos(p_2)\cos(2\phi - \beta_2)\\
B_{\text{h}r} &= B_{\text{h}0}\sin(p_{\text{h}0}) + B_{\text{h}1}\sin(p_{\text{h}1})\cos(\phi - \beta_{\text{h}1}) \\
B_{\text{h}\phi} &=  B_{\text{h}0}\cos(p_{\text{h}0}) + B_{\text{h}1}\cos(p_{\text{h}1})\cos(\phi - \beta_{\text{h}1})
\end{split}
\end{align}
are the components of the large-scale regular field
in cylindrical coordinates, where ($r$, $\phi$) are the radial and azimuthal coordinates in the galaxy plane. 
Equations \eqref{eq:RM_model} and \eqref{eq:B_model} are taken from \citet{Berk97} (Equation\,(A2)), neglecting the field component oriented perpendicular to the galaxy plane, and \citet{Shneider14II} (Equation\,(1)), respectively.
$B_0$, $B_{\text{h}0}$, $B_2$, $B_{\text{h}1}$, and $p_0$, $p_{\text{h}0}$, $p_2$, $p_{\text{h}1}$ are the amplitudes and pitch angles of the corresponding field modes in the disk and halo, respectively. $\beta_2$ and $\beta_{\text{h}1}$ are the azimuthal angles at which the corresponding $m>0$ mode has its maximum.
If the field amplitudes are expressed in units of rad/m$^2$, as in Equation\,(A1) of \citet{Fletcher11}, then Equation\,\eqref{eq:RM_model} above gives RMs directly.


To compare the S-band data to the model by \citet{Fletcher11},
the observed RM in the S-band was averaged in sectors of rings with 1.2\,kpc width (while the resolution is $15''=\,$0.55\,kpc in the S-band), each with an opening angle of $20\degr$. The error bars of RM are given by the standard deviation in each sector. The results are shown in Figure\,\ref{fig:mode_fits}. Our data are found to be consistent with the large-scale RM variation expected from the model by \citet{Fletcher11}. However, fits with constant RM value have similar values of reduced $\chi^2$. Hence, the quality of the present S-band data does not allow to test or improve the \citet{Fletcher11} model.

\section{Revised version of Equation 6.1.4 in \cite{KierdorfThesis2019}}

The amplitudes of the modes in disk and halo are related to the regular fields $B_{\text{d}}$ and $B_{\text{h}}$ in disk and halo derived by the depolarization model (Table\,\ref{tab:parameters}) as follows:

\begin{equation}
\label{eq:thesis}
\begin{split}
B_0 &= B_{\text{d}}\,\,\biggl[1 + \left(\frac{R_2}{R_0}\right)^2 \cos^2(2\phi - \beta_2) \\&\qquad + 2\left(\frac{R_2}{R_0}\right)\,\cos(2 \phi - \beta_2)  \cos(p_0-p_2)\biggr]^{\nicefrac{-1}{2}},\\
B_2 &= B_{\text{d}}\,\,\biggl[\left(\frac{R_0}{R_2}\right)^2 + \cos^2(2\phi - \beta_2) \\&\qquad + 2\left(\frac{R_0}{R_2}\right)\,\cos(2 \phi - \beta_2)  \cos(p_0-p_2)\biggr]^{\nicefrac{-1}{2}},\\
B_{\text{h}0} &= B_{\text{h}}\,\,\biggl[1 + \left(\frac{R_{\text{h}1}}{R_{\text{h}0}}\right)^2 \cos^2(\phi - \beta_{\text{h}1}) \\&\qquad + 2\left(\frac{R_{\text{h}1}}{R_{\text{h}0}}\right)\,\cos(\phi - \beta_{\text{h}1})  \cos(p_{\text{h}0}-p_{\text{h}1})\biggr]^{\nicefrac{-1}{2}},\\
B_{\text{h}1} &=  B_{\text{h}}\,\,\biggl[\left(\frac{R_{\text{h}0}}{R_{\text{h}1}}\right)^2 + \cos^2(\phi - \beta_{\text{h}1}) \\&\qquad+ 2\left(\frac{R_{\text{h}0}}{R_{\text{h}1}}\right)\,\cos(\phi - \beta_{\text{h}1}) \cos(p_{\text{h}0}-p_{\text{h}1})\biggr]^{\nicefrac{-1}{2}}
\end{split}
\end{equation}

\section{Data and additional plots}

\begin{figure}[ht]
\centering
\hspace*{-0.2cm}\includegraphics[scale=0.2, trim=2.4cm 0.8cm 0.0cm 1.2cm, clip]{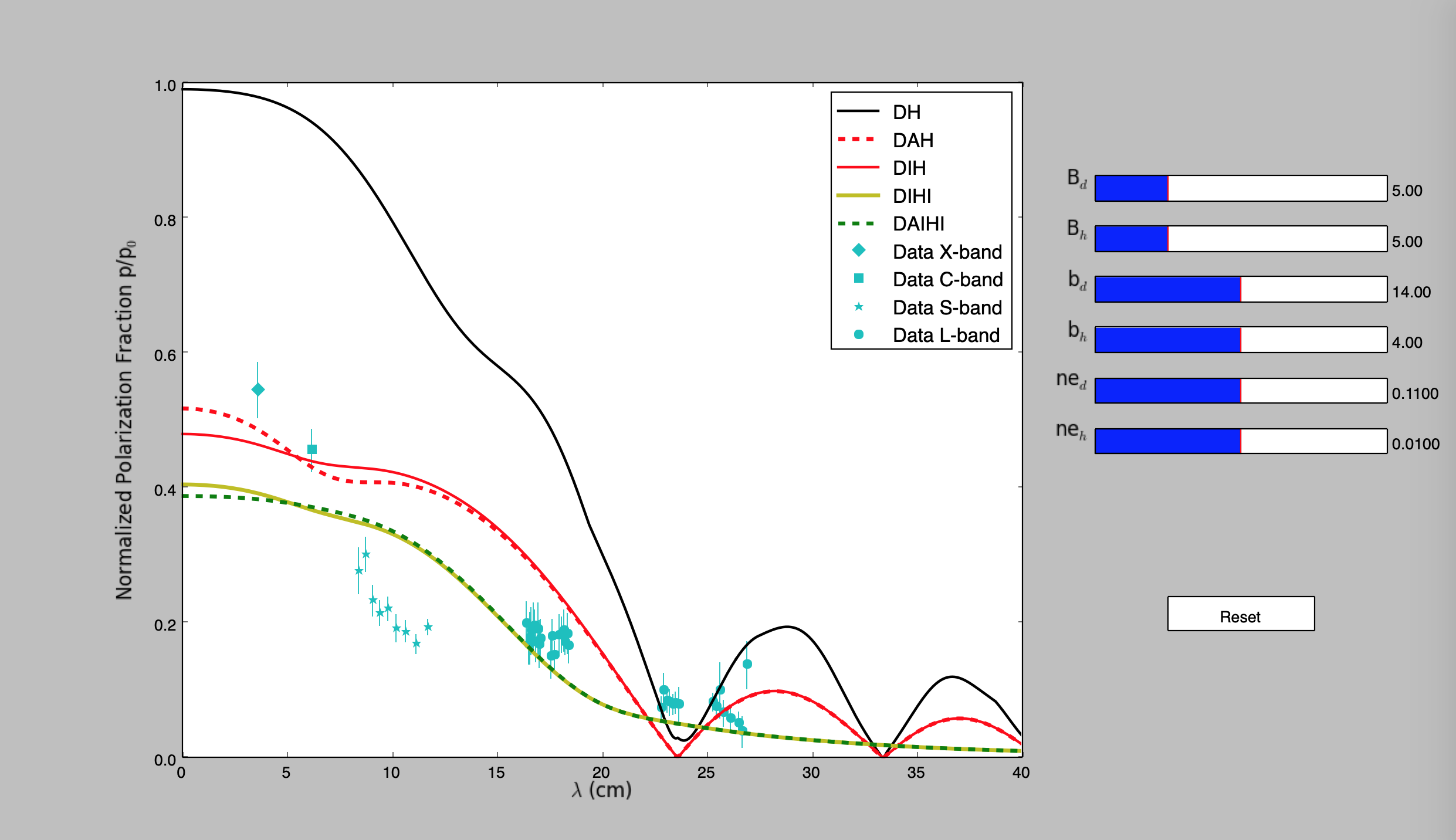}
\caption[Interactive Python tool]{Interactive tool to adjust different model parameters to the data of region ``A'', assuming a two-layer system.} The model configurations are the same as in Figure\,\ref{fig:shneider_plots}. This tool was developed within \textsf{Python} 2.7 Software Foundation (Python Language Reference, version 2.7, available at http://www.python.org) using the module Matplotlib \citep{Hunter:2007}. The code is available on GitHub: \url{https://github.com/MKierdorf/Depoltool.git}.
\label{fig:shneider_interactive}
  \end{figure}

\begin{table*}[]
\centering
\caption{Observed degree of polarization in region ``A'' marked in Figure\,\ref{fig:M51_sectors}.}
\begin{tabular}{lc|lc|lc|lc}
\toprule
\toprule
X-band 		&  			& C-band 	&  			& S-band 	& 	 & L-band & \\ 
$\lambda$ (cm) & $\left(p/p_0\right)$ & $\lambda$ (cm) & $\left(p/p_0\right)$ & $\lambda$ (cm) & $\left(p/p_0\right)$ & $\lambda$ (cm) & $\left(p/p_0\right)$ \\ 
\midrule
3.59 & 0.54$\,\pm\,$0.04 & 6.18 & 0.46$\,\pm\,$0.03 & \phantom{1}8.41 	& 0.28$\,\pm\,$0.04 & 16.41 & 0.20$\,\pm\,$0.04 \\ 
 &&&										& \phantom{1}8.73 	& 0.30$\,\pm\,$0.03 & 16.48 & 0.17$\,\pm\,$0.02 \\ 
 &&&										& \phantom{1}9.07 	& 0.23$\,\pm\,$0.02 & 16.55 & 0.18$\,\pm\,$0.02 \\ 
 &&&										& \phantom{1}9.43 	& 0.21$\,\pm\,$0.02 & 16.63 & 0.19$\,\pm\,$0.02 \\ 
 &&&										& \phantom{1}9.83 	& 0.22$\,\pm\,$0.02 & 16.70 & 0.19$\,\pm\,$0.02 \\ 
 &&&										& 10.17 			& 0.19$\,\pm\,$0.02 & 16.78 & 0.19$\,\pm\,$0.04 \\ 
 &&&										& 10.64 			& 0.19$\,\pm\,$0.02 & 16.85 & 0.17$\,\pm\,$0.02 \\ 
 &&&										& 11.14 			& 0.17$\,\pm\,$0.02 & 16.93 & 0.19$\,\pm\,$0.01 \\ 
 &&&										& 11.70 			& 0.19$\,\pm\,$0.01 & 17.00 & 0.17$\,\pm\,$0.03 \\ 
 &&&										&				&				& 17.08& 0.18$\,\pm\,$0.02 \\ 
 &&&										&				&				& 17.56 & 0.15$\,\pm\,$0.02 \\ 
 &&&										&				&				& 17.65 & 0.18$\,\pm\,$0.02 \\ 
 &&&										&				&				& 17.73 & 0.15$\,\pm\,$0.02 \\ 
 &&&										&				&				& 17.98 & 0.18$\,\pm\,$0.03 \\ 
 &&&										&				&				& 18.07 & 0.18$\,\pm\,$0.02 \\ 
 &&&										&				&				& 18.16 & 0.19$\,\pm\,$0.03 \\ 
 &&&										&				&				& 18.25 & 0.17$\,\pm\,$0.03 \\ 
 &&&										&				&				& 18.34 & 0.18$\,\pm\,$0.02 \\ 
 &&&										&				&				& 18.43 & 0.16$\,\pm\,$0.03 \\ 
 &&&										&				&				& 22.80 & 0.07$\,\pm\,$0.03 \\ 
 &&&										&				&				& 22.94 & 0.10$\,\pm\,$0.03 \\ 
 &&&										&				&				& 23.08 & 0.08$\,\pm\,$0.02 \\ 
 &&&										&				&				& 23.22 & 0.08$\,\pm\,$0.03 \\ 
 &&&										&				&				& 23.37 & 0.08$\,\pm\,$0.03 \\ 
 &&&										&				&				& 23.51 & 0.08$\,\pm\,$0.03 \\ 
 &&&										&				&				& 23.66 & 0.08$\,\pm\,$0.04 \\ 
 &&&										&				&				& 25.26 & 0.08$\,\pm\,$0.04 \\ 
 &&&										&				&				& 25.43 & 0.08$\,\pm\,$0.03 \\ 
 &&&										&				&				& 25.60 & 0.10$\,\pm\,$0.02 \\ 
 &&&										&				&				& 25.78 & 0.07$\,\pm\,$0.04 \\ 
 &&&										&				&				& 26.14 & 0.06$\,\pm\,$0.04 \\ 
 &&&										&				&				& 26.51 & 0.05$\,\pm\,$0.04 \\ 
 &&&										&				&				& 26.70 & 0.04$\,\pm\,$0.03 \\ 
 &&&										&				&				& 26.89 & 0.14$\,\pm\,$0.03 \\ 
 \bottomrule
\end{tabular}
\tablefoot{The non-thermal and polarized intensity values to calculate the degree of polarization were averaged in a region with an azimuthal angle centered at 100$\degr$, an opening angle of 20$\degr$, and radial boundaries of 2.4\,--\,3.6\,kpc (see \citealt{Fletcher11}).}
\label{tab:datapoints}
\end{table*}


\end{document}